\documentclass[prd,12pt,reprint, twocolumn,noshowpacs,nofootinbib,amsmath,amssymb,superscriptaddress,preprintnumbers]{revtex4-1}

\usepackage{graphicx,color}
\usepackage{caption}
\usepackage{amssymb,amsmath,mathrsfs}
\usepackage{cancel}
\usepackage{subfig}
\usepackage{hyperref}
\usepackage{verbatim,slashed}
\usepackage{xspace}

\newcommand{\beq}{\begin{equation}}
\newcommand{\eeq}{\end{equation}}
\newcommand{\be}{\begin{eqnarray}}
\newcommand{\ee}{\end{eqnarray}}

\newcommand{\bi}{\begin{itemize}}
\newcommand{\ei}{\end{itemize}}
\newcommand{\benum}{\begin{enumerate}}
\newcommand{\eenum}{\end{enumerate}}

\definecolor{greenp1}{rgb}{0, 0.8, 0}

\def\lsim{\mathrel{\rlap{\lower4pt\hbox{\hskip1pt$\sim$}}
    \raise1pt\hbox{$<$}}}
\def\gsim{\mathrel{\rlap{\lower4pt\hbox{\hskip1pt$\sim$}}
    \raise1pt\hbox{$>$}}}

\begin{document}
\title{Hidden-Sector  Neutrinos and Freeze-In Leptogenesis}
\author{Ina Flood}
\affiliation{Harvey Mudd College, 301 Platt Blvd., Claremont, CA 91711, USA}
\affiliation{University of Maryland, 1490 Regents Dr., College Park, MD 20742, USA}
\author{Rafael Porto}
\affiliation{Harvey Mudd College, 301 Platt Blvd., Claremont, CA 91711, USA}
\author{Jane Schlesinger}
\affiliation{Harvey Mudd College, 301 Platt Blvd., Claremont, CA 91711, USA}
\author{Brian Shuve}
\affiliation{Harvey Mudd College, 301 Platt Blvd., Claremont, CA 91711, USA}
  \author{Maxwell Thum}
\affiliation{Harvey Mudd College, 301 Platt Blvd., Claremont, CA 91711, USA}

\date{\today}

\begin{abstract}
Sterile neutrinos at the GeV scale can resolve several outstanding problems of the Standard Model (SM), such as the source of neutrino masses and the origin of the baryon asymmetry through freeze-in leptogenesis, but they can be challenging to detect experimentally due to their small couplings to SM particles. In extensions of the SM with new interactions of the sterile neutrinos, they can be produced copiously at accelerators and colliders. We systematically investigate the impact of such novel interactions on the asymmetry from freeze-in leptogenesis. We find that the interactions tend to bring the sterile neutrinos into equilibrium at early times, leading to a significant reduction in the generated asymmetry. We also show that observable rates of several hidden-sector neutrino signatures, such as SM Higgs decays to pairs of sterile neutrinos, can be inconsistent with the observed baryon asymmetry and provide an opportunity to falsify freeze-in leptogenesis.
\end{abstract}
\maketitle

  \section{Introduction}
  \label{sec:introduction}
  
Leptogenesis is a popular solution to the problem of the matter-antimatter asymmetry, in part because it simultaneously accounts for the observed masses of Standard Model (SM) neutrinos along with the baryon asymmetry \cite{Fukugita:1986hr}. In the original proposal, thermal leptogenesis is difficult to test because the  right-handed neutrinos (RHNs) responsible for generating the baryon asymmetry have masses $\gtrsim10^9$ GeV \cite{Davidson:2002qv}, well beyond the reach of current or planned experiments. However, there exist many models, such as the neutrino minimal SM ($\nu$MSM) \cite{Asaka:2005pn} and resonant leptogenesis \cite{Pilaftsis:1997jf,Pilaftsis:2003gt}, that can simultaneously account for neutrino masses, baryogenesis, and even dark matter (DM) with all new states lying at or below the weak scale. This has led to a resurgence of experimental and phenomenological studies of RHNs at the GeV--TeV scales (for recent reviews, see, \emph{e.g.}, Refs.~\cite{Deppisch:2015qwa,Chun:2017spz,Beacham:2019nyx}), including at hadron colliders \cite{LHCb:2014osd,Shuve:2016muy,Curtin:2018mvb,Kling:2018wct,CMS:2018iaf,ATLAS:2019kpx,LHCb:2020wxx}, electron-positron colliders \cite{DELPHI:1996qcc,Liventsev:2013zz}, and accelerators \cite{NA3:1985yvr,CHARM:1985nku,Bernardi:1987ek,Britton:1992xv,NuTeV:1999kej,PIENU:2011aa,E949:2014gsn,Alekhin:2015byh,NA62:2017qcd}.
  
 While GeV-scale RHNs are kinematically accessible at many experiments, the simplest implementation of the see-saw mechanism of neutrino masses \cite{Minkowski:1977sc,Mohapatra:1979ia,GellMann:1980vs,Yanagida:1980xy,Schechter:1980gr,Schechter:1981cv} predicts tiny couplings ($\sim10^{-7}$) between the RHNs and SM particles. Even the highest-intensity upcoming experiments do not have the luminosities needed to probe such tiny couplings \cite{Beacham:2019nyx}. Fortunately, many theories featuring sub-weak-scale RHNs predict new or modified interactions of the RHNs that improve the prospects for discovery. These include models of left-right gauge symmetry \cite{Pati:1974yy,Mohapatra:1974hk,Mohapatra:1974gc,Senjanovic:1975rk,Senjanovic:1978ev}, gauged baryon minus lepton ($B-L$) number \cite{Marshak:1979fm,Mohapatra:1980qe,Wetterich:1981bx,Masiero:1982fi,Buchmuller:1991ce} with spontaneous breaking of $B-L$ to generate the RHN Majorana masses, or more general hidden sectors involving RHNs \cite{Ballett:2019pyw,Ballett:2019cqp}. Even in the $\nu$MSM, a large part of the viable parameter space requires larger Yukawa couplings of the RHNs than is expected from the Type-I see-saw mechanism \cite{Canetti:2012kh,Eijima:2018qke,Klaric:2021tdt}, suggesting some approximate lepton number symmetry \cite{Shaposhnikov:2006nn}. If the breaking of this symmetry is dynamical, we once again expect  interactions between the RHNs and new scalar degrees of freedom. These new interactions modify the phenomenology of RHNs, often providing new avenues for discovering the physics of neutrino mass.
  
  However, new interactions between RHNs and other states change the dynamics of the RHNs in the early universe. In particular, successful theories of baryogenesis require a departure from thermal equilibrium \cite{Sakharov:1967dj,Weinberg:1979bt}. In models with RHNs below the weak scale, the baryon asymmetry originates primarily from the mechanism of freeze-in leptogenesis via RHN oscillations, also known as  Akhmedov-Rubakov-Smirnov (ARS) leptogenesis \cite{Akhmedov:1998qx,Asaka:2005pn}. For freeze-in leptogenesis, it is typically assumed that RHNs are absent after reheating, and they remain out of equilibrium for almost the entire cosmic history down to the weak scale due to their tiny Yukawa couplings with SM neutrinos. This provides a very long time for the RHN production, oscillation, and scattering needed to generate a substantial asymmetry. If, however, there are new interactions involving the RHNs, they can be brought into equilibrium far earlier than they otherwise would, inhibiting the generation of an asymmetry. In other words, the same new couplings that improve the discovery prospects for RHNs could also invalidate their role in generating a lepton asymmetry.
  
In this paper, we analytically and numerically demonstrate that the asymmetry from freeze-in leptogenesis can be severely curtailed depending on the equilibration time, $t_{\rm eq}$, of the RHNs. In particular, the bulk of the asymmetry in ARS leptogenesis is generated on the time scale of oscillations among the RHN mass eigenstates, $t_{\rm osc}\sim E/\Delta M^2$ \cite{Akhmedov:1998qx,Asaka:2005pn}, where $E$ is the energy of a particular coherent superposition of RHN mass eigenstates and $\Delta M^2$ is the squared-mass difference. If $t_{\rm eq} < t_{\rm osc}$, the generation of a lepton asymmetry is greatly suppressed by a fifth-power dependence on the scattering rate of RHNs. This greatly diminishes the possibility of obtaining the observed baryon asymmetry. Conversely, if $t_{\rm eq} > t_{\rm osc}$  then asymmetry generation is not inhibited by the new RHN interactions, and the hidden-sector predictions are identical to the minimal ARS scenario.

Because of the severe suppression of the baryon asymmetry in scenarios where $t_{\rm eq}<t_{\rm osc}$, the couplings between RHNs and hidden-sector couplings must be small to generate a sufficient asymmetry. We study in detail a particular hidden-sector model consisting of a singlet scalar, $\phi$, with a SM-Higgs-portal quartic coupling $\lambda$  and a Yukawa coupling $y$ to the RHNs. Over a range of hidden-sector particle masses and coupling hierarchies, we find that baryogenesis requires $y\sqrt{\lambda}\lesssim2\times10^{-5}$ (see Fig.~\ref{asymmetry_fulleq_optimized} for our final result). This places an upper bound on the magnitude of certain hidden-sector signals of RHNs, including SM Higgs decay to RHN pairs, such that an observable signal at the Large Hadron Collider (LHC) could falsify freeze-in leptogenesis as an explanation for the baryon asymmetry. By contrast, other channels such as $h\rightarrow \phi\phi\rightarrow$ 4 RHNs can be detected without interfering with leptogenesis.  Although we focus  on a particular hidden-sector model, we expect that the analytic results and numerical methods we have developed should extend straightforwardly to any hidden-sector coupling to RHNs.

Several earlier works have considered the impact of hidden-sector interactions on ARS and GeV-scale leptogenesis. Ref.~\cite{Caputo:2018zky} examined the effects of RHN equilibration due to interactions with gauge bosons and scalars in a gauged $\mathrm{U}(1)_{B-L}$ model, finding that it was possible to obtain freeze-in sterile neutrino DM while maintaining the viability of leptogenesis. The authors presented modifications to the quantum kinetic equations for leptogenesis that are analogous to those we derive, provided estimates of equilibration timescales that can be used to identify parameters  for which leptogenesis is unsuppressed, and studied the effects of RHN thermalization on some benchmark points. This study was generally focused on smaller couplings than those we consider in the present work. Ref.~\cite{Heeck:2016oda} similarly considered a gauged $\mathrm{U}(1)_{B-L}$ model and derived relatively simple and conservative bounds on the hidden-sector parameters by requiring sufficient baryon asymmetry from  freeze-out, but not freeze-in, leptogenesis. Finally, Ref.~\cite{Escudero:2021rfi} sketched out some estimates for parameters in the singlet Majoron model that would avoid spoiling leptogenesis due to RHN equilibration, although some of their stated conditions are overly conservative. To our knowledge, we perform the first comprehensive study of the parametric suppression of the freeze-in lepton asymmetry due to RHN equilibration including the effects of equilibration within the hidden sector, which allows us to make definitive statements about the parameters consistent with leptogenesis and the consequent phenomenological implications.

Following a review of the relevant dynamics and time scales of the ARS mechanism, we analytically investigate the suppression of the lepton asymmetry due to RHN equilibration in Sec.~\ref{sec:newinteractions}. We provide details of our scalar-RHN hidden-sector model in Sec.~\ref{sec:darkhiggs}, and in Sec.~\ref{sec:phi_in_eq} we study the effects of RHN equilibration in a particular limit that facilitates comparison with our analytic results from Sec.~\ref{sec:newinteractions}, namely assuming that the dark scalar is always in thermal equilibrium. In Sec.~\ref{sec:phi_not_in_eq}, we provide a full treatment of the equilibration of the hidden sector, allowing us to study leptogenesis for all model parameters. Finally, we turn to the model phenomenology in Sec.~\ref{sec:pheno}, delineating the parts of parameter space in which a  discovery would  imply that the baryon asymmetry could not originate through freeze-in leptogenesis, and the parts of parameter space in which the leptogenesis predictions are equivalent to those of ARS. 
  
  \section{Hidden-Sector Interactions and Freeze-In Leptogenesis}
  \label{sec:newinteractions}

  \subsection{Review of ARS Leptogenesis}
  \label{sec:ARS_review}

  We first review the main results of freeze-in leptogenesis, focusing on the time scales of asymmetry generation that will be important in assessing the effects of new interactions on leptogenesis. In the   $\nu$MSM, the SM is supplemented with two RHNs\footnote{The $\nu$MSM includes three RHNs in total, only two of which play a role in leptogenesis. We therefore focus on the case with  two RHNs. The addition of a third RHN can somewhat expand the parameter space for leptogenesis \cite{Drewes:2012ma,Abada:2018oly,Drewes:2021nqr}.}, $N_I$, each of which has a Majorana mass $M_I$. The new terms in the Lagrangian are
  \be
  \mathcal{L}_N &=& -F_{\alpha I}\, \overline{L}_\alpha (\epsilon H^*) N_I - \frac{M_I}{2} \overline{N}_I^{\rm c} N_I + \mathrm{h.c.},
  \ee
  where $H$ is the SM Higgs field, $L_\alpha$ is the lepton doublet of flavor $\alpha$, and we have written the Lagrangian in the mass-diagonal basis for $N_I$. This Lagrangian implements the Type I see-saw mechanism \cite{Minkowski:1977sc,Mohapatra:1979ia,GellMann:1980vs,Yanagida:1980xy,Schechter:1980gr,Schechter:1981cv} with SM neutrino mass matrix $m_{\alpha\beta} = v^2(FM^{-1}F^{\rm T})_{\alpha\beta}/2$, where $v=246$ GeV is the SM Higgs vacuum expectation value (VEV). We use the Casas-Ibarra parametrization for the $F$ couplings \cite{Casas:2001sr} (see Appendix \ref{app:CI_param} for more details). Our discussion of the asymmetry in ARS leptogenesis closely parallels that of Ref.~\cite{Shuve:2020evk}.
  
\begin{figure}
        \includegraphics[width=\columnwidth]{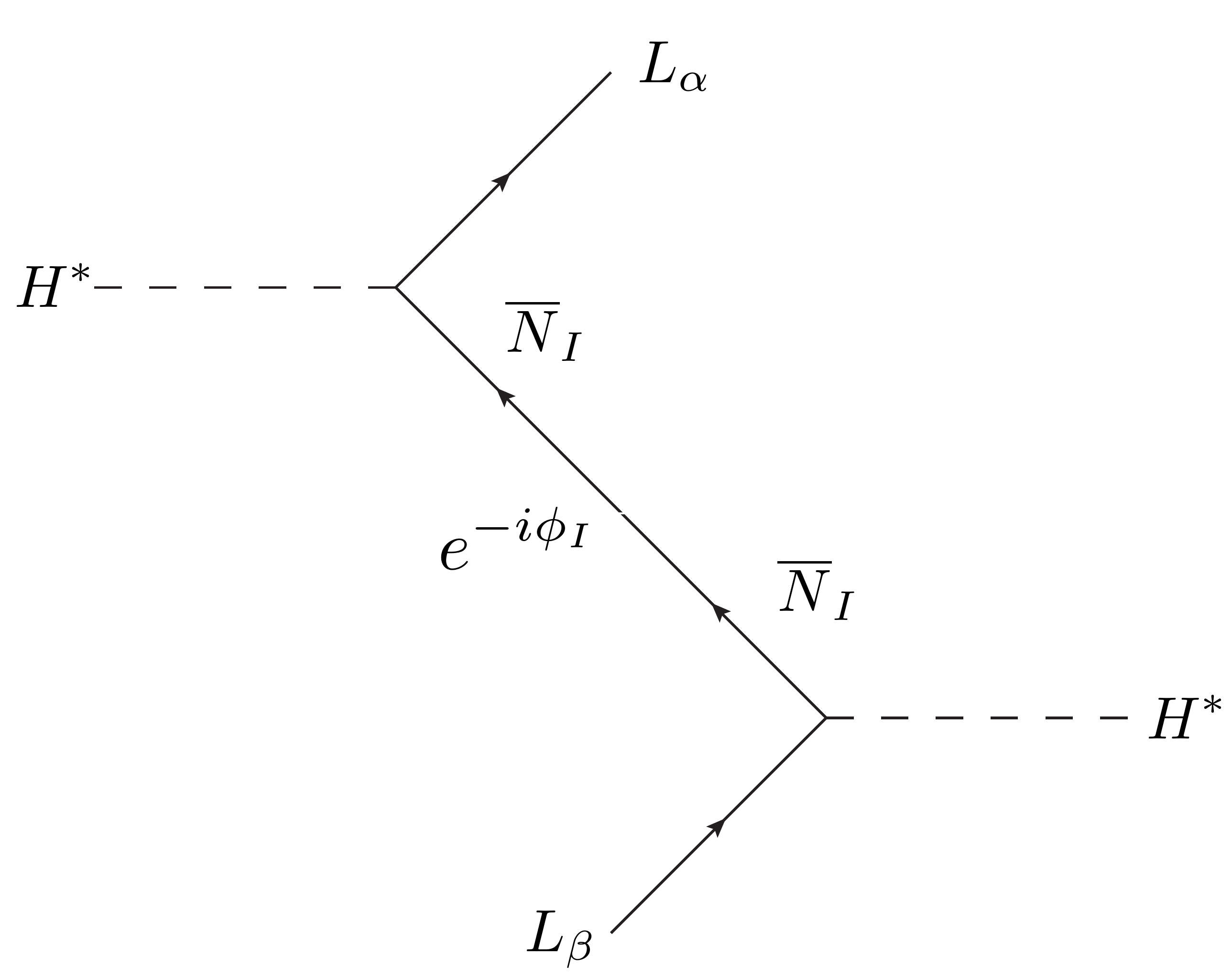}
                     \caption{
Feynman diagram illustrating one of the physical processes underlying freeze-in leptogenesis. The decays of a SM Higgs create coherent superpositions of  RHN mass eigenstates, $N_I$, a subset of which subsequently annihilate back into a SM Higgs. The rate of the net process $L_\beta H^*\rightarrow L_\alpha H^*$ can differ from the $CP$-conjugate process when propagation phases are taken into account, giving rise to asymmetries in individual lepton flavors. These asymmetries are subsequently processed into a total baryon asymmetry by flavor-dependent washout effects.
    }
    \label{fig:feynman_diagram}
\end{figure}
  
  We assume that there are no $N_I$ produced in reheating. Because of the typically small Yukawa couplings $F_{\alpha I}$, the RHNs are slowly produced out of equilibrium in interactions like $H^*\rightarrow \overline{N}_I L_\alpha$. In general, the RHNs are produced in  superpositions of mass eigenstates, and because they are out of equilibrium they propagate coherently, with the amplitude for each mass eigenstate acquiring a phase $e^{-i\phi_I}$, with $\phi_I = \int E_I\,dt$. A small subset of these RHNs inverse decay back into a Higgs field at a later time, $\overline{N}_I L_\beta\rightarrow H^*$, giving rise to a net reaction $L_\beta H^*\rightarrow L_\alpha H^*$ as illustrated in Fig.~\ref{fig:feynman_diagram}. The matrix element of this process goes as
  \small
  \be
  \mathcal{M}(L_\beta H^*\rightarrow L_\alpha H^*) &\propto& F_{\alpha 1}F_{\beta 1}^*e^{-i\phi_1}+F_{\alpha 2}F_{\beta 2}^*e^{-i\phi_2}.
  \ee
  \normalsize
  The matrix element for the $CP$-conjugate process has a change in sign for the phases in the Yukawa couplings, but not for the phases from time evolution because the energy is always positive. The difference in squared matrix elements between the $L_\beta H^*\rightarrow L_\alpha H^*$ and $\overline{L}_\beta H\rightarrow \overline{L}_\alpha H$ processes is
  \be
  |\mathcal{M}(L_\beta H^*\rightarrow L_\alpha H^*)|^2 -   |\mathcal{M}(\overline{L}_\beta H\rightarrow \overline{L}_\alpha H)|^2\nonumber\\
  \propto \mathrm{Im}\left(F_{\alpha 1}F_{\beta 2}F_{\beta 1}^*F_{\alpha 2}^*\right)\sin\left[\int (E_2-E_1)\,dt\right]\label{eq:CPV_ARS},
  \ee
  where the integral is computed between the times of $N_I$ production and subsequent annihilation. We see that, for non-degenerate $N_I$, a $CP$ asymmetry between leptons and antileptons accumulates at $\mathcal{O}(F^4)$ due to RHN oscillations.
  
  The $CP$ asymmetry in scattering rates represented by Eq.~\eqref{eq:CPV_ARS} does not change the total lepton number, since processes like $L_\beta H^*\rightarrow L_\alpha H^*$ conserve overall lepton number. Indeed, summing over all lepton flavors in Eq.~\eqref{eq:CPV_ARS} gives a result of zero. However, the rates are asymmetric in lepton flavor; for example, it may be that $\Gamma(L_e H^*\rightarrow L_\mu H^*) > \Gamma(\overline{L}_e H\rightarrow \overline{L}_\mu H)$, which would result in an excess of muons over antimuons and an equal excess of positrons over electrons. Because of the non-zero SM lepton chemical potentials $\mu_\alpha$ resulting from the lepton flavor asymmetries, subsequent decays $H\rightarrow \overline{L}_\alpha N_I$ occur at a different rate than $H^*\rightarrow L_\alpha \overline{N}_I$ due to differences in Pauli blocking; the net effect is a total lepton number asymmetry arising at $\mathcal{O}(F^6)$, with an equal and opposite asymmetry stored in the RHNs. This total lepton asymmetry is then transferred to a baryon asymmetry via sphalerons, with the final baryon asymmetry determined at the time of sphaleron decoupling at the electroweak phase transition. This is the standard ARS mechanism.
  
The out-of-equilibrium Sakharov condition for baryogenesis must be satisfied in order to generate a $CP$ asymmetry. Therefore, the relevant time scale for our analysis is the time of lepton flavor asymmetry generation established by Eq.~\eqref{eq:CPV_ARS}. The subsequent redistribution of the flavor asymmetries into a total lepton asymmetry occurs even if the $N_I$ are in thermal equilibrium \emph{provided}  the lepton flavor asymmetries have already been generated by the time of RHN equilibration. In other words, if the RHNs come into equilibrium by some new interaction that conserves SM lepton number, the further generation of lepton flavor asymmetries is suppressed after equilibration but any pre-existing flavor asymmetries will not be eradicated. Thus, if we want to determine the impact on leptogenesis of the RHNs coming into equilibrium by some new interaction, the dominant effect is on the generation of lepton flavor asymmetries at $\mathcal{O}(F^4)$ from Eq.~\eqref{eq:CPV_ARS}.

To determine the time scale of flavor asymmetry generation, we assume that the coherent RHN state is ultra-relativistic with momentum $\vec k(t)$ for both mass eigenstates\footnote{This is a standard approximation in the studies of SM neutrino oscillations and gives the correct result provided the initial wave packet is sufficiently broad that the different energy eigenstates do not separate during propagation \cite{Lipkin:2003hj}. Ref.~\cite{Shuve:2020evk} argued that the effects of propagation decoherence are expected to be small for this baryogenesis mechanism.}, and that the oscillation time is long compared to the initial production time of the RHN state. We can then evaluate the oscillation phase, finding
\be\label{eq:osc_phase}
\int_0^t\,dt'\left(E_2-E_1\right) &\approx& \frac{\Delta M_{21}^2}{6k(t)H(t)},
\ee
where $\Delta M_{21}^2 \equiv M_2^2-M_1^2$ and $H(t)$ is the Hubble expansion rate. Since $H = (2t)^{-1}$ in a radiation-dominated universe, we see that the oscillation frequency is $\Delta M_{21}^2/ (3k)$, and hence the oscillation frequency increases as a function of time due to momentum redshift.  

An important time scale for asymmetry generation is the time it takes for the RHN state to oscillate once. It is convenient to change to a dimensionless time variable, 
\be
z\equiv T_{\rm ew}/T(t),
\ee
 where $T$ is the temperature and $T_{\rm ew}\approx131\,\,\mathrm{GeV}$ is the temperature at sphaleron decoupling. To get an estimate for the oscillation time, we perform a thermal average over RHN momenta, using $\langle T/k\rangle = 1/2$ for Maxwell-Boltzmann statistics\footnote{The quantity for Fermi-Dirac statistics differs by 10\%, but we use Maxwell-Boltzmann statistics throughout for consistency in various rates and abundances that we calculate.}.  We then define $z_{\rm osc}$ as the dimensionless time at which Eq.~\eqref{eq:osc_phase} is 1, corresponding to approximately one oscillation:
\be\label{eq:zosc_averaged}
z_{\rm osc} &=& \left(\frac{12T_{\rm ew}^3}{\Delta M_{21}^2M_0}\right)^{1/3},
\ee
where $M_0\approx 7\times10^{17}\,\,\mathrm{GeV}$ is defined so that the Hubble rate is $H(T) = T^2/M_0$. Because the flavor asymmetry generation rate is proportional to the sine of the oscillation phase, the sign of the asymmetry being created changes  when the oscillation phase crosses an integer multiple of $\pi$.
  
 At early times, $z\ll z_{\rm osc}$, the oscillation phase is small and the rate of asymmetry generation is consequently very slow. At late times, $z\gg z_{\rm osc}$, the oscillations become very rapid, and the positive and negative contributions to the asymmetry average to zero. Therefore, the bulk of the asymmetry is created during times $z\sim z_{\rm osc}$. Since the baryon asymmetry is fixed at the time of sphaleron decoupling, $z=1$, the asymmetry is largest if $z_{\rm osc}\sim1$. In this case, the asymmetry generation rate accumulates over the entire age of the universe to the point of sphaleron decoupling. The condition $z_{\rm osc}=1$ implies an optimal squared-mass splitting $\Delta M_{21}^2 \sim\left(10\,\,\mathrm{keV}\right)^2$. For RHNs in the GeV range, this corresponds to a mass degeneracy $\Delta M\equiv M_2-M_1\sim 10^{-10}\,\,\mathrm{GeV}$, which is the reason why most (although not all) implementations of the ARS mechanism feature highly degenerate RHNs\footnote{A mass-degenerate RHN spectrum is not needed if $M_N\approx 10\,\,\mathrm{keV}$. However, if the RHNs are at the keV scale it is impossible to simultaneously obtain the observed baryon asymmetry and SM neutrino masses because the Yukawa couplings are too small. In models where the oscillation states are not RHNs, a non-degenerate spectrum of keV-scale singlets can give the correct baryon asymmetry \cite{Shuve:2020evk}.  }.
  
    \subsection{Right-Handed Neutrino Equilibration}
  \label{sec:RHN_eq}
  
According to the above discussion, the bulk of asymmetry generation occurs in the vicinity of a particular dimensionless time $z_{\rm osc}$. In ARS leptogenesis, the only couplings of the RHNs are to  SM leptons and the Yukawa couplings are sufficiently small that the RHNs do not come into equilibrium until $z\sim1$. As a result,  RHN equilibration does not substantially affect the generation of the asymmetry in the minimal RHN model.

Because ARS leptogenesis is a freeze-in implementation of leptogenesis\footnote{Note that the $\nu$MSM can also generate a lepton asymmetry via freeze out, \emph{i.e.}, in the departure of RHNs from equilibrium due to finite-mass effects \cite{Hernandez:2016kel,Hambye:2016sby,Antusch:2017pkq,Hambye:2017elz,Granelli:2020ysj,Klaric:2021tdt,Drewes:2021nqr}. We return to this possibility in Appendix \ref{sec:freezeout},  discussing the implications of new hidden-sector interactions for low-scale freeze-out leptogenesis.}, its success depends crucially on the fact that the RHNs are not brought into equilibrium before $z_{\rm osc}$. If there exists a new interaction that produces RHNs with rate $\Gamma_N$, then the RHNs come into equilibrium around a dimensionless time $z_{\rm eq}$ defined  by
\be
\frac{\Gamma_N(z_{\rm eq})}{H(z_{\rm eq})} &\equiv& 1.
\ee

In the following sections, we will consider specific models of RHNs coupled to scalar degrees of freedom in which we can compute $z_{\rm eq}$ in terms of model parameters. This allows us to compare $z_{\rm eq}$ and $z_{\rm osc}$ to determine whether the lepton asymmetry is suppressed. For now, however, we will take a more generic approach and stipulate a coupling $\xi\ll 1$ that connects the  hidden sector, including RHNs, to the SM. Because the RHNs have masses well below the electroweak scale and the only dimensionful scale in the SM before electroweak symmetry breaking is the temperature, $T$, we can argue on general grounds that 
\be
\Gamma_N \equiv a_N\,\xi^2 T,
\ee
 where $a_N$ is some dimensionless number that can be calculated from the full theory. For example, the production rate of RHNs from Higgs decays and $2\leftrightarrow2$ scattering in ARS leptogenesis gives a value of $a_N \sim 5\times10^{-3}$ \cite{Besak:2012qm,Drewes:2012ma}. We can now compute the equilibration time,
\be
z_{\rm eq} &=& \frac{T_{\rm ew}}{a_N\xi^2 M_0},
\ee
and from this we determine the ratio of equilibration to oscillation times,
\be\label{eq:ratio_of_times}
\frac{z_{\rm eq}}{z_{\rm osc}} &=& \frac{1}{a_N\xi^2}\left(\frac{\Delta M_{21}^2}{12M_0^2}\right)^{1/3}.
\ee
For the asymmetry to be unaffected by RHN equilibration, we need $z_{\rm eq}/z_{\rm osc}\gtrsim 1$. We thus find
\be
\xi^2 &\lesssim& \frac{1}{a_N}\left(\frac{\Delta M_{21}^2}{12M_0^2}\right)^{1/3}.
\ee
If we take a typical  squared-mass splitting of $\Delta M_{21}^2=(10\,\,\mathrm{keV})^2$ and $a_N=10^{-2}$, the coupling constraint is $\xi\lesssim10^{-7}$, a value so small as to render the hidden sector states nearly unobservable in current experiments. 

The constraint is surprisingly robust:~even if we increase the squared mass splitting to $\Delta M_{21}^2=1\,\,\mathrm{GeV}^2$, which is typically too large to obtain the correct baryon asymmetry in the minimal ARS setup, the constraint only relaxes to $\xi\lesssim10^{-5}$! This preliminary analysis suggests that for almost any new hidden-sector coupled to RHNs, an experimental discovery of the new particle's couplings to RHNs could cause the RHNs to equilibrate before the oscillation time scale, suppressing the lepton asymmetry.

\subsection{Asymmetry Suppression}
\label{sec:asym_suppr}

The equilibration of the RHNs halts the generation of the lepton flavor asymmetries for two  reasons. First, unitarity and $CPT$ conservation dictate that the generation of any $CP$-asymmetry requires at least one species of particle to have a distribution that deviates from the equilibrium value \cite{Weinberg:1979bt}. This property is manifest in the $CP$-violating source terms that result from the calculations of  lepton flavor asymmetries in both the Closed-Time Path (CTP) \cite{Schwinger:1960qe,Keldysh:1964ud,Drewes:2012ma} and density-matrix formalisms \cite{Akhmedov:1998qx,Asaka:2005pn} of non-equilibrium quantum field theory applied to freeze-in leptogenesis. Second, the creation and annihilation of RHNs from hidden-sector interactions leads to decoherence that suppresses the  oscillation phase needed to generate an asymmetry. The deviation of RHN abundances from equilibrium is exponentially damped  with characteristic time scale $\Gamma_N^{-1}$. Consequently, lepton flavor asymmetry generation shuts off rapidly after $z_{\rm eq}$.

To obtain an analytic estimate of the magnitude of asymmetry suppression, we take the quantum kinetic equations for the RHN density matrices (see Section \ref{sec:be_phi_in_eq} and Appendix \ref{app:be}) and employ a perturbative calculation in the Yukawa coupling, $F$, that is valid when the RHNs are far from equilibrium \cite{Asaka:2005pn,Hambye:2017elz,Shuve:2020evk}. The lepton flavor asymmetries at dimensionless time $z$ are proportional to the factor\footnote{For the interested reader, the coefficient between the asymmetry in flavor $\alpha$ and the function $\mathcal{A}(z)$ is $\Delta Y_{\alpha}/\mathcal{A}(z) = 4(\pi^2M_0a_N/T_{\rm ew})^2Y_N^{\rm eq}\mathrm{Im}[F_{\alpha 1}^*F_{\alpha 2}(F^\dagger F)_{21}]$ \cite{Hambye:2017elz}. However, this constant is not important for our parametric study in this section.}
\be\label{eq:perturbative_asymmetry}
\mathcal{A}(z) &=& \int_0^z\,dz_2\int_0^{z_2}\,dz_1\,\sin\left(\frac{z_2^3-z_1^3}{z_{\rm osc}^3}\right),
\ee
which is the imaginary part of the time evolution phase integrated over all RHN production times ($z_1$) and  annihilation times ($z_2>z_1$). This is simply an integral of the collision terms for production and annihilation of RHNs, which are independent of $z$, dressed by the oscillation phase. This perturbative calculation has been performed in detail in the literature \cite{Asaka:2005pn,Hambye:2017elz,Shuve:2020evk} and we refer the reader to these references for a more thorough derivation of Eq.~\eqref{eq:perturbative_asymmetry}.

The argument of the sine function is simply  Eq.~\eqref{eq:osc_phase} recast in dimensionless form. At very late times compared to the oscillation time, $z\gg z_{\rm osc}$, this provides the usual ARS result 
\be\label{eq:osc_ARS}
\mathcal{A}_{\rm ARS}(z\gg z_{\rm osc}) \approx 0.67z_{\rm osc}^{2/3} = 0.67\left(\frac{12T_{\rm ew}^3}{\Delta M_{21}^2M_0}\right)^{2/3},
\ee
such that the lepton flavor asymmetries scale like $(\Delta M_{21}^2)^{-2/3}$. This explains why, in ARS leptogenesis, delaying the onset of oscillations through smaller RHN mass splittings leads to a larger asymmetry:~a longer integration time before the onset of rapid oscillations leads to a larger total asymmetry.

If, however, the RHNs come into equilibrium at $z_{\rm eq}\ll z_{\rm osc}$ the asymmetry generation gets cut off. We employ an instantaneous decoupling approximation, in which we assume asymmetry generation is unaffected by new RHN interactions prior to $z_{\rm eq}$ and completely stopped after $z_{\rm eq}$. For $z_{\rm eq}\gg z_{\rm osc}$, the asymmetry has already saturated and the cutoff has no effect, giving the standard ARS result. By contrast, the lepton flavor asymmetries for $z_{\rm eq} \ll z_{\rm osc}$ are proportional to: 
\be
\mathcal{A}(z_{\rm eq}) &=&  \int_0^{z_{\rm eq}}\,dz_2\int_0^{z_2}\,dz_1\,\sin\left(\frac{z_2^3-z_1^3}{z_{\rm osc}^3}\right)\\
&\approx & \int_0^{z_{\rm eq}}\,dz_2\int_0^{z_2}\,dz_1\left(\frac{z_2^3-z_1^3}{z_{\rm osc}^3}\right)\\
&=& \frac{3z_{\rm eq}^5}{20z_{\rm osc}^3}.
\ee
This can be expressed in terms of physical parameters as
\be\label{eq:ARS_cutoff}
\mathcal{A}(z_{\rm eq}) &=& \frac{T_{\rm ew}^2}{80a_N^5M_0^4}\,\frac{\Delta M_{21}^2}{\xi^{10}}.
\ee
We see that the asymmetry can be suppressed by the tenth power of the coupling $\xi$ between the hidden sector and the SM! Even if the coupling is only slightly larger than the value satisfying $z_{\rm eq}\sim z_{\rm osc}$, there is an enormous suppression of the asymmetry, rendering leptogenesis non-viable. 

We also see that the dependence of the asymmetry on the squared-mass splitting is the opposite of the usual ARS case, Eq.~\eqref{eq:osc_ARS}:~because the RHNs come into equilibrium before oscillations occur, the asymmetry is enhanced by having oscillations occur earlier in time. 
Indeed, the optimal mass splitting is the one giving $z_{\rm osc} \approx z_{\rm eq}$. Using Eq.~\eqref{eq:ratio_of_times}, treating the coupling $\xi$ as a given and setting $z_{\rm osc}=z_{\rm eq}$ predicts an optimal mass-squared splitting of
\be\label{eq:optimal_mass_splitting}
{\Delta M_{21}^2}^{(\rm optimized)} \approx 12 M_0^2 a_N^3 \xi^6.
\ee
Substituting into our expression for the asymmetry, Eq.~\eqref{eq:ARS_cutoff}, gives
\be\label{eq:optimized_asymmetry}
\mathcal{A}(z_{\rm eq})^{\rm(optimized)} &=& \frac{3T_{\rm ew}^2}{20a_N^2M_0^2\xi^4}.
\ee
Even in the best-case scenario where the mass splitting is optimally configured to get the largest asymmetry, we still get a quartic suppression of the asymmetry in the coupling $\xi$.

In deriving these results, we have thermally averaged the oscillation phase prior to calculating the asymmetry. This is typically done to make the quantum kinetic equations for leptogenesis  simpler to solve, and it leads to a single oscillation time, $z_{\rm osc}$, for all RHNs. In reality, there exists a distribution of RHN momenta, each of which oscillates at its own frequency. Consequently, the total asymmetry should be calculated by convolving the momentum-dependent asymmetry with the RHN phase-space distribution. In Appendix \ref{app:momentum_averaging}, we perform a numerical study of the effects of the different momentum-averaging prescriptions. We find that the asymmetry changes by roughly 15\% for the optimized baryon asymmetry with $z_{\rm osc} = z_{\rm eq}$, but varies by a factor of up to 7.5 when $z_{\rm osc}\gg z_{\rm eq}$. However, since the asymmetry has a $\xi^{-10}$ coupling dependence in this limit, our numerical estimates of the couplings needed for successful leptogenesis are only off by about 20\% if we take the simpler approach. Because the quantitative effect is minimal, for the remainder of this work we retain the momentum-averaging prescription for the RHN energies outlined in this section.

\subsection{Summary}

To summarize the results of the last several sections:

  \bi

  \item In ARS leptogenesis, most of the lepton flavor asymmetries are generated on a typical dimensionless time scale $z_{\rm osc}$, which is inversely correlated with the RHN squared-mass splitting:~a smaller splitting gives rise to a later oscillation time.
  
  \item If any new interactions bring the RHNs into equilibrium at a time $z_{\rm eq}$, then the interactions have no effect if the RHN equilibration occurs after the oscillation time, but significantly suppress the asymmetry if equilibration occurs before the oscillation time. To avoid RHN equilibration prior to oscillation,  hidden-sector couplings need to be sufficiently small that experimental detection of the new RHN interactions would be challenging. 
  
  \item Quantitatively, the lepton flavor asymmetry is suppressed by the tenth power of the coupling connecting the hidden sector to the SM for fixed mass splittings, which means that leptogenesis is no longer viable if the RHNs equilibrate prior to the oscillation time. The asymmetry is enhanced for larger RHN squared-mass splittings because this makes the oscillation time earlier and closer to the equilibration time; even if the RHN mass splitting is tuned to maximize the asymmetry, the asymmetry is still suppressed by the fourth power of the coupling connecting the hidden sector and the SM.
  
  \ei
  
  \section{Dark Scalar Model}
  \label{sec:darkhiggs}
  
Many simple and popular models of hidden sectors contain a new dark scalar. This could, for example, be a scalar associated with the spontaneous breaking of lepton number, giving Majorana masses to the RHNs. We will be somewhat agnostic about the full theory, considering a real scalar $\phi$ that couples to the SM and to RHNs according to the following Lagrangian:
\be
\mathcal{L}_\phi &=& -\frac{\lambda}{2}\phi^2|H|^2 - \frac{y_{IJ}}{2} \,\phi \overline{N}^{\rm c}_I N_J\nonumber \\
&&{}-F_{\alpha I}\, \overline{L}_\alpha (\epsilon H^*) N_I  + \mathrm{h.c.}\label{eq:phi_interactions}
\ee
For completeness, we have repeated the Yukawa couplings $F$ between the SM Higgs and the RHNs so that all the relevant model interactions can be found in Eq.~\eqref{eq:phi_interactions}. In the early universe, we now have two main production modes of RHNs:~in SM Higgs boson decays \& $2\leftrightarrow2$ scattering via the  couplings $F_{\alpha I}$, and in $\phi$ decays \& annihilations via the couplings $y_{IJ}$. 

We take $\phi$ to have a tree-level mass, $M_\phi$, as well as thermal contributions to the self-energy from the surrounding medium. For now, we neglect finite-density corrections from the $\phi$ coupling to RHNs since we are most interested in the $\phi$ properties prior to RHN equilibration when there is a negligible RHN density. We also treat the $\phi^4$ self-quartic coupling as small:~because a larger thermal mass for $\phi$ only serves to increase the RHN production rate, taking the self quartic to be small allows us to get conservative bounds on the parameter space\footnote{We have also performed numerical studies verifying that varying the $\phi^4$ quartic coupling does not appreciably change our results.}. We therefore use the following thermally corrected $\phi$ mass,
\be
\overline{M}_\phi^2(T) &=& M_\phi^2 + \frac{\lambda}{6}T^2,
\ee
in calculations where it is relevant, such as the $\phi\rightarrow N_IN_I$ decay rate.

We take the most pessimistic scenario in which no additional asymmetry is produced through the couplings $y_{IJ}$. This is most easily realized by taking the couplings to be real and flavor universal,  $y_{IJ} = y\delta_{IJ}$.

The parameter space of this theory is expansive, and there are several interesting limits. If we subscribe to the hidden-sector paradigm, in which $\phi$ and the RHNs both belong to the hidden sector, we generically expect the coupling $y$ to be large because it connects particles within the hidden sector, whereas the mediator couplings $\lambda$ and $F$ are expected to be tiny. Alternatively, we can imagine a scenario in which the RHNs are truly sterile and have tiny couplings to all fields. In this case, we expect $\lambda$ to be larger than $y$.

 If $\lambda \ll F$, then the RHNs are the dominant mediators between the SM and the hidden sector and $\phi$ plays a negligible role, both cosmologically and phenomenologically. Since this limit simply reduces to the conventional ARS scenario (which typically does not bring the RHNs into equilibrium until close to the time of electroweak symmetry breaking), we do not consider it further. Of much more interest is the scenario $\lambda\gg F$, in which case $\phi$ comes into equilibrium and provides a significant new source of RHN production. The equilibration of RHNs from $\phi$ decays can then suppress the lepton asymmetry according to the arguments of Sec.~\ref{sec:newinteractions}.

Before embarking on our study, we summarize the principal finding of our study:~realizing successful leptogenesis requires that the hidden-sector couplings satisfy $y\sqrt{\lambda}\lesssim2\times10^{-5}$ over a vast swathe of parameter space. While the precise coupling bound depends on the particular scenario, this is a handy rule of thumb for determining the viability of leptogenesis. The optimal baryon asymmetries for a  range of couplings $\lambda$ and $y$  are presented in Fig.~\ref{asymmetry_fulleq_optimized}. 

We begin our numerical explorations of the dark scalar model in Sec.~\ref{sec:phi_in_eq} with a simpler case:~we assume that $\lambda$ is sufficiently large that $\phi$ is always in equilibrium, and $y$ is very small. We  use this relatively simpler scenario to show numerical agreement with the parametric asymmetry suppression predictions from Sec.~\ref{sec:newinteractions}. We then move on in Sec.~\ref{sec:phi_not_in_eq} to the more interesting scenario where both $\phi$ and the RHNs are out of equilibrium at early times.

Throughout, we assume that the RHN masses, $M_I$, are constant. If the RHN masses originate from spontaneous symmetry breaking induced by the scalar $\phi$, then there may be a period of time in the early Universe when $M_I=0$ at tree level. We study this case in Appendix \ref{sec:thermal_masses}, finding that the final results for the optimal baryon asymmetry as a function of the hidden-sector couplings (summarized in Fig.~\ref{asymmetry_fulleq_optimized}) persist even for non-trivial thermal histories of the RHN masses.

\begin{figure*}[t]
        \includegraphics[width=3.05in]{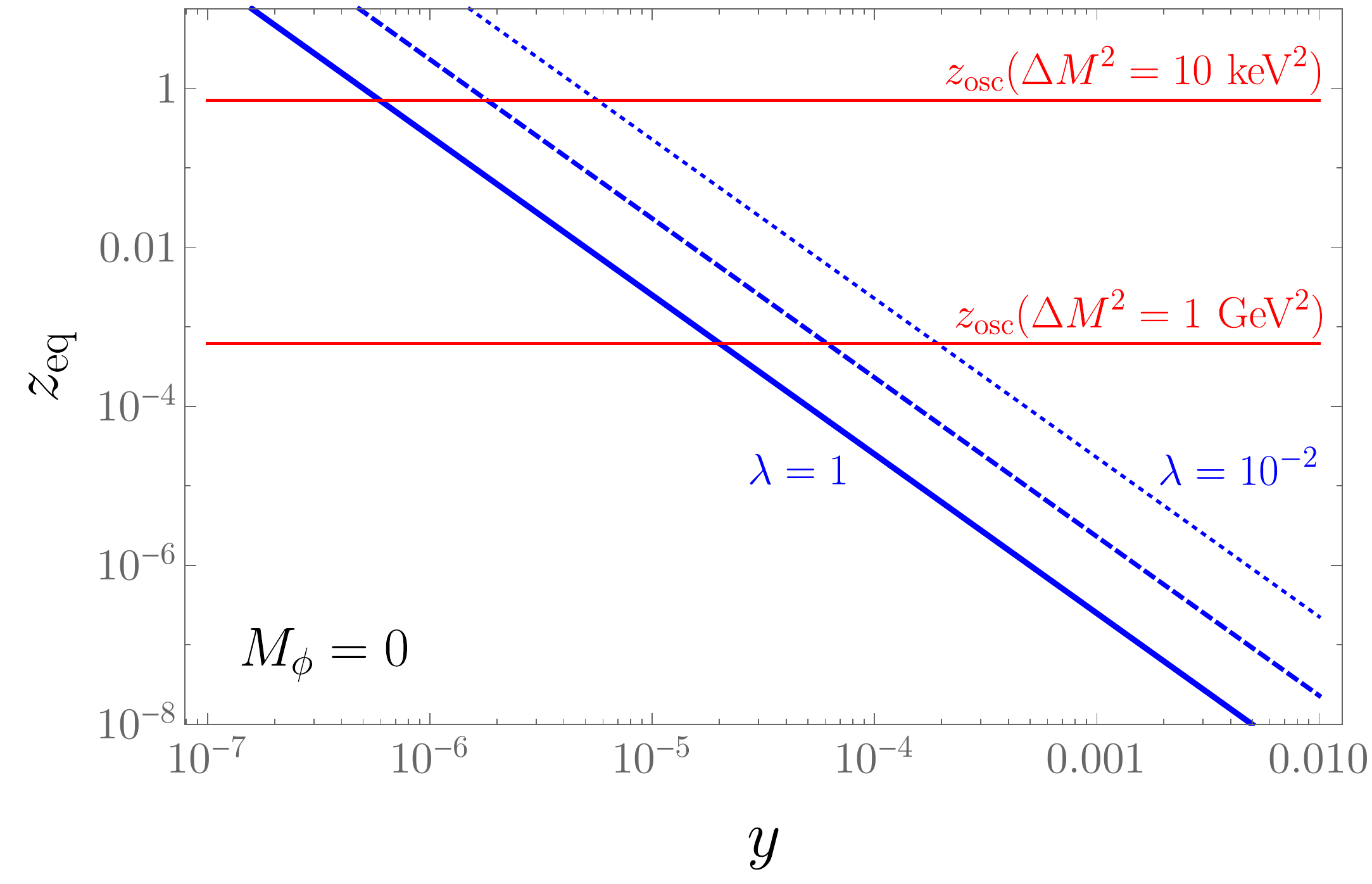}
        \quad\quad\quad
                  \includegraphics[width=3.05in]{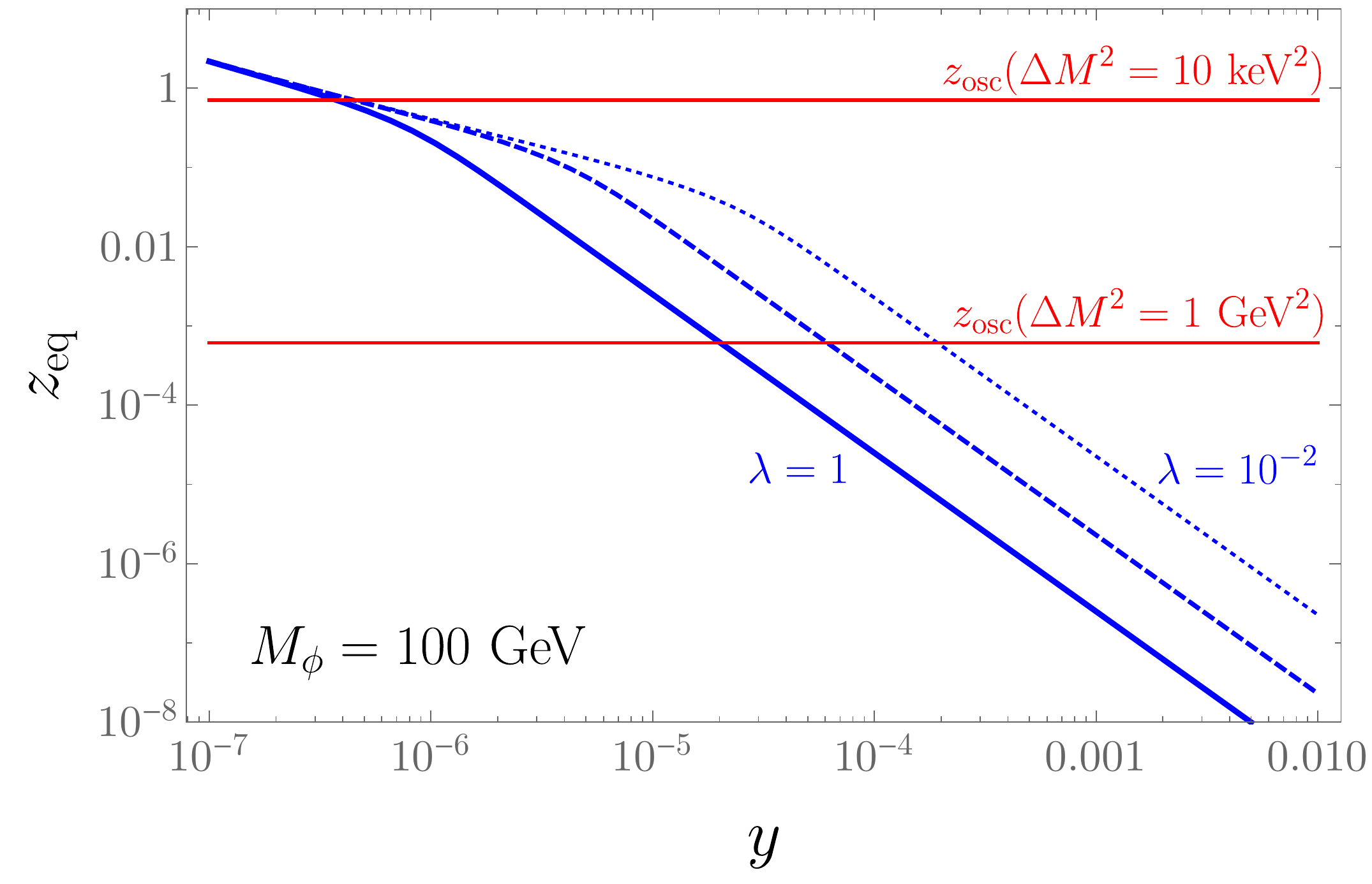}
                     \caption{
Dimensionless equilibration times for the RHNs, $z_{\rm eq}$, as a function of the dark Higgs coupling to RHNs, $y$ and (left panel) $M_\phi=0$; (right panel) $M_\phi=100\,\,\mathrm{GeV}$. We have assumed that $\phi$ is always in equilibrium with the SM. The different blue contours represent different values of the dark Higgs-SM Higgs quartic coupling:~(solid) $\lambda=1$; (dashed) $\lambda=10^{-1}$; (dotted) $\lambda=10^{-2}$. To facilitate comparison with the oscillation time scale $z_{\rm osc}$, we have indicated in red the values of $z_{\rm osc}$ corresponding to RHN squared mass splittings of (top) $\Delta M_{21}^2=10\,\,\mathrm{keV}^2$; (bottom) $\Delta M_{21}^2=1\,\,\mathrm{GeV}^2$. We see that couplings $y\gtrsim10^{-4}$ cause the RHNs to come into equilibrium prior to oscillation and asymmetry generation for the indicated values of $\Delta M_{21}^2$.
    }
    \label{fig:rates_phieq}
\end{figure*}
  
\section{Asymmetry with $\phi$ in Equilibrium}\label{sec:phi_in_eq}
\subsection{Rates \& Quantum Kinetic Equations}\label{sec:be_phi_in_eq}
In this section, we assume that $\phi$ is always in equilibrium with the SM in computing the equilibration of RHNs and determining the resulting effect on leptogenesis. While this approximation does not incorporate the full effects of equilibration within the hidden sector, it allows us to study the effects of RHN equilibration on leptogenesis in a manner that facilitates comparison with analytic results. 

When $\phi$ is abundant in the early universe, there are two significant modes of RHN production:~$\phi\rightarrow N_I N_I$ and $\phi\phi\rightarrow \overline{N}_I N_I$. The first process depends on the $\phi$ mass and can be suppressed if $\overline{M}_\phi$ is small:
\be\label{eq:width_thermalavg}
\left\langle \Gamma_{\phi\rightarrow N_IN_I} \right\rangle&=& \frac{y^2\overline{M}_\phi}{32\pi}\,\frac{K_1(\overline{M}_\phi/T)}{K_2(\overline{M}_\phi/T)},
\ee
where $K_i$ are modified Bessel functions of the second kind and we have taken the limit $\overline{M}_\phi\gg M_I$. Eq.~\eqref{eq:width_thermalavg} gives the partial width into a single RHN flavor and helicity; we separately track the decay rates into $N_I N_I$ and $\overline{N}_I\overline{N}_I$ because the $N_I$ and $\overline{N}_I$ density matrices evolve according to separate quantum kinetic equations. The annihilation of $\phi$ pairs into RHNs is independent of $\overline{M}_\phi$ (up to a logarithmic enhancement in $T/\overline{M}_\phi$ due to forward scattering) but is of higher order in the coupling $y$.  

Using Maxwell-Boltzmann statistics and assuming $T\gg \overline{M}_\phi$, which is expected for dark scalar masses at or below the weak scale, we find the following thermally averaged rates:
\be
\left\langle \Gamma_{\phi\rightarrow N_IN_I} \right\rangle&=& \frac{y^2\overline{M}_\phi^2}{64\pi T},\\
\left\langle \Gamma_{\phi\phi\rightarrow \overline{N}_IN_I}\right\rangle &=& \frac{1.50y^4 T}{64\pi^3}\log\left(\frac{0.850T}{\overline{M}_\phi}\right),
\ee
where $\left\langle \Gamma_{\phi\phi\rightarrow \overline{N}_IN_I}\right\rangle  \equiv n_\phi^{\rm eq}\left\langle \sigma_{\phi\phi\rightarrow \overline{N}_IN_I}v\right\rangle$.

The $\phi$ decay rate is smallest in the limit where the tree-level mass vanishes, $M_\phi=0$, and $\overline{M}_\phi$ is dominated by the thermal correction. In this case, the rates are strictly functions of temperature and we have:
\be
\left\langle \Gamma_{\phi\rightarrow N_IN_I} \right\rangle&\approx& \frac{y^2\lambda T}{384\pi},\label{eq:thermalwidth_N}\\
\left\langle \Gamma_{\phi\phi\rightarrow \overline{N}_IN_I}\right\rangle &\approx& \frac{1.50 y^4 T}{64\pi^3}\log\left(\frac{2.08}{\sqrt{\lambda}}\right).
\ee

Using these rates, we have determined the dimensionless time of equilibration of the RHNs, $z_{\rm eq}$, which is defined as the time at which the total RHN production rate is equal to the Hubble expansion rate. We show these equilibration times in Fig.~\ref{fig:rates_phieq} for two values of the tree-level dark scalar mass, $M_\phi=0$ and 100 GeV, comparing them to  oscillation times $z_{\rm osc}$ spanning the range from Sec.~\ref{sec:ARS_review}. We see that for $\lambda\gg y$, the RHNs can equilibrate before oscillations begin for couplings as small as $y\sim10^{-6}$, in approximate agreement with our arguments in Sec.~\ref{sec:RHN_eq}. Also, the tree-level mass is typically irrelevant for determining the equilibration time scale of the RHNs when $M_\phi\lesssim100$ GeV.   Due to the smallness of $y$, we  find that the $2\leftrightarrow2$ processes are irrelevant compared to the dominant $1\leftrightarrow2$ process for $N$ equilibration, and we do not include them further in this part of our analysis.

The evolution of the lepton asymmetry is determined by solving a set of coupled quantum kinetic equations that simultaneously track the RHN density matrices along with the lepton asymmetry. It is convenient to write the RHN density matrices as $n_N(t)_{IJ} = n_N^{\rm eq}\,R_N(t)_{IJ}$, where $R_N(t)_{IJ}=\delta_{IJ}$ corresponds to $N_I$ being in equilibrium; we assume in this section that the RHNs are sufficiently relativistic that $n_N^{\rm eq}$ can be treated as time independent, although we consider the time-dependence of $n_N^{\rm eq}$ in our study of freeze-out leptogenesis in Appendix \ref{sec:freezeout}.

The initial conditions for the quantum kinetic equations are taken to be $R_N=R_{\overline{N}}=0$, consistent with freeze in,  and vanishing lepton asymmetries. Because we assume that $\phi$ is in equilibrium in this section, the momentum-averaged Boltzmann equations for the RHN density matrices and lepton asymmetry are of the standard form for ARS leptogenesis \cite{Asaka:2005pn,Hernandez:2016kel,Hambye:2017elz,Abada:2018oly}, along with additional terms in the $R_N$ and $R_{\overline N}$ equations that account for RHN production and destruction from $\phi$ (inverse) decays:
\be
\frac{dR_N}{dt} &=& -i\left[\langle H\rangle,R_N\right] - \frac{1}{2}\langle\tilde\Gamma_h\rangle\left\{F^\dagger F,R_N-\mathbb{I}\right\}\nonumber\\
&&{}-\frac{2Y_\phi^{\rm eq}}{Y_N^{\rm eq}}\langle\Gamma_{\phi\rightarrow N_IN_I}\rangle\left(R_N^2-\mathbb{I}\right) + \langle\tilde\Gamma_{\mathrm{w.o.1}}\rangle F^\dagger\mu F \nonumber \\
&&{} - \frac{1}{2}\langle\tilde\Gamma_{\mathrm{w.o.2}}\rangle
\left\{F^\dagger \mu F,R_N\right\},\label{eq:qke_for_lepto}
\ee
where $\langle H\rangle$ is the momentum-averaged RHN Hamiltonian with finite-temperature corrections, and $\langle \tilde \Gamma_{h,\mathrm{w.o.1},\mathrm{w.o.2}}\rangle$ are thermally averaged rates of RHN production from SM Higgs decay/scattering, stripped of coupling constants. These reaction rates take into account both $1\leftrightarrow2$ and $2\leftrightarrow2$ processes involving the RHN coupling $F$ to the SM Higgs and neutrinos \cite{Hernandez:2016kel,Abada:2018oly}, including washout terms\footnote{The two washout rates come from the collision terms that depend on the lepton chemical potential, and differ in whether or not they have a RHN distribution function in the initial state.} that depend on the SM lepton chemical potential normalized to temperature, $\mu$. 

We have derived the term incorporating $\phi\leftrightarrow N_I N_I$ using the CTP formalism \cite{Beneke:2010wd,Drewes:2012ma}, neglecting for simplicity quantum statistics and assuming flavor universality of the $y$ coupling\footnote{The density-matrix dependence of our collision term agrees with that in Ref.~\cite{Caputo:2018zky} in the  limit of flavor-universal coupling and neglecting quantum statistics.}. The factor of two in the collision term for $\phi\rightarrow N_IN_I$ accounts for the fact that two $N_I$ quanta  are produced or destroyed in each collision. 

The quantum kinetic equation for $R_{\overline{N}}$ is the same as that for $R_N$, but with $\mu\rightarrow -\mu$, and $F\rightarrow F^*$. The full set of quantum kinetic equations that we solve, including the equation for the lepton asymmetry and the form of the rates  $\langle \tilde \Gamma_{h,\mathrm{w.o.1},\mathrm{w.o.2}}\rangle$, are provided in Appendix \ref{app:be}. 

\begin{figure*}[t]
        \includegraphics[width=3.05in]{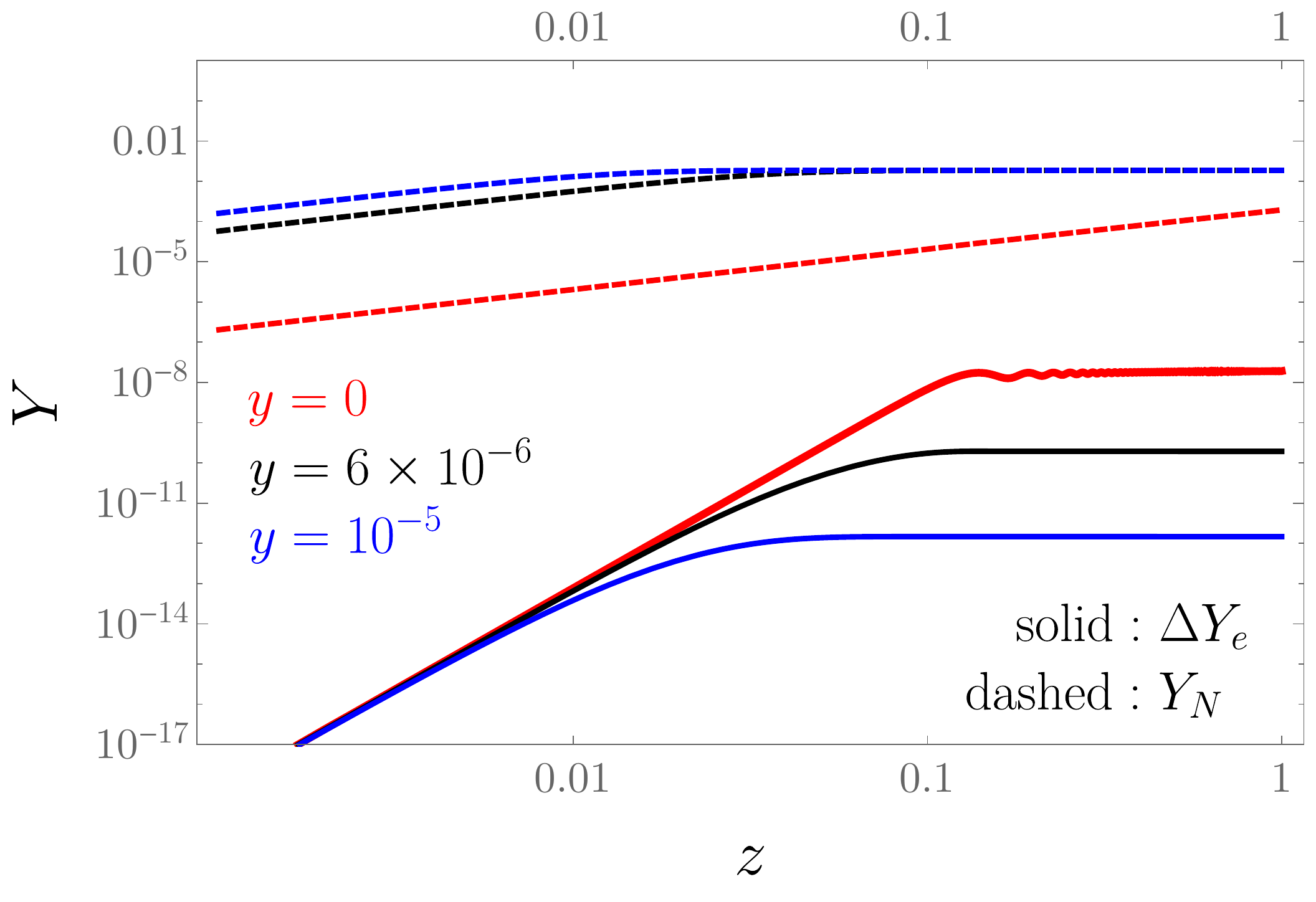}
        \quad\quad\quad
                  \includegraphics[width=3.05in]{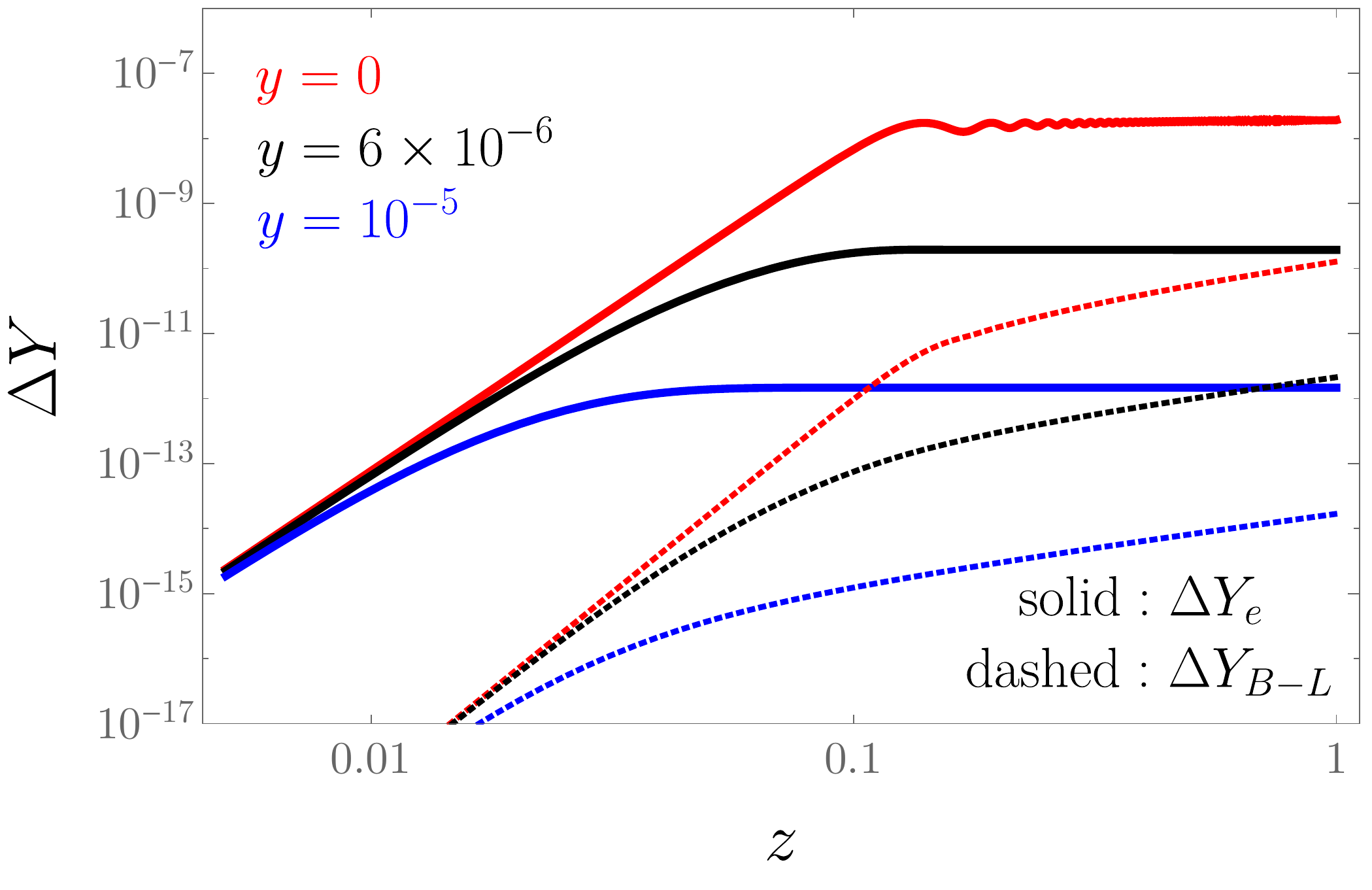}
                     \caption{
Abundances and asymmetries as a function of dimensionless time, $z$, for various couplings $y$ between RHNs and the dark scalar, $\phi$. We have fixed $\lambda=0.1$, $M_\phi=1\,\,\mathrm{GeV}$, $M_1=1\,\,\mathrm{GeV}$, $\Delta M=3\times10^{-8}\,\,\mathrm{GeV}$, and the Yukawa couplings $F^{(I)}$ indicated in Eq.~\eqref{eq:first_yukawa}. 
(Left) The dashed lines indicate $\mathrm{max}(Y_N)$, and the solid lines indicate the asymmetry in the anomaly-free electron number, $\Delta Y_e\equiv \Delta Y_{B/3-L_e}$. We take the following values of $y$:~(red) $y=0$, corresponding to the standard ARS scenario; (black) $y=6\times10^{-6}$; (blue) $y=10^{-5}$. It is evident that  asymmetry generation is suppressed once $Y_N$ comes into equilibrium.
(Right)  Electron flavor asymmetry $\Delta Y_e$ (solid) and total asymmetry $\Delta Y_{B-L}$ (dotted) for various values of $y$:~(top, red) $y=0$; (middle, black) $y=6\times10^{-6}$; (bottom, blue) $y=10^{-5}$. While the flavor asymmetries stop growing when the RHNs equilibrate, the re-processing of the flavor asymmetries into a total $B-L$ asymmetry persists at later times.
    }
    \label{asymmetry_time}
\end{figure*}

In the above treatment of the RHN density matrix evolution, we have neglected lepton-number-violating terms which are typically subdominant over the parameters of interest to us. We examine the effects of lepton number violation and freeze-out leptogenesis in Appendix \ref{sec:freezeout}.

\subsection{Numerical Results}

We begin our numerical study by examining the relationship between RHN equilibration, and the evolution of anomaly-free flavor ($\Delta Y_\alpha \equiv \Delta Y_{B/3-L_\alpha}$) and total $B-L$ number $(\Delta Y_{B-L})$ asymmetries. To do this, we select a benchmark point with $M_\phi=1$ GeV, $\lambda=0.1$, $M_1=1$ GeV, and RHN couplings to the SM Higgs given by
\be\label{eq:first_yukawa}
F^{(I)} &=&
\footnotesize \left(\begin{array}{cc} -0.575+0.608i & -1.141-0.725i \\
-3.461 - 1.047i & 1.909 - 1.764i \\
-1.421 - 1.089i & 2.867 - 0.421i
 \end{array}\right)\times 10^{-8},
\ee
\normalsize
which is consistent with the observed SM neutrino masses and mixings for $M_1\approx M_2\approx 1\,\,\mathrm{GeV}$. We  obtained this coupling matrix using the Casas-Ibarra parametrization \cite{Casas:2001sr} (see Appendix \ref{app:CI_param})\footnote{The parameters going into this matrix are $m_1=0$, $m_2=8.6$ meV, $m_3=58$ meV, $\delta=221^\circ$, $\eta=60^\circ-\delta$.}. We solve the Boltzmann equations for different values of the $\phi$--RHN coupling, $y$, and RHN mass splitting, $\Delta M\equiv M_2-M_1$.

\begin{figure}[t]
                  \includegraphics[width=\columnwidth]{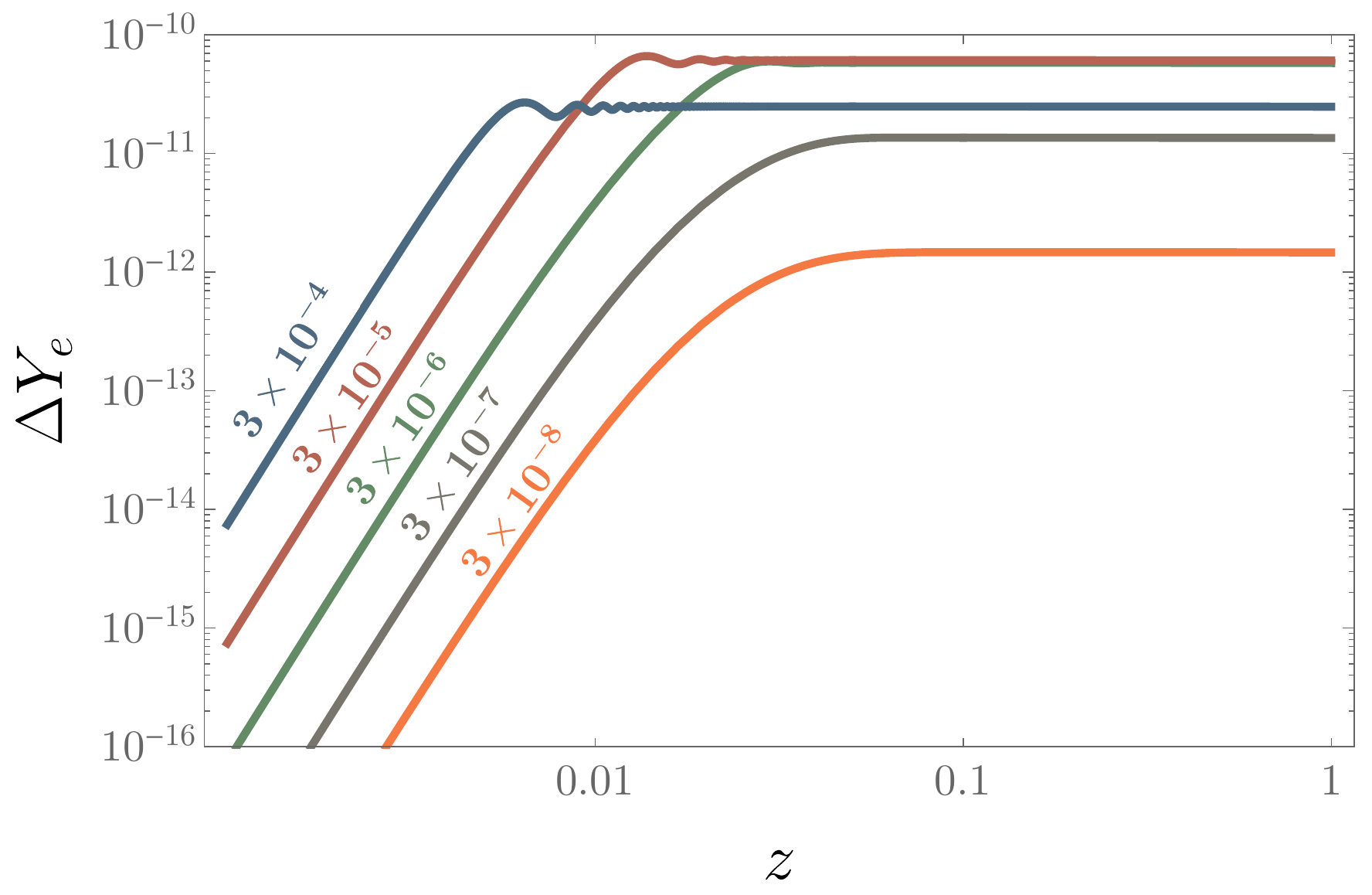}
                     \caption{
Electron flavor asymmetry, $\Delta Y_e$, as a function of dimensionless time, $z$, for various RHN mass splittings $\Delta M$ in GeV as indicated in the figure. We take $y=10^{-5}$ with other parameters the same as in Fig.~\ref{asymmetry_time}. We see that increasing $\Delta M$ enhances the asymmetry at earlier times, which leads to a larger final asymmetry when oscillations are interrupted by RHN equilibration. When $\Delta M$ is sufficiently large that $z_{\rm osc}<z_{\rm eq}$, however,  the RHNs oscillate too early and the asymmetry saturates at smaller values for larger splittings.
    }
    \label{asymmetry_mass_time}
\end{figure}

We show our results in Fig.~\ref{asymmetry_time}. We express abundances in terms of the dimensionless yield, $Y = n/s$, where $s$ is the entropy density. In the left panel, we show the time evolution of $Y_N$ and the electron-flavor asymmetry, $\Delta Y_e$, for $\Delta M = 3\times10^{-8}\,\,\mathrm{GeV}$ and various values of $y$. We choose the electron flavor for concreteness, but the effect on the other flavor asymmetries is analogous. 

We see that when $y=0$, corresponding to minimal ARS leptogenesis, the RHNs are out of equilibrium for all times before the electroweak phase transition, and the  flavor asymmetry saturates around the oscillation time, $z_{\rm osc}$. For larger values of $y$, the RHNs come into equilibrium earlier, and the generation of the flavor asymmetry is suppressed once the RHNs are close to equilibrium. As expected, we  see an absence of oscillation in $\Delta Y_{e}$ when $z_{\rm eq}<z_{\rm osc}$.

In the right panel of Fig.~\ref{asymmetry_time}, we compare the time evolution of the  flavor and total  asymmetries for the same parameters as above. It is evident that the total  asymmetries are smaller than the flavor asymmetries because they arise at higher order in the couplings $F$. We  see that the total asymmetry continues to accumulate after the RHNs equilibrate because the total asymmetry results from a re-processing of the existing lepton flavor asymmetries rather than from a direct source of $CP$-violation. However, the suppression in the flavor asymmetry from RHN equilibration carries over to the  overall normalization of the total asymmetry, giving rise to a comparable reduction in the total $B-L$ asymmetry.

\begin{figure*}
        \includegraphics[width=3.05in]{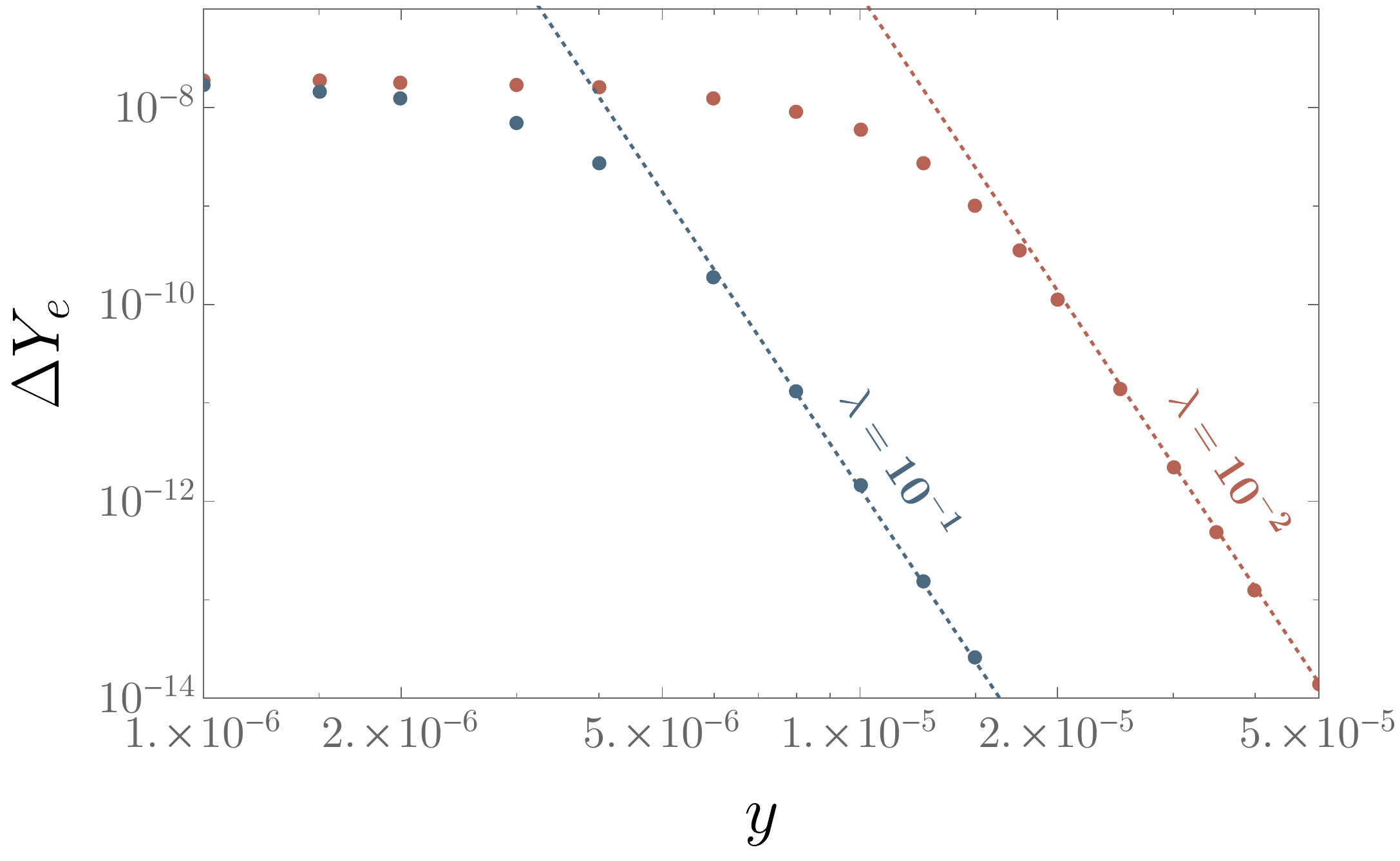}
        \quad\quad\quad
                  \includegraphics[width=3.05in]{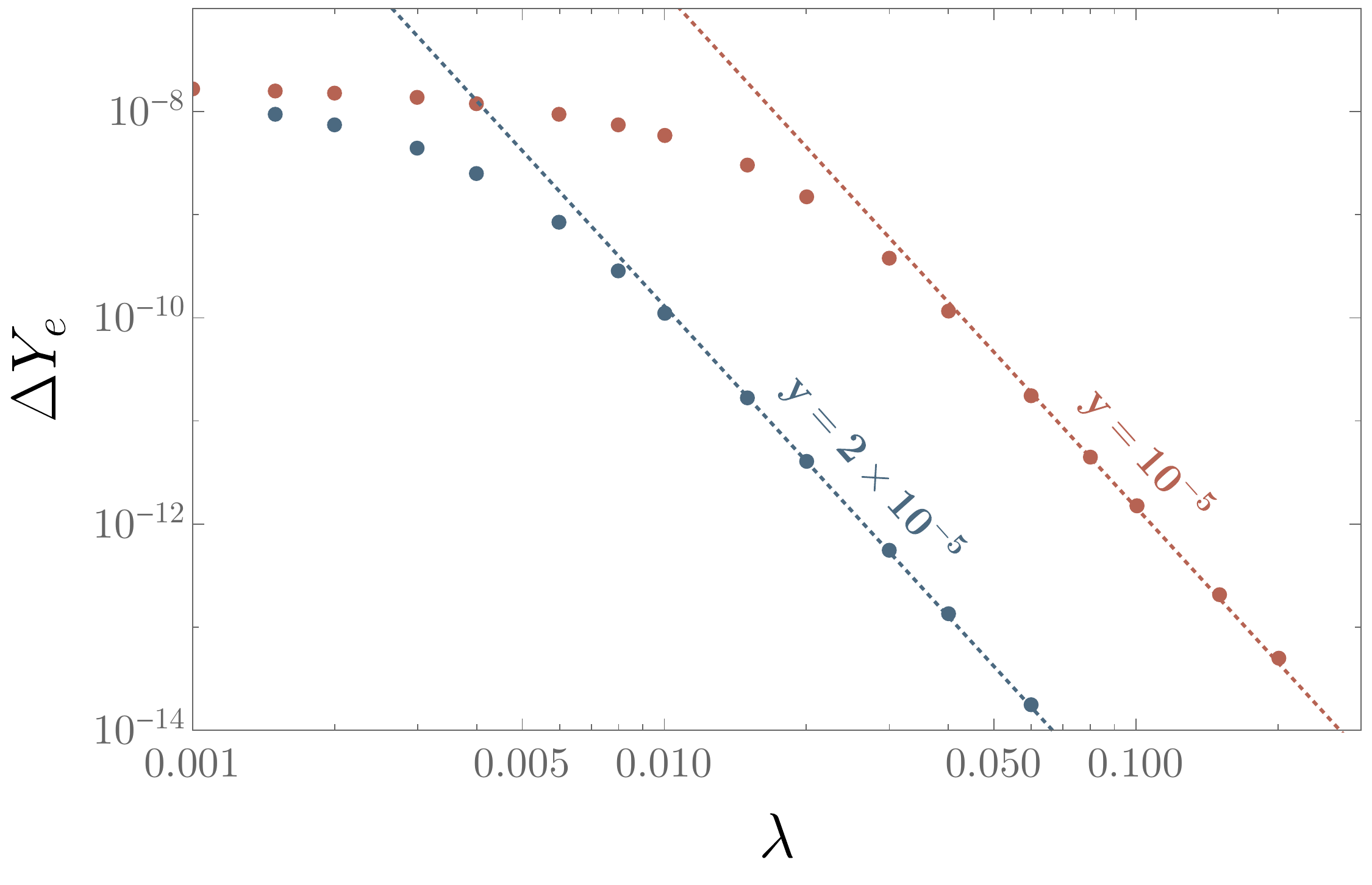}
                     \caption{
Electron flavor asymmetry, $\Delta Y_e$,  shown as a function of (left) $y$ for the indicated values of $\lambda$; (right) $\lambda$ for the indicated values of $y$. Other parameters are the same as in Fig.~\ref{asymmetry_time}. The points are the lepton flavor asymmetries obtained from numerically solving the quantum kinetic equations, while the dashed lines indicate a (left) $y^{-10}$ power-law dependence; (right) $\lambda^{-5}$ power-law dependence.  }
    \label{asymmetry_parameter_dependence}
\end{figure*}

We show the effects of RHN mass splitting on the  lepton flavor asymmetries  in Fig.~\ref{asymmetry_mass_time}. For the smallest mass splittings, oscillation is delayed until after RHN equilibration, $z_{\rm osc}>z_{\rm eq}$, suppressing the asymmetry. As the mass splitting increases, corresponding to an earlier oscillation time, the suppression is less pronounced. This is in qualitative agreement with our estimate in Eq.~\eqref{eq:ARS_cutoff}, where we found that the asymmetry was larger with increased $\Delta M_{21}^2 = M_2^2-M_1^2 \approx 2M_1\Delta M$. Once the mass splitting is sufficiently large that $z_{\rm osc}< z_{\rm eq}$, then we recover the typical ARS scaling where the asymmetry is optimized by delaying oscillations. The optimal asymmetry occurs for $z_{\rm osc}\sim z_{\rm eq}$, which in the case of the benchmark shown in Fig.~\ref{asymmetry_mass_time} corresponds to $\Delta M\sim10^{-5}\,\,\mathrm{GeV}$. We see that the larger mass splitting partly mitigates the suppression of the asymmetry from RHN equilibration, but the optimal electron flavor asymmetry is still orders of magnitude lower than the optimal  asymmetry in the ARS limit from Fig.~\ref{asymmetry_time}.

\begin{figure*}
        \includegraphics[width=3.05in]{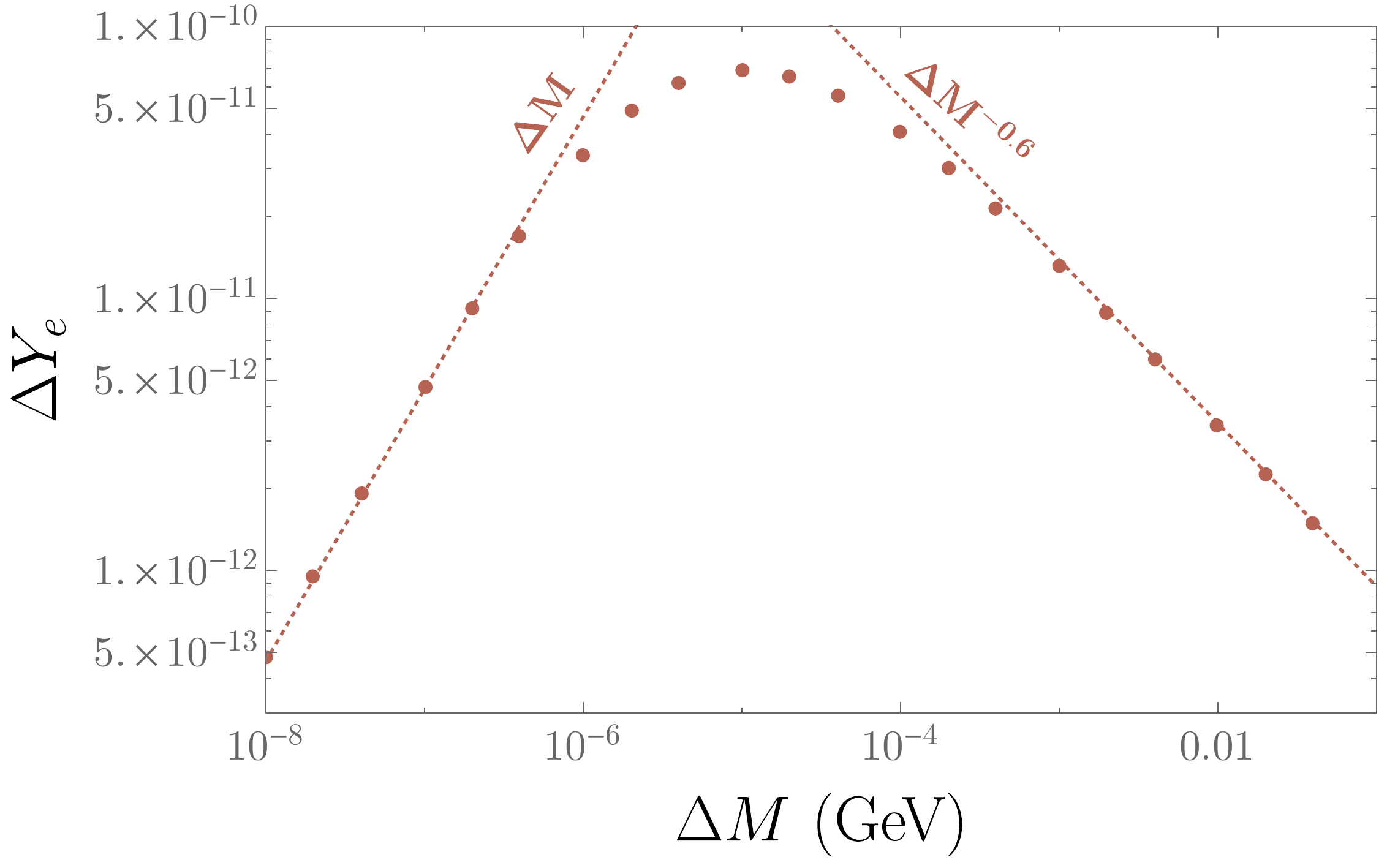}
                \quad\quad\quad
                  \includegraphics[width=3.05in]{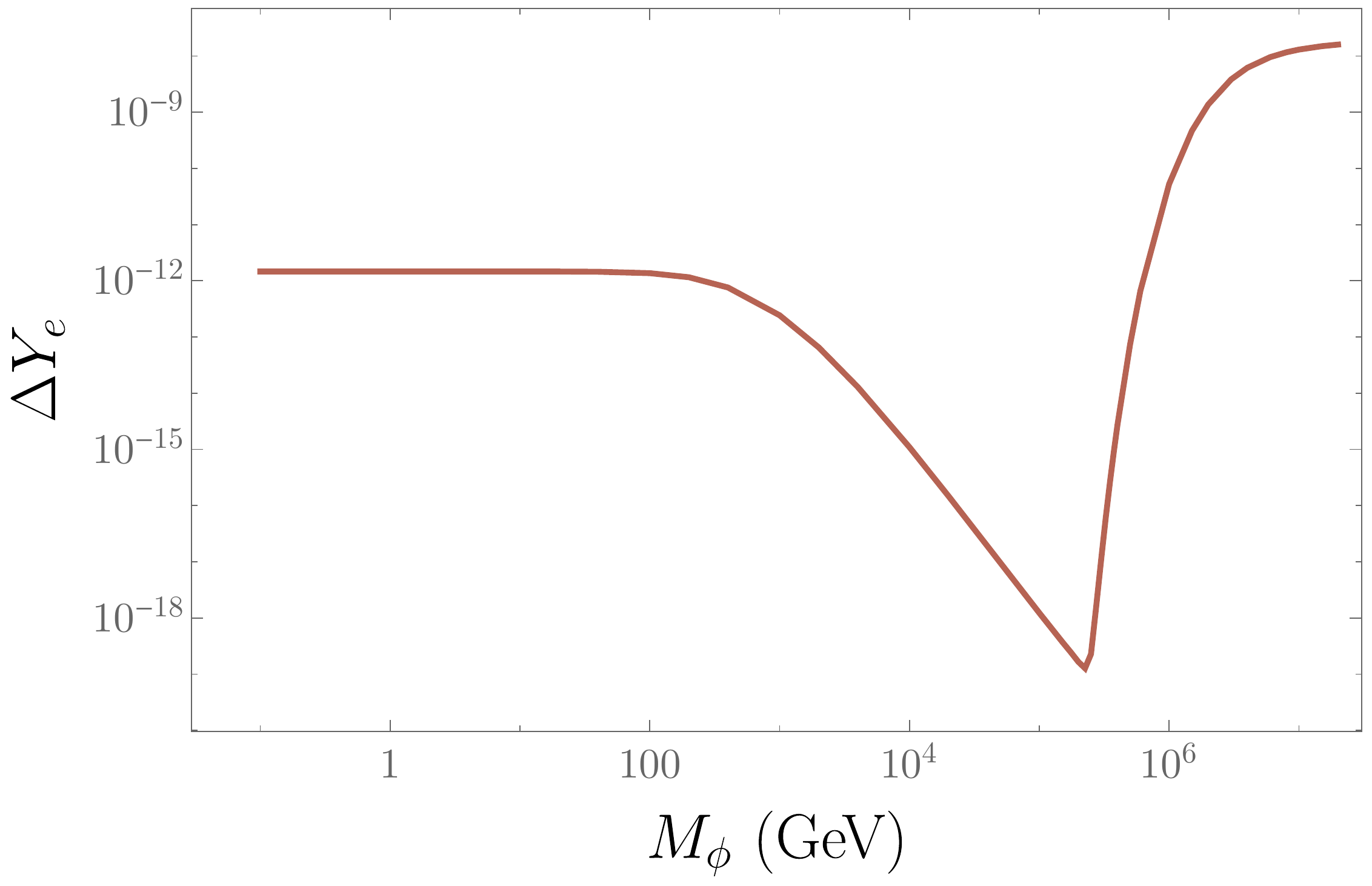}
                     \caption{
(Left) Electron flavor asymmetry, $\Delta Y_e$,  shown as a function of $\Delta M$. We have fixed $M_\phi=1\,\,\mathrm{GeV}$, $M_1=1\,\,\mathrm{GeV}$, $\lambda=0.1$, $y=10^{-5}$, and the Yukawa couplings $F^{(I)}$ indicated in Eq.~\eqref{eq:first_yukawa}. The points are the flavor asymmetries obtained from numerically solving the quantum kinetic equations, while the dashed lines indicate  $\Delta M$ and $\Delta M^{-0.6}$ power law dependences to facilitate comparison with analytic results. (Right) Dependence of $\Delta Y_e$ on tree-level dark scalar mass, $M_\phi$. We set $\Delta M = 3\times10^{-8}\,\,\mathrm{GeV}$ and otherwise keep all other parameters apart from $M_\phi$ the same.    }
    \label{asymmetry_mass_dependence}
\end{figure*}

Having shown that the qualitative suppression of asymmetries due to RHN equilibration is in accordance with the discussion in Sec.~\ref{sec:asym_suppr}, we now turn to a quantitative comparison. In particular, from Eq.~\eqref{eq:ARS_cutoff} we expect that flaavor asymmetries should be inversely proportional to the tenth power of the effective coupling between the SM and the hidden sector, $\xi^{-10}$, and linearly proportional to $\Delta M$. For the dominant RHN production rate in Eq.~\eqref{eq:thermalwidth_N}, we see that the effective squared coupling is $\xi^2 = y^2\lambda$, where the $\lambda$-dependence comes from its contribution to the $\phi$ thermal mass. We therefore expect that asymmetries should scale like $y^{-10}\lambda^{-5}$ provided $z_{\rm eq}<z_{\rm osc}$. In Fig.~\ref{asymmetry_parameter_dependence} we hold all parameters fixed except for the couplings $y$ and $\lambda$, and we see that the numerical solutions to the quantum kinetic equations indeed show a $y^{-10}\lambda^{-5}$ dependence on the asymmetry in the limit of ultra-relativistic $\phi$. This is an extremely severe suppression of the asymmetry as a function of the couplings:~for $y\sqrt{\lambda}\gtrsim2\times10^{-6}$, flavor asymmetries are below $10^{-10}$.

We now investigate the dependence of the lepton flavor asymmetries on the RHN mass splitting, $\Delta M$. We show our results in the left panel of Fig.~\ref{asymmetry_mass_dependence} for a representative benchmark point. As predicted, when $z_{\rm osc}>z_{\rm eq}$ the asymmetry grows linearly with $\Delta M$, reaches a maximum for mass splittings giving $z_{\rm osc}\sim z_{\rm eq}$, and then decreases again. In the limit $z_{\rm osc}\ll z_{\rm eq}$, we expect to reproduce the ARS result which predicts an asymmetry dependence of $\Delta M^{-2/3}$. Instead, we see a slightly shallower power-law dependence of approximately $\Delta M^{-0.6}$. We suspect that this minor deviation from the ARS relation is due to the fact that the convergence of the integration of many oscillations is somewhat slow, and  we do not have a sufficient separation of oscillation and equilibration scales to give exactly the ARS prediction.

We also  examine the effect of the tree-level dark scalar mass, $M_\phi$, on the asymmetries. Earlier, we argued that the RHN production rate is minimized when $M_\phi$ is as small as possible; in other words, leptogenesis is most viable when the mass of $\phi$ is dominated by the irreducible thermal mass from its coupling to SM Higgs. We show our results in the right panel of Fig.~\ref{asymmetry_mass_dependence}, finding that indeed the asymmetry is largest for tree-level masses $M_\phi\lesssim 100\,\,\mathrm{GeV}$. For larger masses, the asymmetry suppression due to RHN equilibration is even more pronounced because the decay rate is dominated by the tree-level $\phi$ mass. If the $\phi$ is sufficiently heavy, its abundance is Boltzmann-suppressed prior to RHN equilibration and the asymmetry approaches the ARS value; however, depending on the hidden-sector couplings we see that this requires a very heavy  mass ($M_\phi\sim\mathrm{PeV})$, which would put its effects far outside of the reach of even colliders like the LHC or FCC.

\subsection{Viable Baryogenesis}\label{sec:viable_baryo_phiequilibrium}

\begin{figure}[t]
        \includegraphics[width=\columnwidth]{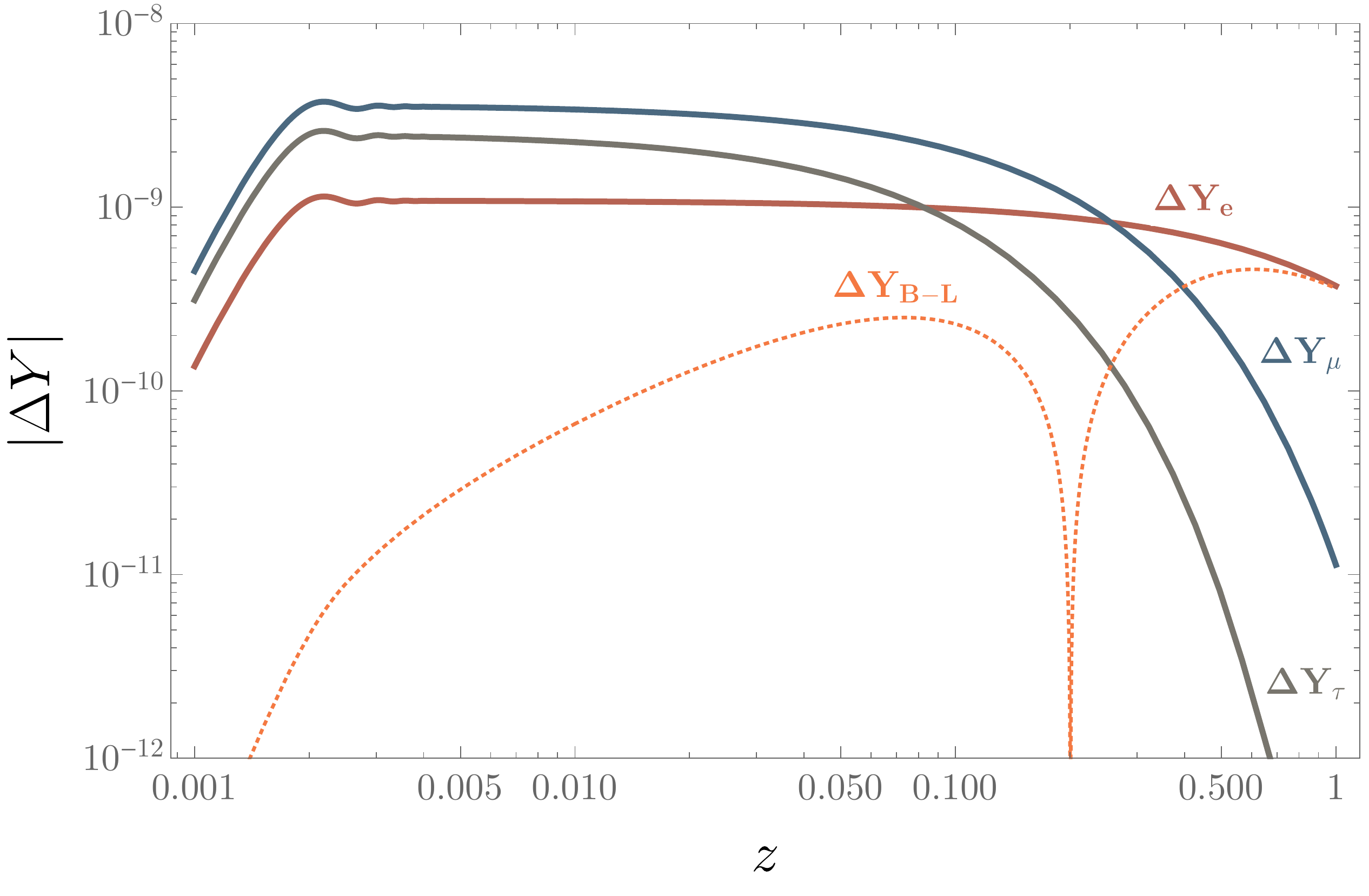}
                     \caption{
Flavor and total $B-L$ asymmetries with $F^{(II)}$ defined in Eq.~\eqref{eq:second_yukawa}, and other parameters set to $\Delta M = 1.5\times10^{-3}\,\,\mathrm{GeV}$, $M_\phi=1\,\,\mathrm{GeV}$, $\lambda=0.1$, and $y=3\times10^{-5}$. The final muon and tau asymmetries are suppressed by washout but the electron asymmetry is protected, leading to a large $B-L$ asymmetry of comparable size to $\Delta Y_e$. This maximizes the asymmetry when the Yukawa couplings, $F$, are large enough to be in the strong washout regime.  }
    \label{regime3}
\end{figure}

\begin{figure*}[t]
        \includegraphics[width=3.05in]{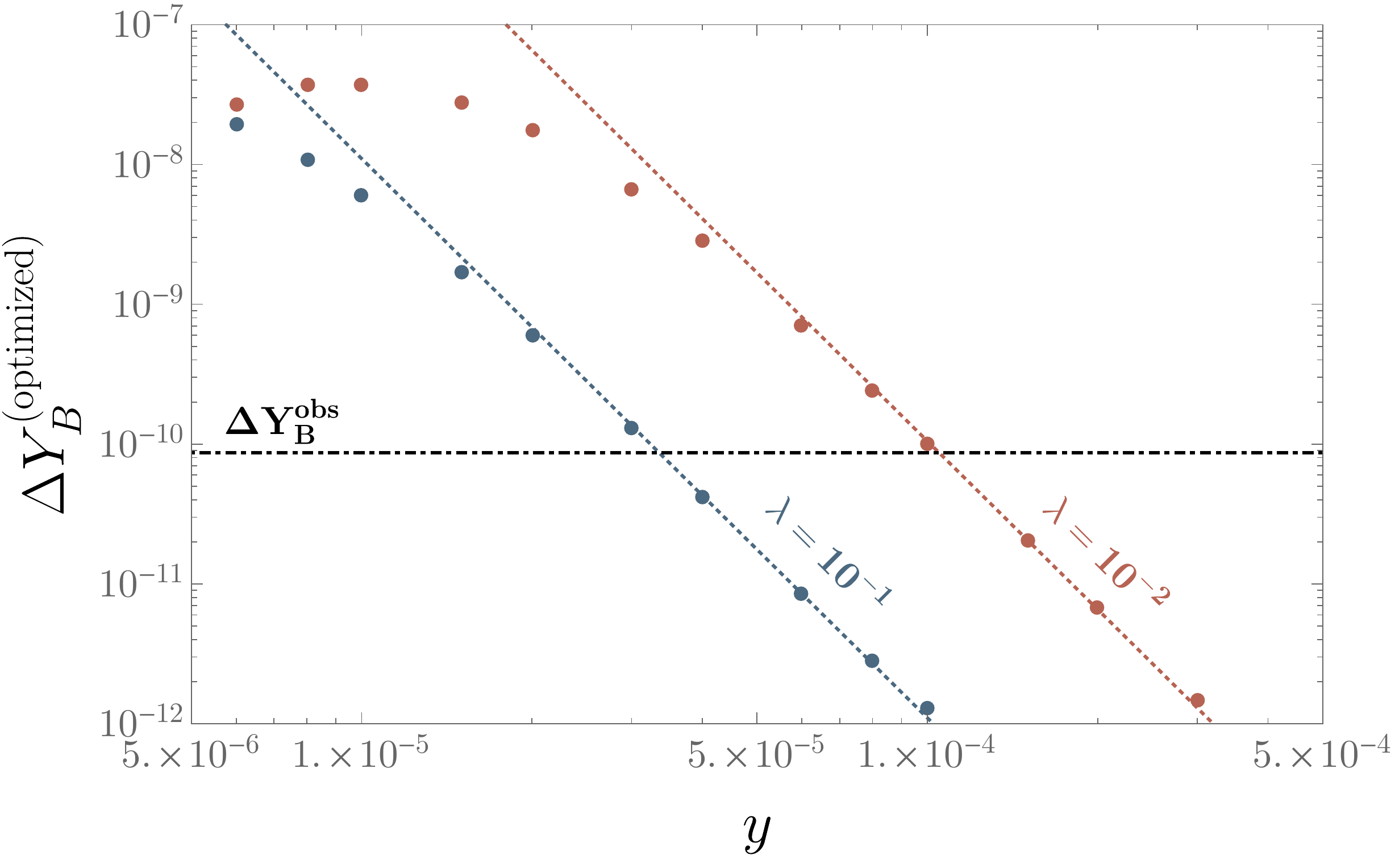}
        \quad\quad\quad
                  \includegraphics[width=3.05in]{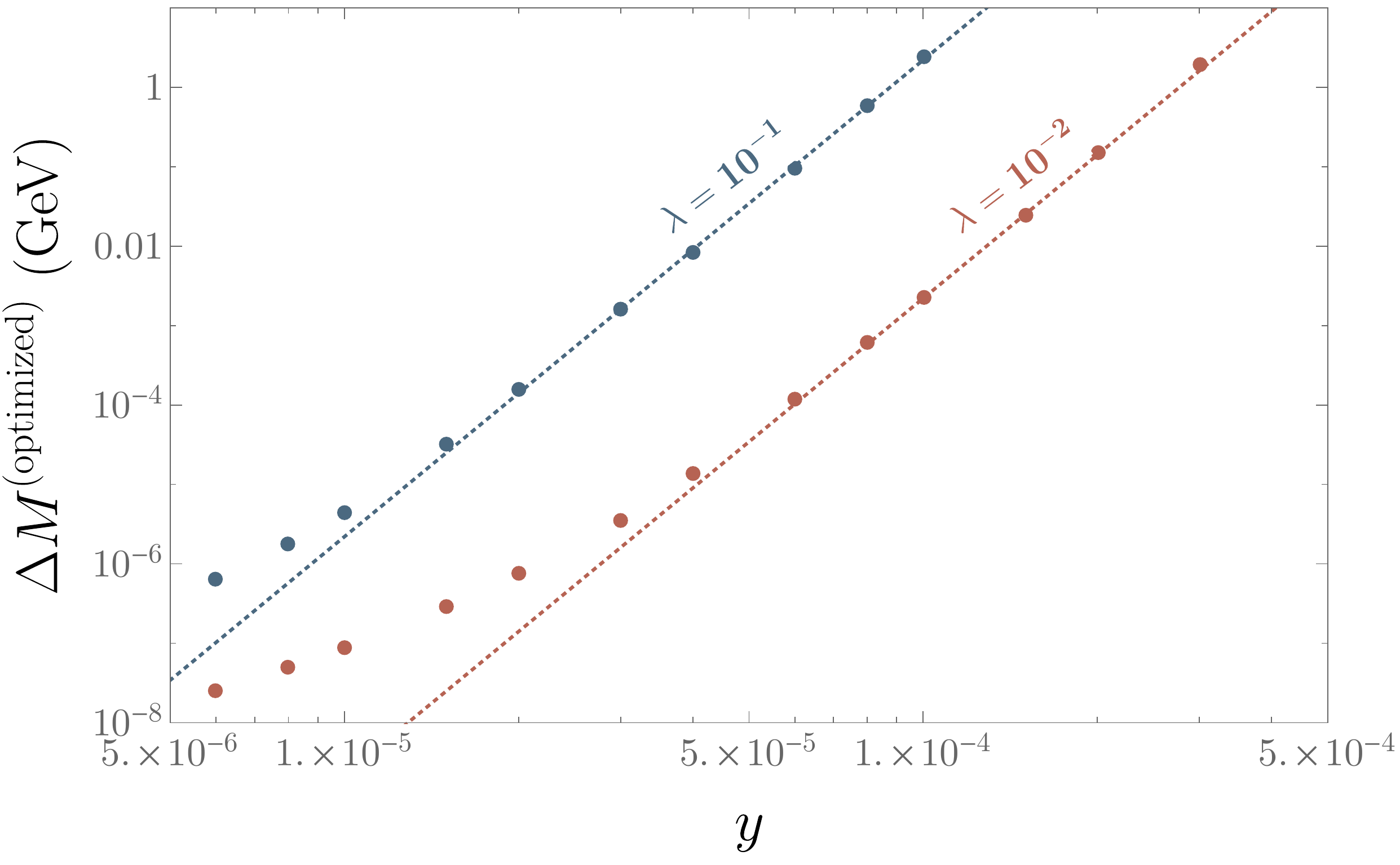}
                     \caption{
(Left) Baryon asymmetry as a function of $y$ obtained using the Yukawa texture approximately equal to $F^{(II)}$ from Eq.~\eqref{eq:second_yukawa} (although with $\mathrm{Im}\,\omega$ adjusted slightly for each point), which gives the largest asymmetry in the strong washout regime by protecting the asymmetry in a single lepton flavor. We have set $M_\phi=1\,\,\mathrm{GeV}$, and for each value of $y$ and $\lambda$ we optimized the parameters $\Delta M$ and $\mathrm{Im}\,\omega$ to give the maximum baryon asymmetry. The dashed lines indicate a $y^{-4}$ power-law dependence, and we find good agreement with the analytic prediction of Eq.~\eqref{eq:optimized_asymmetry}. The observed baryon asymmetry is indicated with the dot-dashed line. For comparison, the optimized ARS asymmetry for this benchmark is $\Delta Y_B = 2\times10^{-7}$, and the dip seen around $y=5\times10^{-6}$ is due to flavor effects. (Right) Optimal $\Delta M$ as a function of $y$, demonstrating that larger mass splittings (and earlier oscillation times) are preferred for larger couplings. The dashed lines indicate a $y^6$ power-law dependence to compare numerical results with the analytic prediction of Eq.~\eqref{eq:optimal_mass_splitting}.    }
    \label{asymmetry_dependence_optimized}
\end{figure*}

Finally, we wish to address the most pressing question:~given a particular set of couplings $\lambda$ and $y$, and a set of masses $M_1$ and $M_\phi$, what is the largest asymmetry that can be obtained? Is it compatible with the observed baryon asymmetry of $\Delta Y_B =8.65\times10^{-11}$ \cite{Aghanim:2018eyx}? To make progress in answering this question, we identify parameters for leptogenesis that give the largest asymmetry in the ARS limit and then determine the impact of RHN equilibration on the asymmetry. Because ARS is a freeze-in leptogenesis mechanism, the  flavor and total  asymmetries increase with larger Yukawa couplings $F$, provided the couplings are not so large that they bring the RHNs fully into equilibrium and wash out the asymmetry. 

In fact, the largest ARS asymmetry results when two flavors of SM lepton come into equilibrium with RHNs but one flavor does not due to suppressed couplings with the RHNs. For example, if two of the lepton numbers (such as muon and tau) come into equilibrium with the RHNs, a net baryon asymmetry still results if the electron asymmetry is protected \cite{Drewes:2012ma,Shuve:2014zua,Eijima:2018qke}. The time evolution of flavor and total asymmetries is shown in Fig.~\ref{regime3} for a benchmark point with this behavior. We see that the relatively large Yukawa couplings $F$ lead to a more substantial asymmetry at early times. Even though the muon and tau flavors come into equilibrium with the RHNs and their asymmetries are exponentially damped, the electron flavor stays out of equilibrium, preserving a net $B-L$ asymmetry. This particular limit is possible when $\mathrm{Im}\,\omega\gg1$ in the Casas-Ibarra parametrization, corresponding to an enhancement in the magnitude of $F$ relative to the na\"ive see-saw prediction \cite{Asaka:2011pb}.

The particular $F$ matrix used in Fig.~\ref{regime3} is
\small
\be\label{eq:second_yukawa}
F^{(II)} &=& \left(\begin{array}{cc} 1.33+0.930i & 0.947-1.34i \\
-2.08 +3.15i & 3.21 +2.04i \\
-4.52 + 2.80i & 2.81 + 4.49i
 \end{array}\right)\times 10^{-7}.
\ee
\normalsize
This corresponds to a particular alignment such that the Dirac and Majorana phases of the Pontecorvo-Maki-Nakagawa-Sakata (PMNS) matrix sum to $-\pi/2$ and $\mathrm{Re}\,\omega=\pi/4$ \cite{Asaka:2011pb,Shuve:2014zua,Eijima:2018qke}\footnote{Note that Refs.~\cite{Shuve:2014zua} and \cite{Eijima:2018qke} use opposite sign conventions for the Majorana phase, but the physical parameters in this limit are the same for both.}. We take the lightest RHN neutrino mass to be $M_1=5$ GeV, which gives rise to larger $F$ couplings than $M_1=1$ GeV while still allowing $\mathrm{Im}\,\omega > 1$ so that the electron asymmetry is protected from washout. We further assume for concreteness that the SM neutrinos have a normal mass hierarchy, and for the results to follow we optimize the asymmetry over the mass splitting $\Delta M$ and  $\mathrm{Im}\,\omega$, the latter of which can enhance the overall magnitude of $F$. 

We now vary $y$ and $\lambda$, using the optimal values of $\Delta M$ and the magnitude of $F$ (via adjustments to $\mathrm{Im}\,\omega$) to give the largest possible asymmetry.  We report our results in terms of the baryon asymmetry for each set of parameters, which can be obtained from the $B-L$ asymmetry  after taking into account spectator effects \cite{Harvey:1990qw}: 
\be
\Delta Y_{B} &=& \frac{28}{79} \Delta Y_{B-L}.
\ee
We show our final results in Fig.~\ref{asymmetry_dependence_optimized} for $M_\phi=1\,\,\mathrm{GeV}$. In the left panel, we show the optimal baryon asymmetry as a function of $y$ for two values of the quartic coupling $\lambda$. For comparison, we also show a $y^{-4}$ power-law dependence with dashed lines. When the hidden-sector couplings are sufficiently large that the RHNs equilibrate well before the electroweak phase transition, we see that our numerical solutions for the optimal baryon asymmetry closely follow the $y^{-4}$ power law, in agreement with our analytic arguments in Eq.~\eqref{eq:optimized_asymmetry} with $\xi=\lambda y^2$. In the right panel of Fig.~\ref{asymmetry_dependence_optimized}, we show the mass splitting which optimizes the asymmetry for each value of $y$ and $\lambda$. As the RHNs equilibrate earlier, the optimal mass splitting tends to cause oscillations to begin earlier as well such that $z_{\rm osc}\sim z_{\rm eq}$. The dependence of the optimal mass splitting on $y$ is approximately $y^6$, again in agreement with our earlier arguments in Eq.~\eqref{eq:optimal_mass_splitting}.

To summarize, we have shown that couplings of new hidden-sector particles to RHNs can bring the RHNs into equilibrium earlier than they otherwise would, significantly suppressing the asymmetry from freeze-in leptogenesis. We studied a concrete dark scalar-RHN model, and having assumed the new dark scalar to always be in equilibrium, we confirmed that the asymmetry can suffer from a tenth-power dependence on the coupling bringing the RHNs into equilibrium. We find that the couplings must satisfy $y\sqrt{\lambda}\lesssim10^{-5}$ to obtain the observed baryon asymmetry, even when other parameters have been optimized. Our numerical results agree very well with our earlier analytic arguments based on an  asymmetry cutoff at $z_{\rm eq}$ from Sec.~\ref{sec:asym_suppr}. We therefore expect the results from this section to readily generalize to any model where the RHNs are brought into equilibrium via an interaction involving fields in thermal equilibrium with the SM.

In the next section, we study the effects on leptogenesis of a hidden sector with multiple fields that can all be out of equilibrium simultaneously (with the SM and one another).

\section{Full Treatment of Hidden-Sector Equilibration}\label{sec:phi_not_in_eq}

The quantum kinetic equations in the previous section were simplified by assuming that the dark scalar, $\phi$, is always in equilibrium with the SM. In this limit,  both $\phi\rightarrow N_I N_I$ and $H\rightarrow \overline{L}_\alpha N_I$ produce RHNs with typical momentum $\sim T$. This is no longer the case when $\phi$ is out of equilibrium. To illustrate why this is the case, consider the production of two dark scalars via the process $HH^*\rightarrow \phi\phi$. These $\phi$ quanta have momenta $\sim T$. Now, imagine that one undergoes the $\phi\rightarrow N_I N_I$ decay and the other undergoes the $\phi \rightarrow \overline{N}_I\overline{N}_I$ decay, followed by two instances of the scattering $\overline{N}_I N_I\rightarrow \phi\phi$. This net process has taken two $\phi$ quanta and turned them into four $\phi$ quanta, now with a characteristic energy $\sim T/2$. This process can occur repeatedly:~in the limit that interactions within the hidden sector are in equilibrium (but both $\phi$ and RHNs are out of equilibrium with the SM), this results in a rapid cooling of the hidden sector down to a dark temperature $T_{\rm d} \sim \rho_{\rm d}^{1/4}$, where $\rho_{\rm d}$ is the hidden-sector energy density. There are now multiple possible equilibration time scales corresponding to the establishment of kinetic and chemical equilibrium within the hidden sector as well as between the hidden sector and the SM.

We first describe in Sec.~\ref{sec:hseq_formalism} our treatment of the Boltzmann equations modelling the equilibration of the hidden sector. This allows us to derive quantitative results in Sec.~\ref{sec:hseq_results} for the evolution of the hidden-sector abundances and temperatures. Finally, we incorporate these results into the calculation of the lepton asymmetry and numerically determine the implications for leptogenesis in Secs.~\ref{sec:phi_neq_lepto} and \ref{sec:phi_neq_lepto_results}, respectively.

\subsection{Hidden-Sector Equilibration:~Formalism \& Boltzmann Equations} \label{sec:hseq_formalism}
In principle, when $\phi$ and $N$ are far from equilibrium we need to solve the full momentum-dependent Boltzmann equations for the  distribution functions $f_\phi$ and $f_N$. This requires solving a very large system of coupled differential equations with one equation for each momentum mode, and the quantum kinetic equations for leptogenesis similarly need to be solved for a large number of momenta. 

We pursue a computationally simpler approach that allows for the treatment of both chemical and kinetic equilibrium. We take the following ans\"atze for the statistical distribution functions:
\be\label{eq:noeq_ansatz}
f_N(E,t) &=& \frac{n_N(t)}{n_N^{\rm eq}[T_N(t)]}e^{-E/T_N(t)},\\
f_\phi(E,t) &=& \frac{n_\phi(t)}{n_\phi^{\rm eq}[T_\phi(t)]}e^{-E/T_\phi(t)},
\ee
where we have characterized the $\phi$  ($N$) field by some characteristic temperature $T_\phi$ ($T_N$). To determine the time evolution of the hidden sector distributions, we need four Boltzmann equations to solve for all of $n_N(t)$, $n_\phi(t)$, $T_N(t)$, and $T_\phi(t)$. This approach does not capture the possible deviation of the hidden-sector distribution functions from the Maxwell-Boltzmann form, but does allow us to model leading-order effects of the different typical momenta of $\phi$ and $N$. We  neglect any back-reaction on the SM temperature, $T$, as a result of hidden-sector equilibration due to the much larger number of degrees of freedom in the SM.

The evolution of the $\phi$ number density originates from production and annihilation with the SM Higgs field, as well as (inverse) decays and scattering with RHNs. The RHN number density is, to leading order, only affected by its interactions with $\phi$. The Boltzmann equations for the number densities are derived in the usual fashion by doing an average over the momentum-dependent Boltzmann equations \cite{Gondolo:1990dk}, although care must be taken in determining the temperatures used in the thermal average. The resulting Boltzmann equations are:
\begin{widetext}
\small
\be
\dot{n}_\phi + 3Hn_\phi &=& -2\left[\langle\sigma(\phi\phi\rightarrow HH^*)v\rangle_{T_\phi}\, n_\phi(t)^2 - \langle\sigma(\phi\phi\rightarrow HH^*)v\rangle_T \,n_\phi^{\rm eq}(T)^2 \right] \label{eq:phi_number_be}\\
&&{}-2\sum_I\left[\langle\Gamma_{\phi\rightarrow N_IN_I}\rangle_{T_\phi}n_\phi(t) - \langle\Gamma_{\phi\rightarrow N_IN_I}\rangle_{T_N}\,n_\phi^{\rm eq}(T_N)\left(\frac{n_{N_I}(t)}{n_N^{\rm eq}(T_N)}\right)^2\right]\nonumber\\
&& {}-2\sum_I\left[\langle\sigma(\phi\phi\rightarrow \overline{N}_IN_I)v\rangle_{T_\phi}\,n_\phi(t)^2-\langle\sigma(\phi\phi\rightarrow \overline{N}_IN_I)v\rangle_{T_N}\,n_\phi^{\rm eq}(T_N)^2\left(\frac{n_{N_I}(t)}{n_{N_I}^{\rm eq}(T_N)}\right)^2\right]\nonumber,\\
\dot{n}_{N_I} + 3Hn_{N_I} &=& 2\left[\langle\Gamma_{\phi\rightarrow N_IN_I}\rangle_{T_\phi}n_\phi(t) - \langle\Gamma_{\phi\rightarrow N_IN_I}\rangle_{T_N}\,n_\phi^{\rm eq}(T_N)\left(\frac{n_{N_I}(t)}{n_N^{\rm eq}(T_N)}\right)^2\right]\label{eq:N_number_be}\\
&&{}+\left[\langle\sigma(\phi\phi\rightarrow \overline{N}_IN_I)v\rangle_{T_\phi}\,n_\phi(t)^2-\langle\sigma(\phi\phi\rightarrow \overline{N}_IN_I)v\rangle_{T_N}\,n_\phi^{\rm eq}(T_N)^2\left(\frac{n_{N_I}(t)}{n_N^{\rm eq}(T_N)}\right)^2\right]\nonumber,
\ee
\normalsize
\end{widetext}
where $\langle\cdots\rangle_{T_X}$ denotes a thermal average over temperature $T_X$. Note that all thermally averaged quantities with identical initial or final particles include appropriate symmetry factors. The factors of two in the first and third lines of the $n_\phi$ equation result from two $\phi$ particles being produced or destroyed in each collision, while the factor of two in the second line results from summing over decays to both $N_I$ and $\overline{N}_I$. The factor of two in the first line of the $n_{N_I}$ equation similarly results from the production or destruction of two $N_I$ in each $\phi$ decay or inverse decay, and there is no sum over RHN flavors in the $N_I$ equation because we assume the couplings $y_{IJ}$ are flavor-diagonal (and, in fact, universal). The Boltzmann equation for $n_{\overline{N}_I}$ is the same as for $n_{N_I}$ because of an assumed lack of $CP$-violation in the hidden sector; however, we keep them separate here because their quantum kinetic equations for leptogenesis are ultimately different.

To determine the evolution of the temperatures $T_N$ and $T_\phi$, we  determine differential equations for the evolution of the energy density $\rho_\phi$ ($\rho_N$) by first multiplying the momentum-dependent Boltzmann equation by $E_\phi$ ($E_N$) and then integrating over momentum. When there are identical particles in the initial or final state, we appropriately symmetrize each integral so that the energy-weighted collision term tracks the net inflow or outflow of energy for the species under consideration (for more details,  Appendix \ref{app:be_full_hseq}). We then use the ansatz Eq.~\eqref{eq:noeq_ansatz} to relate $\rho_\phi$ to $n_\phi$ and $T_\phi$ (and similarly for $\rho_N$), which allows us to determine the time evolution of the temperature. Unlike for the number-density Boltzmann equations, we also need to take into account elastic scattering processes that change the momentum of the various species involved in the collision. Due to the complexity of the collision terms, we assume that $\phi$ and $N$ are always relativistic:~as we will see in Sec.~\ref{sec:hseq_results}, $T_\phi$ and $T_N$ do not differ from $T$ by more than about an order of magnitude at any point in time, and as we found earlier the asymmetry is largest for $M_\phi\lesssim100\,\,\mathrm{GeV}$, in which case $\phi$ is always highly relativistic even when taking into account cooling within the hidden sector.

Assuming $\phi$ and $N$ are relativistic, the energy-weighted Boltzmann equations  are:
\begin{widetext}
\small
\be
\dot{\rho}_\phi + 4H\rho_\phi &=& -\left[\langle\sigma(\phi\phi\rightarrow HH^*)vE_\phi\rangle_{T_\phi}\,n_\phi(t)^2 - \langle\sigma(\phi\phi\rightarrow HH^*)vE_\phi\rangle_{T}\,n_\phi^{\rm eq}(T)^2\right]\\
&&{}-n_H^{\rm eq}(T)n_\phi(t)\langle\sigma(\phi H\rightarrow\phi H)vE_\phi\rangle_{T_\phi}\left(\frac{T_\phi}{T}-1\right)\nonumber\\
&& {}-2\overline{M}_\phi\sum_I\,\Gamma_{\phi\rightarrow N_IN_I}\left[n_\phi(t) - n_\phi^{\rm eq}(T_N)\left(\frac{n_{N_I}(t)}{n_N^{\rm eq}(T_N)}\right)^2\right]\nonumber\\
&&{}-\sum_I\left[\langle\sigma(\phi\phi\rightarrow \overline{N}_IN_I)vE_\phi\rangle_{T_\phi}\,n_\phi(t)^2 - \langle\sigma(\phi\phi\rightarrow \overline{N}_IN_I)vE_\phi\rangle_{T_N}\,n_\phi^{\rm eq}(T_N)^2\left(\frac{n_{N_I}(t)}{n_N^{\rm eq}(T_N)}\right)^2\right]\nonumber\\
&&{}-\frac{2}{3}n_\phi(t)\sum_I\,n_{N_I}(t)\langle\sigma(\phi N_I\rightarrow \phi N_I)vE_\phi\rangle_{T_\phi}\left(\frac{T_\phi}{T_N}-1\right)\nonumber,\\
\dot{\rho}_{N_I}+4H\rho_{N_I} &=& \overline{M}_\phi\Gamma_{\phi\rightarrow N_IN_I}\left[n_\phi(t) - n_\phi^{\rm eq}(T_N)\left(\frac{n_{N_I}(t)}{n_N^{\rm eq}(T_N)}\right)^2\right] \label{eq:N_energy_be}\\
&&{}+\frac{1}{2}\left[\langle\sigma(\phi\phi\rightarrow \overline{N}_IN_I)vE_\phi\rangle_{T_\phi}\,n_\phi(t)^2 - \langle\sigma(\phi\phi\rightarrow \overline{N}_IN_I)vE_\phi\rangle_{T_N}\,n_\phi^{\rm eq}(T_N)^2\left(\frac{n_{N_I}(t)}{n_N^{\rm eq}(T_N)}\right)^2\right]\nonumber\\
&&{}+\frac{1}{3}n_\phi(t)\sum_I\,n_{N_I}(t)\langle\sigma(\phi N_I\rightarrow \phi N_I)vE_\phi\rangle_{T_\phi}\left(\frac{T_\phi}{T_N}-1\right)\nonumber,
\ee
\normalsize
\end{widetext}
where $\langle \sigma vE\rangle_{T_X}$ are energy-weighted thermally averaged cross sections. The collision terms for $\rho_N$ are half the magnitude of those for $\rho_\phi$:~in all cases, this is due to the fact that $\phi$ can decay into and scatter off of both $N_I$ and $\overline{N}_I$, but $\rho_N$ only counts the energy density in the particle $N_I$ and not $\overline{N}_I$ (of course, in the absence of $CP$-violation, $\rho_{N_I}=\rho_{\overline{N}_I}$). Similarly, in the second line we have summed over both $\phi H\rightarrow \phi H$ and $\phi H^*\rightarrow \phi H^*$ elastic scattering, although the energy-weighted cross section in Eq.~\eqref{eq:N_energy_be} is calculated with respect to only of these Higgs states\footnote{In evaluating the SM Higgs number density, $n_H$, we assume that $n_H$ and $n_{H^*}$ separately count the number of Higgs and anti-Higgs states, respectively. Thus, $g_H=g_{H^*}=2$ because of the $\mathrm{SU}(2)$ multiplicity.}. Note that there is no thermal average for the $\phi$ decay width in the energy-weighted Boltzmann equation because the energy-weighting factor of $E_\phi$ in the numerator of the thermal average integral cancels the denominator of the time dilation factor, $\overline{M}_\phi/E_\phi$, and as a result the energy-weighted thermal average is independent of temperature. The precise definitions of all terms and rates in the Boltzmann equations, as well as the dimensionless versions of the Boltzmann equations that we use for our numerical studies, are presented in Appendix \ref{app:be_full_hseq}.

Finally, we must modify our expression for the thermally-corrected $\phi$ mass to account for the fact that in the $\lambda\ll y$ limit the dominant contribution to the $\phi$ mass can potentially come from RHNs. We computed this contribution following the method of Ref.~\cite{Weldon:1982bn} using our ansatz Eq.~\eqref{eq:noeq_ansatz}, and the  finite-temperature mass $\overline{M}_\phi$ with this correction is
\be
\overline{M}_\phi^2(T,T_N) &=& M_\phi^2 + \frac{\lambda}{6}T^2 + \sum_I\,\frac{y_I^2}{12}T_N^2\,\frac{n_{N_I}(t)}{n_N^{\rm eq}(T_N)},
\ee
where we have already summed over contributions from both $N_I$ and $\overline{N}_I$ states. For flavor-universal couplings, this sum just gives a factor of the multiplicity of RHNs.

We can gain some analytic understanding of the early stages of hidden-sector equilibration from Eqs.~\eqref{eq:phi_number_be}--\eqref{eq:N_energy_be}. At the earliest times, the dominant production of hidden-sector particles proceeds through $HH^*\rightarrow \phi\phi$ and $\phi\rightarrow N_IN_I$. The average $\phi$ energy produced from SM Higgs scattering is the energy transfer rate divided by the particle production rate,
\be
\langle E_\phi\rangle = \frac{\langle \sigma(\phi\phi\rightarrow HH^*)vE_\phi\rangle_T}{2\langle\sigma(\phi\phi\rightarrow HH^*)v\rangle_T}= 2T.
\ee
Using the relativistic Maxwell-Boltzmann relation $\langle E_\phi\rangle = 3T_\phi$, this gives an early-time relation
\be\label{eq:phi_temp_init}
T_\phi &=& \frac{2T}{3},
\ee
which we take to be the initial condition for $T_\phi$. The reason why $\phi$ is initially colder than the SM Higgs is because of the $1/s$ dependence of the scattering cross section, which tends to deplete the lowest-energy $H$ states. At later times, elastic scattering $\phi H\rightarrow \phi H$ re-distributes kinetic energy and drives the temperatures to be equal.

Similarly, the initial production of $N_I$ is dominated by $\phi\rightarrow N_IN_I$ decays, and so the average $N_I$ energy produced from $\phi$ decays is
\be
\langle E_{N_I}\rangle =\frac{\overline{M}_\phi\Gamma_{\phi\rightarrow N_IN_I}}{2\langle\Gamma_{\phi\rightarrow N_I N_I}\rangle_{T_\phi}}  = T_\phi.
\ee
This gives the early-time relation
\be\label{eq:N_temp_init}
T_N = \frac{T_\phi}{3} = \frac{2T}{9},
\ee
which is independent of the flavor $I$ and which we take to be the initial condition for $T_N$. Once again, the $N_I$ population is colder than the originating $\phi$ population. This can be understood by the time dilation factor in $\phi$ decays:~because the  $\phi$ particles that decay are predominantly from the coldest part of the statistical distribution, this leads to $T_N<T_\phi$. Furthermore, the  energy from a single $\phi$ is divided among two RHNs.

We can also obtain an analytic expression for the dark energy density and temperature in the limit where $\phi$ and $N_I$ have established equilibrium amongst themselves but not with the SM. In this case, $\rho_{\rm d}=\rho_\phi + \sum_I (\rho_{N_I}+\rho_{\overline{N}_I})$ and $T_{\rm d} = T_\phi = T_N$ are related by the usual equilibrium relations. Integrating the energy-density equation in the limit of negligible initial hidden-sector energy gives
\be
\rho_{\rm d} &=& \frac{\lambda^2}{32\pi^5}M_0 T^3,
\ee
from which we can use the Maxwell-Boltzmann relation between energy density and temperature to express the dark energy density in terms of $T_{\rm d}$,
\be\label{eq:hs_equilibration_temp}
\frac{T_{\rm d}}{T} = \left(\frac{\lambda^2 M_0}{96g_{\rm d}\pi^3 T}\right)^{1/4}=\left(\frac{\lambda^2 M_0z}{96g_{\rm d}\pi^3 T_{\rm ew}}\right)^{1/4},
\ee
where $g_{\rm d}=5$ is the number of degrees of freedom in the hidden sector for two RHNs. We see that the temperature of the hidden sector increases relative to the SM temperature as $z^{1/4}$.

Finally, we can estimate the time at which equilibration occurs within the hidden sector. This approximately corresponds to the conditions $n_\phi\langle\sigma(\phi\phi\rightarrow \overline{N}_IN_I)v\rangle \approx H$ and $n_{N_I}\langle\sigma(\overline{N}_IN_I\rightarrow \phi\phi)v\rangle \approx H$, where the thermal averages are computed over either $T_\phi$ or $T_N$ (they are anyway the same once local equilibrium is reached in the hidden sector). The number densities of $\phi$ and $N_I$ can be computed analytically at early times by ignoring the back-reaction terms, and the parametric scaling of the time of equilibration within the hidden sector is
\be\label{eq:analytic_hseq_internal}
z_{\rm h.s.\,eq}\propto \frac{1}{\lambda y^2}.
\ee

\subsection{Hidden-Sector Equilibration:~Results} \label{sec:hseq_results}

\begin{figure}[t]
        \includegraphics[width=\columnwidth]{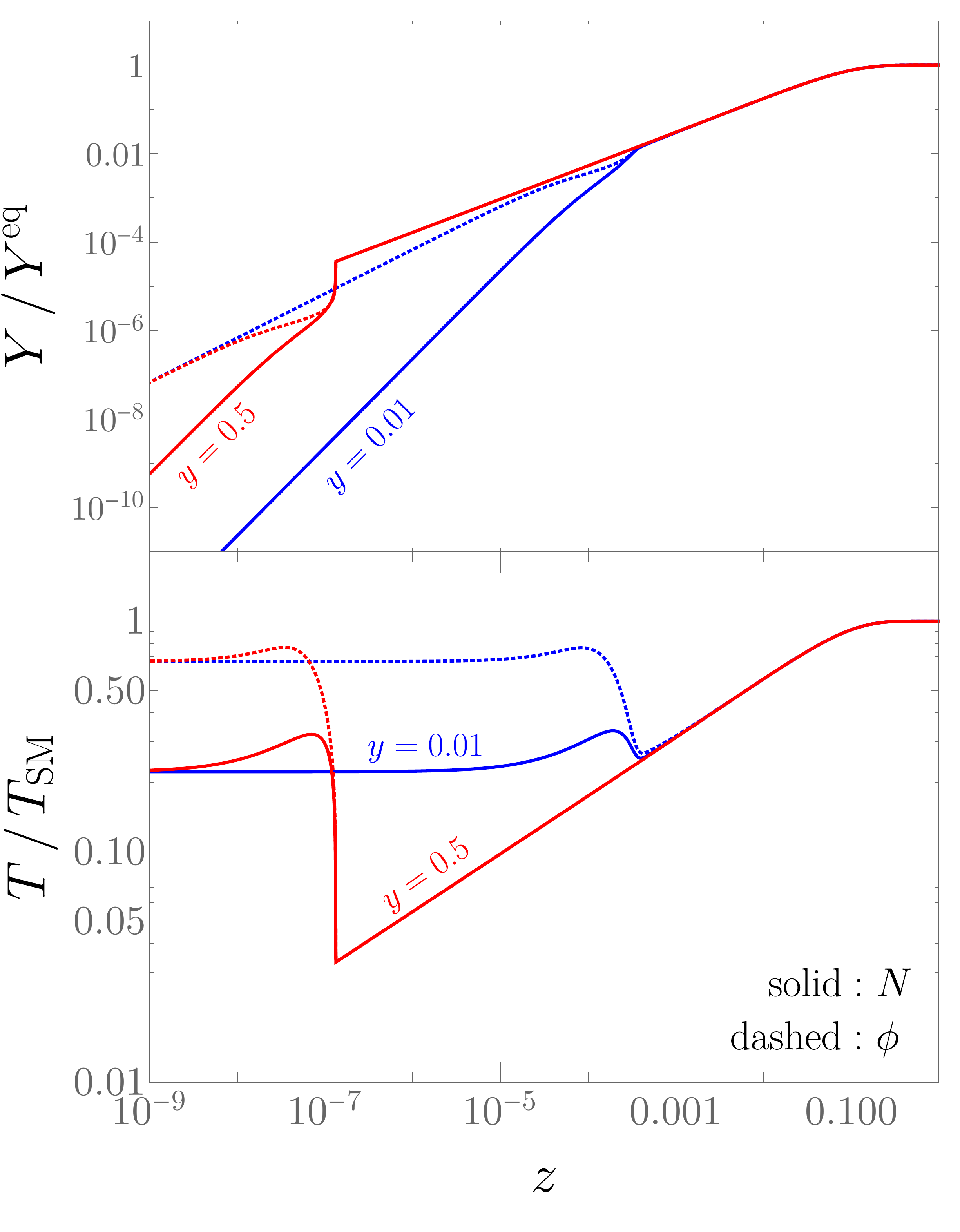}
                     \caption{
Abundances (upper plot) and temperatures (lower plot) as functions of dimensionless time $z$, expressed as ratios to the values when in equilibrium with the SM, for $\phi$ (dashed lines) and $N$ (solid lines). Here, we show benchmarks for which the hidden sector comes to local equilibrium before equilibrating with the SM. We take $\lambda=5\times10^{-6}$, $M_\phi=M_N=1\,\,\mathrm{GeV}$, and consider two values of $y$:~0.01 (blue) and 0.5 (red). Note that when the hidden sector establishes local equilibrium, the abundances and temperatures rapidly approach the values predicted by the dark temperature Eq.~\eqref{eq:hs_equilibration_temp}.
    }
    \label{fig:thermalization_hseq}
\end{figure}

We begin by showing the evolution of the hidden sector abundances and temperatures for some benchmark points. In solving the Boltzmann equations, we take as our initial conditions $Y_\phi=Y_{N_I}=10^{-20}$ and initial hidden-sector temperatures given by Eqs.~\eqref{eq:phi_temp_init} and \eqref{eq:N_temp_init}. We consider a system with two RHNs.\\

\noindent {\bf Evolution of Hidden-Sector Temperatures and Abundances:}~First, we consider benchmarks with $\lambda\ll y$ such that the coupling within the hidden sector is much stronger than the coupling between the hidden sector and the SM. In Fig.~\ref{fig:thermalization_hseq}, we show the time evolution of the $N$ and $\phi$ abundances and temperatures, both taken as ratios with respect to the values when fully in equilibrium with the SM. At early times, the $N$ and $\phi$ temperatures stay fixed at the values derived in Eqs.~\eqref{eq:phi_temp_init} and \eqref{eq:N_temp_init}, and the abundances grow according to a na\"ive integration of the number-density Boltzmann equation with no back-reaction effects. With the sufficient accumulation of $\phi$ and $N$ particles, however, $2\rightarrow2$ processes become important within the hidden sector, leading to rapid equilibration within the hidden sector to the temperature predicted in Eq.~\eqref{eq:hs_equilibration_temp}. If $T_{\rm d}$ is below the initial values of $T_\phi$ and $T_N$, then the hidden sector rapidly evolves to a colder, higher-multiplicity state as dictated by the hidden-sector equilibrium condition. The whole sector then evolves towards equilibrium with the SM. In both cases, we see relatively rapid changes to the hidden-sector temperature and abundances at the time of local equilibration. Note that even with the rapid cooling, the hidden-sector temperature is never more than about an order of magnitude colder than the SM temperature for $z\ge10^{-5}$ and $\lambda\lesssim10^{-5}$, and hence the relativistic assumptions for $\phi$ and $N$ are reasonable for masses of phenomenological interest.

Our analysis shows that, for $y\sim1$, local equilibration inside the hidden sector occurs very early ($z_{\rm h.s.\,eq}\sim10^{-7}$ for $\lambda\sim10^{-5}$). This suggests that large couplings to RHNs even within a thermally decoupled hidden sector can be problematic from the point of view of RHN equilibration and decoherence effects.

\begin{figure}[t]
        \includegraphics[width=\columnwidth]{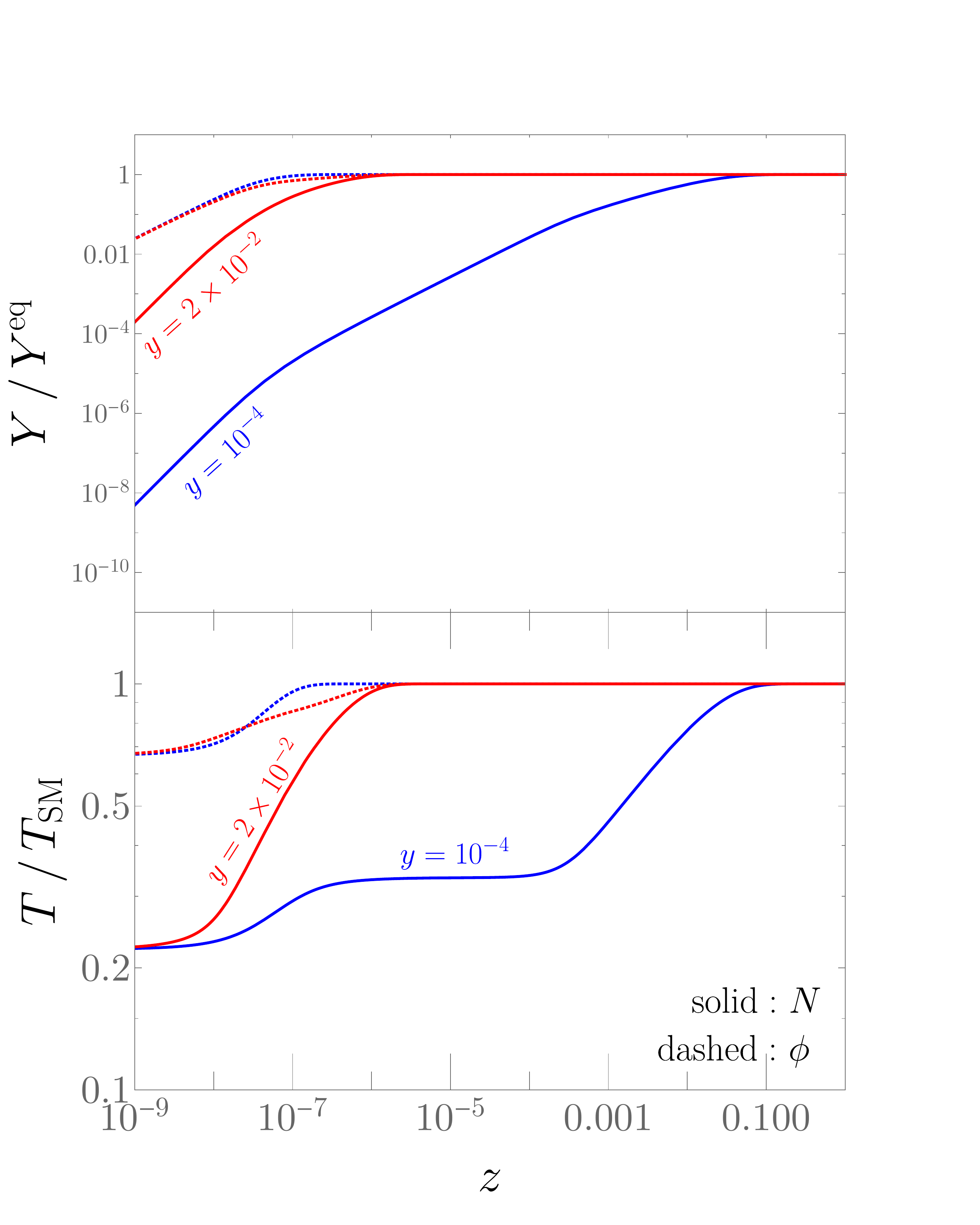}
                     \caption{
Abundances (upper plot) and temperatures (lower plot) as functions of dimensionless time $z$, expressed as ratios to the values when in equilibrium with the SM, for $\phi$ (dashed lines) and $N$ (solid lines). Here, we show benchmarks for which $N$ equilibrates with $\phi$ after $\phi$ is already in equilibrium with the SM. We take $\lambda=3\times10^{-3}$, $M_\phi=M_N=1\,\,\mathrm{GeV}$, and consider two values of $y$:~$10^{-4}$ (blue) and $2\times10^{-2}$ (red).
    }
    \label{fig:thermalization_hseq_point2}
\end{figure}

We next consider the opposite limit, namely $\lambda\gtrsim y$, in which case we expect $\phi$ to equilibrate with the SM prior to RHNs equilibrating with $\phi$. For the extreme case $\lambda\gg y$, we expect to recover the results of Sec.~\ref{sec:phi_in_eq}. We show the results for two benchmark points in Fig.~\ref{fig:thermalization_hseq_point2}. We note that in both cases, $Y_\phi$ is linearly proportional to $z$ as $\phi$ comes into equilibrium; since the source for $Y_N$ production is proportional to $Y_\phi$, this gives $Y_N$ a $z^2$ dependence at early times. For $y=2\times10^{-2}$, the RHNs come into equilibrium at around the same time as $\phi$. For $y=10^{-4}$, however, we see that the temperature rises to a new plateau of $T/3$, which corresponds to Eq.~\eqref{eq:N_temp_init} evaluated with $T_\phi = T$ since $\phi$ has come into equilibrium with the SM. Furthermore, since $Y_\phi$ is constant after it equilibrates with the SM, there is a break in the $Y_N$ evolution and it transitions to a time dependence that is linear in $z$. It takes much longer for $N_I$ to come into equilibrium with this linear dependence in $z$ than in the regime where the time dependence is $z^2$; for $\lambda\ll y$ this linear time dependence dominates the RHN production history and reproduces the qualitative features found in Sec.~\ref{sec:phi_in_eq}. We see that, for both benchmarks, $T_\phi$ and $T_N$ are both an $\mathcal{O}(1)$ factor different from the SM temperature, $T$. \\

\begin{figure}[t]
        \includegraphics[width=\columnwidth]{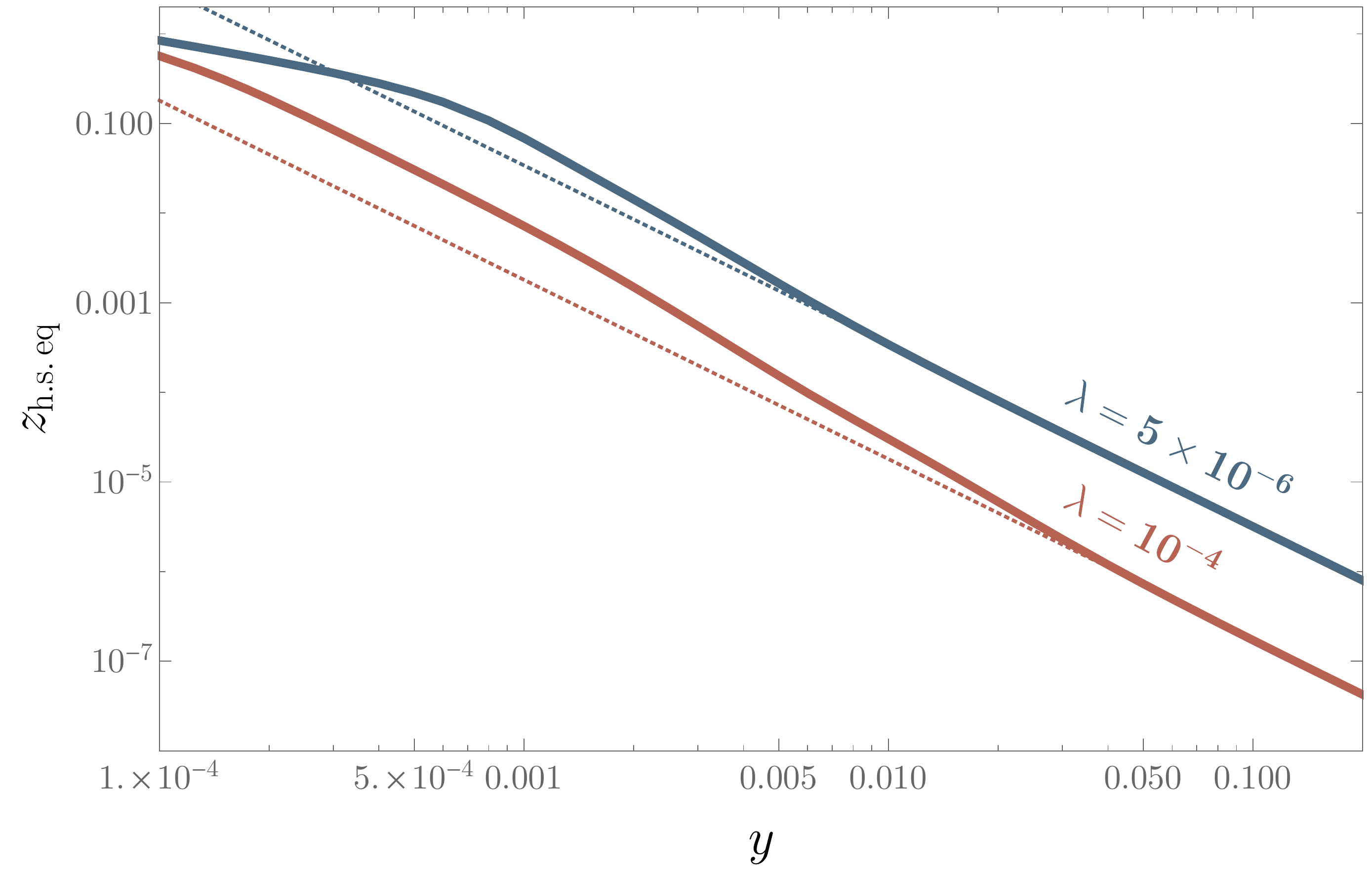}
                     \caption{
Dimensionless time at which $N$ equilibrates with $\phi$, $z_{\rm h.s.\,eq}$, defined as the time at which $T_N = 0.9 T_\phi$. The solid curves show  numerical results for the indicated values of $\lambda$, while the dashed lines show a $y^{-2}$ power-law dependence to facilitate comparison with analytic arguments provided in the text.
    }
    \label{fig:eqtime_hiddensector}
\end{figure}

\noindent {\bf Coupling Dependence of Equilibration Time:}~Next, we study the quantitative dependence of the time of RHN equilibration on the couplings $y$ and $\lambda$. When the $N_I$ come into equilibrium, they first enter equilibrium with $\phi$ (and may simultaneously equilibrate with the SM if $\phi$ and the SM are in equilibrium). A hallmark of equilibrium is that $T_\phi = T_N$. We  therefore define a hidden-sector equilibration time, $z_{\rm h.s.\,eq}$, as the time at which the ratio $T_N/T_\phi =0.9$. In Fig.~\ref{fig:eqtime_hiddensector}, we show $z_{\rm h.s.\,eq}$ as a function of $y$ for two values of $\lambda$:~$10^{-4}$ and $5\times10^{-6}$.

When $y\gg \lambda$, we are in the regime where the RHNs and $\phi$ first come to local equilibrium within the hidden sector, and they subsequently evolve as a whole towards equilibrium with the SM. For both values of $\lambda$, the equilibrium time  scales as $y^{-2}\lambda^{-1}$ in accordance with Eq.~\eqref{eq:analytic_hseq_internal} (a quadratic power-law dependence is indicated by the dashed lines in Fig.~\ref{fig:eqtime_hiddensector}). For $\lambda=10^{-4}$, we also see a $y^{-2}$ power law in the $\lambda\gg y$ limit, which agrees with the arguments from Sec.~\ref{sec:phi_in_eq}. The result is somewhat different for smaller values of $\lambda$, as seen in the $\lambda=5\times10^{-6}$ curve. The reason is as follows:~while $\phi$ is out of equilibrium, the abundance of RHNs grows as $z^2$, whereas it only grows as $z$ once $\phi$ is in equilibrium with the SM. For sufficiently small values of $\lambda$, the RHNs come into equilibrium around the same time as $\phi$, and the parametric scaling goes as  $y^{-1}$ instead of $y^{-2}$. In Fig.~\ref{fig:eqtime_hiddensector}, the equilibration time scaling for $\lambda=5\times10^{-6}$ and small $y$ is indeed  $z_{\rm h.s.\,eq}\propto y^{-1}$.\\

\noindent {\bf Comparison with Previous Results:}~Finally, it is instructive to compare the results taking into account full hidden-sector equilibration with our findings from Sec.~\ref{sec:phi_in_eq}, where we assumed that $\phi$ was always in equilibrium and  that $N$ had a common temperature with the SM. To facilitate the comparison, we consider $\lambda=0.5\gg y$, which is the same limit as Sec.~\ref{sec:phi_in_eq} and for which it is valid to assume that $\phi$ is in equilibrium with the SM throughout the cosmological production of RHNs. We solve the Boltzmann equations from Sec.~\ref{sec:hseq_formalism} twice:~first, we solve the full Boltzmann equations, and second we solve them imposing the conditions  that $T_N=T_\phi=T$ and $Y_\phi=Y_\phi^{\rm eq}$. 

\begin{figure}[t]
        \includegraphics[width=\columnwidth]{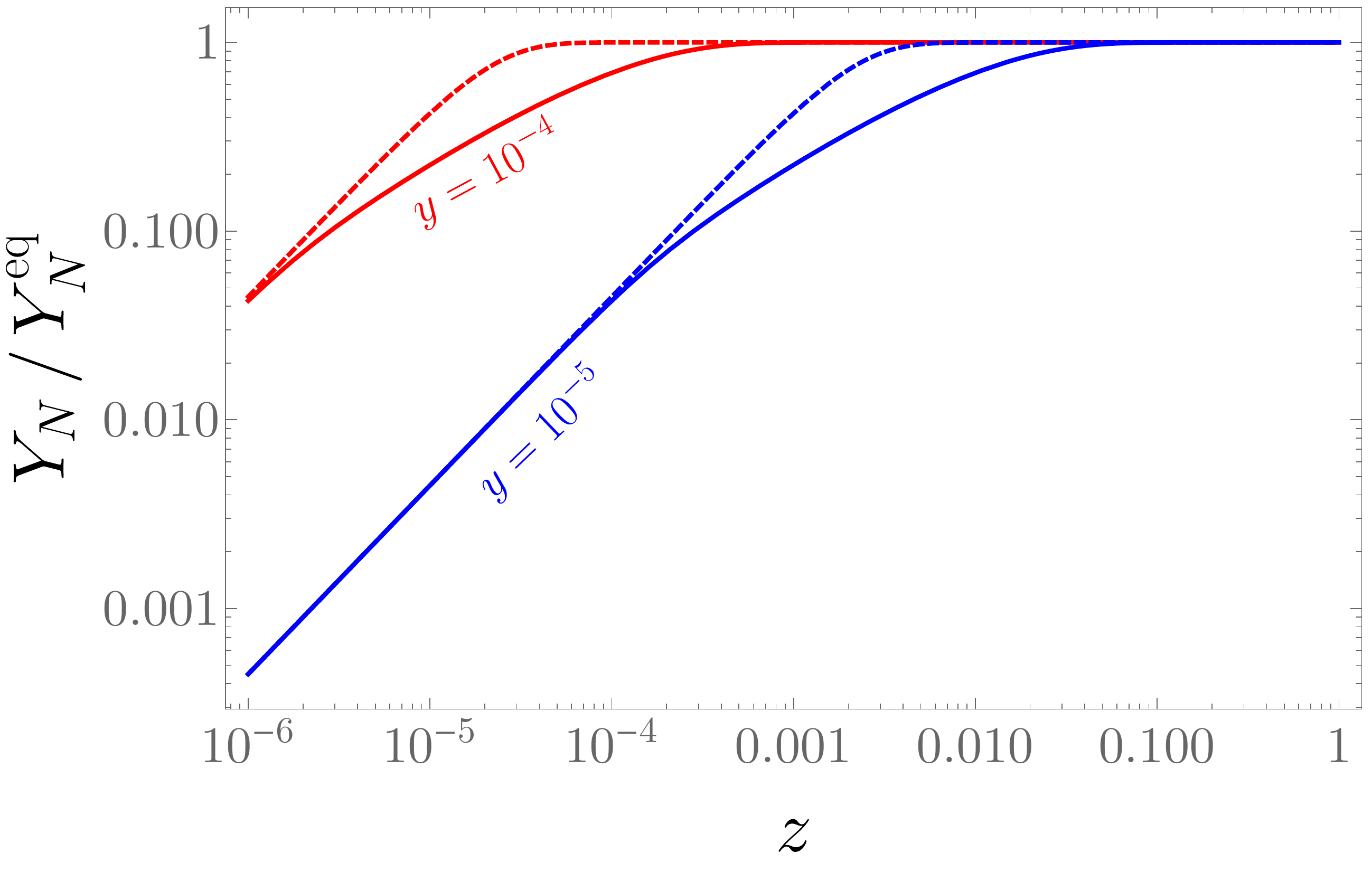}
                     \caption{
Dimensionless-time-dependence of the RHN abundance, $Y_N$, for $\lambda=0.5$ and the indicated values of $y$. The solid lines show the full solutions of the Boltzmann equations from Sec.~\ref{sec:hseq_formalism}, while the dashed lines show the solutions to the Boltzmann equations where we have constrained $T_N=T_\phi=T$ and $Y_\phi = Y_\phi^{\rm eq}$.
    }
    \label{fig:eqtime_comparison}
\end{figure}

The time-evolution of $Y_N$ is shown for both  solutions in Fig.~\ref{fig:eqtime_comparison}. It is evident that, at early times, the two methods closely agree. In this epoch, $NN\rightarrow\phi$ and $N\overline{N}\rightarrow\phi\phi$ processes are negligible and so the RHN abundance is independent of $T_N$. However, as the RHN abundance grows, the inverse processes become more important, and we see that the full Boltzmann equations generally predict a slower approach to equilibrium. One way of understanding this is that the typical RHN momentum is given by $T_N<T$, and hence the annihilation cross section is larger than if it had the same temperature as the SM. 

\begin{figure}[t]
        \includegraphics[width=\columnwidth]{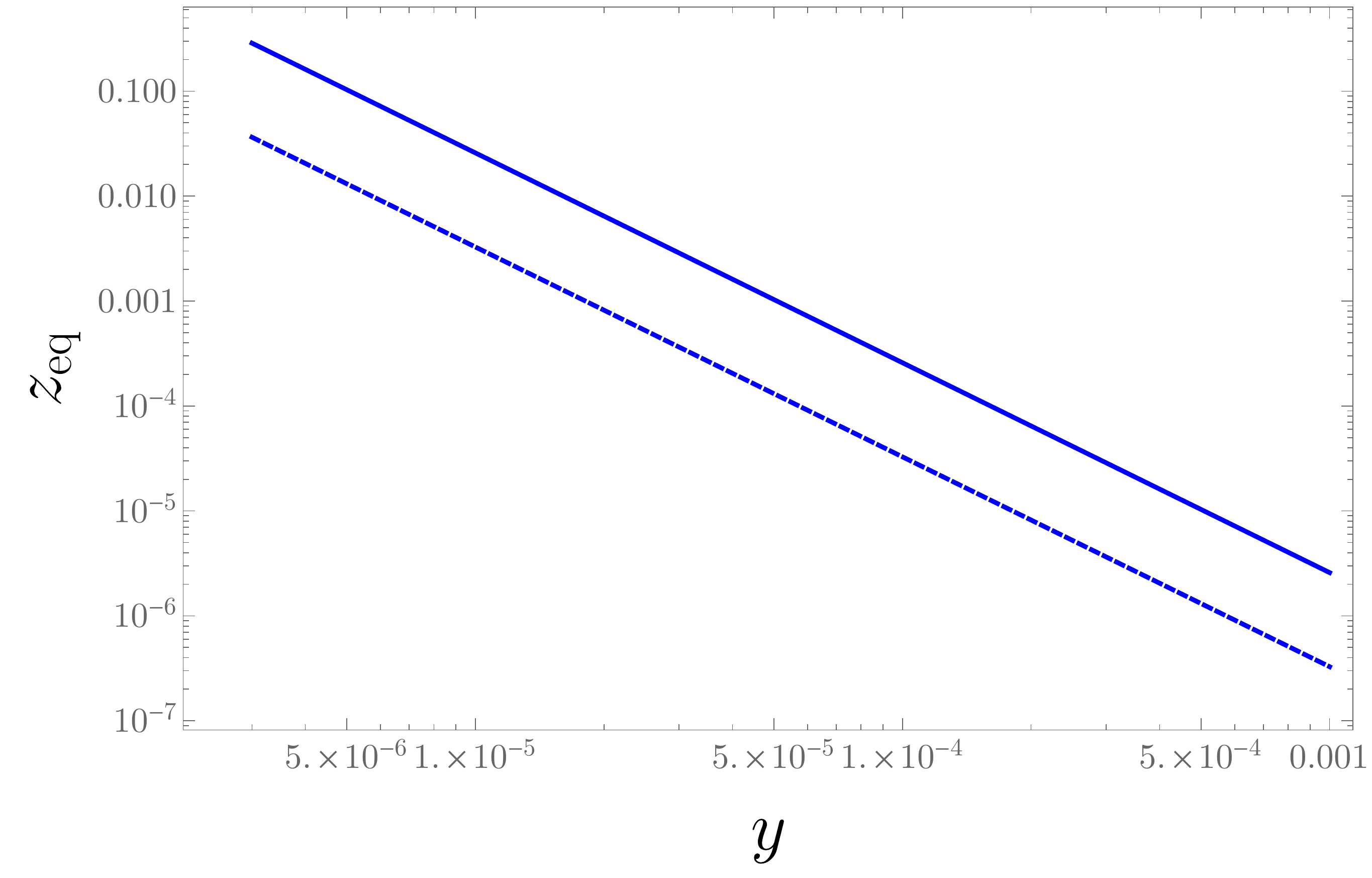}
                     \caption{
Dimensionless RHN equilibration time, $z_{\rm eq}$, defined such that $Y_N(z_{\rm eq}) = 0.9Y_N^{\rm eq}$ as a function of $y$ with $\lambda=0.5$. The solid lines show the full solutions of the Boltzmann equations from Sec.~\ref{sec:hseq_formalism}, while the dashed lines show the solutions to the Boltzmann equations where we have constrained $T_N=T_\phi=T$ and $Y_\phi = Y_\phi^{\rm eq}$. Both solutions exhibit a $z_{\rm eq}\propto y^{-2}$ dependence.
    }
    \label{fig:eqtime_comparison_eqtime}
\end{figure}

Because the approach to equilibrium is delayed relative to the findings in Sec.~\ref{sec:phi_in_eq}, we expect that the results from that section are overly pessimistic with respect to the effects of RHN equilibration on baryogenesis. However, the parametric dependence of the equilibration time continues to hold:~in Fig.~\ref{fig:eqtime_comparison_eqtime}, we show the dimensionless time at which the RHNs come into equilibrium for each of the two methods, where for concreteness we define the equilibration time as the time at which $Y_N=0.9Y_N^{\rm eq}$. It is evident that both the full solution to the Boltzmann equations and the solution with $T_N=T$ have the same parametric dependence $z_{\rm eq}\propto y^{-2}$; the delay in equilibration predicted by the full Boltzmann equations is a constant across all couplings. Thus, all of our earlier results should hold in the $\lambda\gg y$ limit, although the actual equilibration time is somewhat delayed (and the lepton flavor asymmetries consequently larger) by properly considering kinetic and chemical equilibration of the hidden sector. \\

\noindent {\bf Summary:}~We have studied the equilibration of $\phi$ and RHNs using Boltzmann equations that track both number- and energy-changing processes. We find that, over a significant range of parameters, the RHN equilibration time scales as $y^{-2}$, in agreement with earlier arguments. However, there can be deviations from this power-law scaling when RHN and $\phi$ equilibration occur on comparable time scales. We find that, at early times, the $\phi$ and $N_I$ temperatures are held at constant values determined by the ratio of energy-weighted and regular thermally averaged rates, while in the limit of large coupling within the hidden sector, $\phi$ and $N$ establish local equilibrium with abundance and temperature evolution dictated by Eq.~\eqref{eq:hs_equilibration_temp}. In the opposite limit, $\phi$ comes into equilibrium well before $N_I$ and the results of Sec.~\ref{sec:phi_in_eq}  give the qualitatively correct time scales, although the true equilibration time is somewhat delayed with respect to our earlier findings.

\subsection{Quantum Kinetic Equations for Leptogenesis} \label{sec:phi_neq_lepto}

We wish to incorporate the results of Sec.~\ref{sec:hseq_formalism} into our quantum kinetic equations for leptogenesis. However, there is a complication:~the formalism of Sec.~\ref{sec:hseq_formalism} assumes that all RHNs have a temperature $T_N$, whereas the RHNs produced from decay and scattering of SM Higgs have a typical momentum $\sim T$. Indeed, the quantum kinetic equations of Sec.~\ref{sec:be_phi_in_eq} are thermally averaged over distributions with temperature $T$.

To resolve this without solving the complicated full momentum-dependent Boltzmann equations, we instead consider the separate evolution of \emph{two} separate RHN populations:~the hidden-sector population of RHNs as described by the formalism in Sec.~\ref{sec:hseq_formalism} with a temperature $T_N$ (which we denote as a background RHN density, $Y_{\tilde N_I}$), and the population of RHNs produced in specific, coherent superpositions of RHN mass eigenstates from SM Higgs processes at temperature $T$ (which we describe by the  matrices $R_{N_I}$ and $R_{\overline{N}_I}$). It is the coherently propagating, out-of-equilibrium $R_{N_I},R_{\overline{N}_I}$ populations that can generate a lepton asymmetry. We assume that the RHNs predominantly equilibrate through the hidden-sector interactions involving $\phi$; in other words, $Y_{\tilde N}(t) \gg Y_N^{\rm eq}(T) R_N$. This tells us, for example, that processes like $H\rightarrow \overline{\ell} N$ have a negligible impact on the abundance of RHNs at temperature $T_N$, and the evolution of this population $Y_{\tilde N}$ is given purely by solving Eq.~\eqref{eq:N_number_be} above.

The density matrix for $N$ can now be written as
\be
\left(Y_N\right)_{IJ} &=& Y_N^{\rm eq}(T) \left(R_N\right)_{IJ} + Y_{\tilde N}(t)\,\delta_{IJ}.
\ee
We  substitute this into the usual ARS quantum kinetic equations for leptogenesis, supplemented with  collision terms representing $\phi\leftrightarrow N_I N_I$ and $\phi\phi\leftrightarrow N_I\overline{N}_I$ processes. The quantum kinetic equations give the time evolution of the full density matrix, $Y_N$, and we can relate this to the evolution of  $R_N$ via
\be\label{eq:rn_evol_equation}
\frac{dR_N}{dt} &=& \frac{1}{Y_N^{\rm eq}(T)}\left(\frac{dY_N}{dt} - \frac{dY_{\tilde N}}{dt}\right),
\ee
where we again assume that the RHNs are sufficiently relativistic that $Y_N^{\rm eq}(T)$ is approximately independent of time. We now consider the various parts of the quantum kinetic equations:\\

\noindent {\bf Oscillation terms:}~the oscillation terms, proportional to $\left[ H,Y_N\right]$, are identical to the minimal ARS model. For the population $Y_{\tilde N}$, the density matrix is diagonal and the commutator vanishes.\\

\noindent {\bf ARS collision terms:}~the ARS collision terms representing $H\leftrightarrow \overline\ell N_I$ and associated SM $2\leftrightarrow2$ processes, given schematically in Eq.~\eqref{eq:qke_for_lepto} in terms of thermally averaged rates $\langle\tilde\Gamma_h\rangle$ and $\langle\tilde\Gamma_{\rm w.o.}\rangle$, are the same as before except with the replacement $(R_N)_{IJ}\rightarrow (R_N)_{IJ} + Y_{\tilde N} \delta_{IJ} / Y_N^{\rm eq}(T)$. Additionally, we have to account for the fact that the terms proportional to $Y_{\tilde N}$ represent processes like $\tilde{N_I}(T_N) \overline{\ell}(T)\rightarrow H$ for which the colliding species have different temperatures. We re-evaluate the thermal averaging procedure for this case, finding in the limit of Maxwell-Boltzmann statistics that the term in the quantum kinetic equation is exactly the same as before \emph{except} the thermally averaged cross section is computed using the geometric mean temperature,
\be
\overline{T} &\equiv& \sqrt{TT_N}.
\ee
Compared to the  non-dimensionalized quantum kinetic equations we used before (see Appendix \ref{app:standard_ARS}), the ARS collision terms proportional to $Y_{\tilde N}$ are multiplied by a factor of $T/T_N$. In the limit where $T_N\rightarrow T$, this trivially reduces to the usual ARS collision term.  The full form of the quantum kinetic equations we use is provided in Appendix \ref{app:qke_full_hseq}.  \\

\noindent {\bf Hidden-Sector Collision Terms:}~given the two distinct populations of RHNs, there are three categories of hidden-sector annihilation modes into one or more $\phi$ particles:~$N_I N_I$, $\tilde{N}_I \tilde{N}_I$, and $N_I \tilde{N}_I$ annihilation. We are most interested in the limit where the hidden sector is equilibrated through $\phi$ interactions and not through SM Higgs decays, and consequently we can assume $Y_{\tilde N_I}\gg Y_{N_I}^{\rm eq}(T) R_{N_I}$.This suppressed abundance of RHNs from SM Higgs decays renders the $N_I N_I$ annihilation rate negligible. In this same limit, we find that the $\tilde{N}_I \tilde{N}_I$ collision terms sum (by definition) to $dY_{\tilde N}/dt$, and this same quantity is then immediately subtracted in Eq.~\eqref{eq:rn_evol_equation}; in other words, these collision are internal to the hidden sector and are irrelevant for the evolution of the RHN abundances $R_N,R_{\overline{N}}$ responsible for leptogenesis. 

Finally, we are left to compute the collision term for $N_I \tilde{N}_I$ annihilation. As with the ARS collision terms, we must thermally average over an annihilation process where the species have different temperature, and we reach the same conclusion that the thermally averaged cross section must be computed with respect to the geometric mean temperature, $\overline{T}$. In other words, the reaction rates are the same as those in Eq.~\eqref{eq:N_number_be} but with $T\rightarrow \overline{T}$. We also need to generalize the collision terms from Eq.~\eqref{eq:N_number_be} to include off-diagonal density-matrix elements. Because the hidden-sector couplings are universal, the density matrix for $\tilde N$ is $(Y_{\tilde N_I})_{IJ} = Y_{\tilde{N}}\delta_{IJ}$. Thus, the $\phi$ collision term in the quantum kinetic equation Eq.~\eqref{eq:qke_for_lepto} is modified to:
\be\label{eq:qke_including_evap}
\frac{dR_N}{dt} &=& -2\langle\Gamma_{\phi\rightarrow N_IN_I}\rangle_{\overline{T}}\,\frac{Y_\phi^{\rm eq}(\overline{T}) Y_{\tilde{N}}}{Y_N^{\rm eq}(\overline{T})^2}R_N\\
&& {}-s\langle\sigma(\phi\phi\rightarrow N_I\overline{N}_I)v\rangle_{\overline{T}}\,\frac{Y_\phi^{\rm eq}(\overline{T})^2Y_{\tilde N}}{Y_N^{\rm eq}(\overline{T})^2}R_N\nonumber\\
&&{}+\mathrm{ARS\,\,terms}\nonumber.
\ee
A similar modification is made to the quantum kinetic equation for $R_{\overline{N}}$. 

In Eq.~\eqref{eq:qke_including_evap}, we have assumed that the abundance of RHNs from hidden-sector interactions, $\tilde{N}$, is $CP$-symmetric and the result of solving Eqs.~\eqref{eq:phi_number_be}--\eqref{eq:N_energy_be}. We also assume that the rate of $N_I\tilde{N}_I$ annihilation is sufficiently small that it does not appreciably modify the temperature of the scalars, $T_\phi$. We do not need to include the reverse reaction, $\phi\phi\rightarrow \tilde{N}_I N_I$, since by definition $R_N$ is separately  tracking  the out-of-equilibrium RHNs produced from SM Higgs decay and not the internal dynamics of the hidden sector. In other words, Eq.~\eqref{eq:qke_including_evap} describes the absorption of the out-of-equilibrium RHN population $R_N$ into the hidden sector. 

The absorption of RHNs responsible for leptogenesis into the rapidly interacting hidden sector occurs on time scales given by the inverse decay and $2\rightarrow 2$ scattering rate into $\phi$. This leads to an exponential damping of the population of $R_N$ when these processes occur faster than Hubble expansion and the hidden-sector neutrinos are in equilibrium with the SM. Asymmetry generation is suppressed in this limit because the asymmetry depends on phases from the coherent propagation of RHNs between the time of production and destruction, which are encoded in the phases in $R_N$. If the RHNs rapidly annihilate into $\phi$, which subsequently scatter and decay in various ways, the states rapidly become entangled with the environment and the phase information is effectively lost (indeed, if  $\phi$--$H$ scattering is rapid, the $\phi$ produced from RHN annihilation can turn into SM Higgses, quarks, etc.~and not even return to an RHN state!). Because of the flavor universality of $\phi$ decays, the RHNs produced by hidden-sector interactions cannot give rise to the specific coherent superpositions of mass eigenstates needed to generate a net asymmetry. Thus, the process of asymmetry generation is suppressed when the RHNs responsible for leptogenesis begin to rapidly interact with other hidden-sector RHNs.

\begin{figure}[t]
        \includegraphics[width=\columnwidth]{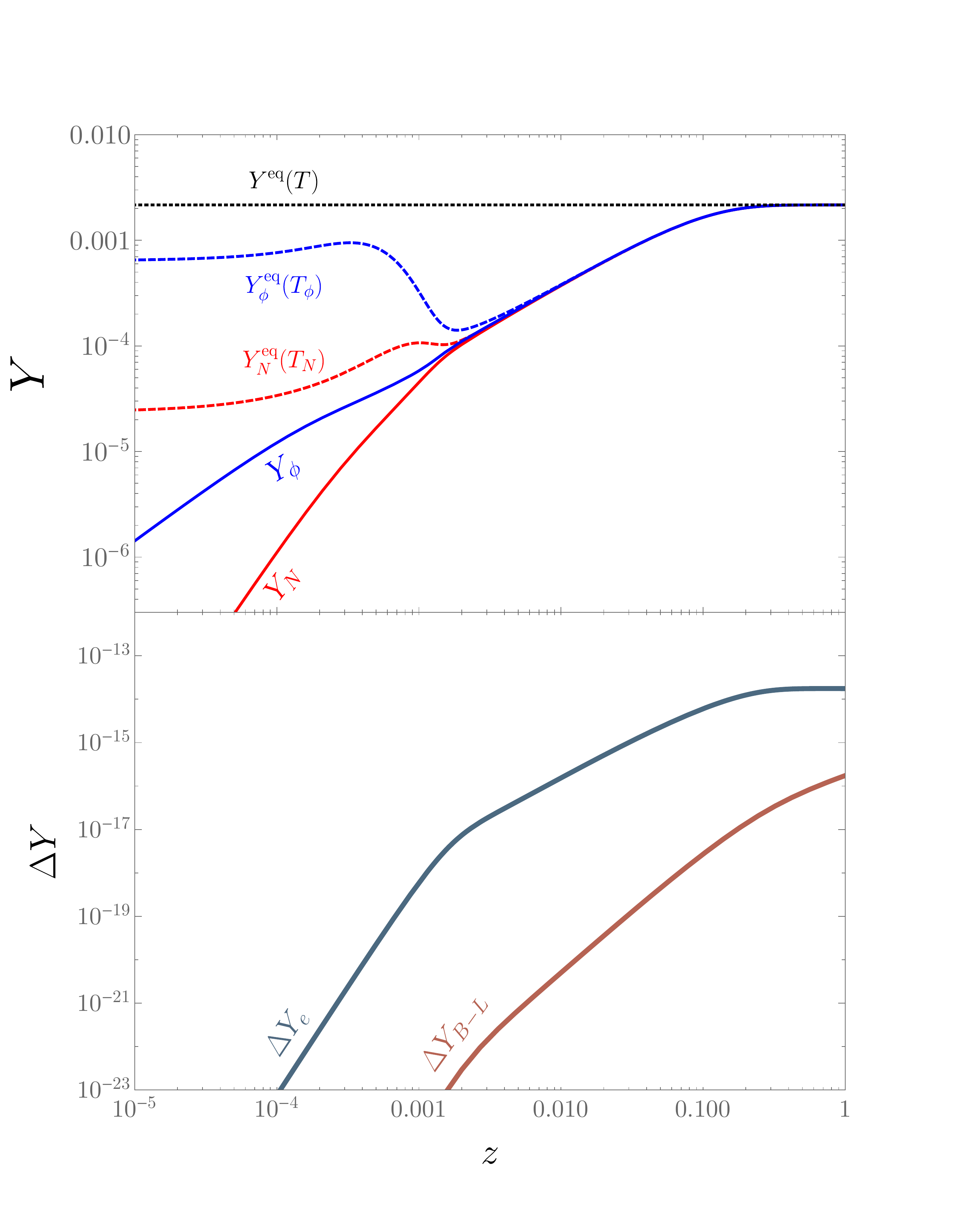}
                     \caption{
Dimensionless-time evolution of (top panel) hidden-sector abundances, (bottom panel) electron flavor and $B-L$ asymmetries, for $M_\phi=1$ GeV, $\Delta M=3\times10^{-8}$ GeV, $\lambda=5\times10^{-6}$, $y=5\times10^{-3}$, and SM Higgs coupling $F^{(I)}$ from Eq.~\eqref{eq:first_yukawa}.
    }
    \label{fig:fullequlibration_asymmetry_compare}
\end{figure}

\subsection{Leptogenesis Results} \label{sec:phi_neq_lepto_results}
We begin our numerical study of the full effects of hidden-sector equilibration on leptogenesis by examining the evolution of the lepton asymmetries as a function of time for scenarios with $y\gtrsim\lambda$. For concreteness, we take $M_\phi=1\,\,\mathrm{GeV}$ and use the benchmark Yukawa coupling $F^{(I)}$ from Eq.~\eqref{eq:first_yukawa} and RHN mass splitting $\Delta M = 3\times10^{-8}\,\,\mathrm{GeV}$. We show the time evolution of the hidden-sector abundances and lepton asymmetries in Fig.~\ref{fig:fullequlibration_asymmetry_compare} for the case $\lambda=5\times10^{-6}$, $y=5\times10^{-3}$. For $z\lesssim10^{-3}$, the hidden sector has not reached equilibrium and the asymmetry is generated as usual. For $z\gtrsim 0.1$, the RHNs have equilibrated with the SM and flavor asymmetry generation halts entirely. For intermediate values of $z$, we see that the hidden sector has reached internal equilibrium but is \emph{not} in equilibrium with the SM, which suppresses but does not entirely stop the generation of lepton flavor asymmetries.

We can understand the intermediate suppression of the asymmetry as follows:~scattering within the hidden sector tends to drive the RHN density matrix to be $(Y_N)_{IJ}=Y_N^{\rm eq}(T_N)\delta_{IJ}$. However, because $T\neq T_N$ we have $(Y_N)_{IJ} \neq Y_N^{\rm eq}(T) \delta_{IJ}$, and there is still net production of RHNs from SM Higgs decays. These two processes reach a quasi-steady state with each $R_N$ element having magnitude $\sim \left[1-Y_{\tilde N}/ u Y_N^{\rm eq}(T)\right]\langle\Gamma_{\phi\rightarrow N_I N_I}\rangle^{-1}$.
 This gives rise to a polynomial suppression of the off-diagonal elements of $(R_N)_{IJ}$ (and, consequently, the $CP$-asymmetry source) in inverse powers of $y$. By contrast, for $z\gtrsim0.1$, $N$ reaches the same temperature and abundance as the SM and there is no more net production of RHNs, leading to an exponential decay of the off-diagonal elements of the density matrix. This explains the much sharper turnoff of asymmetry generation once the RHNs come into full equilibrium with the SM.

\begin{figure}[t]
        \includegraphics[width=\columnwidth]{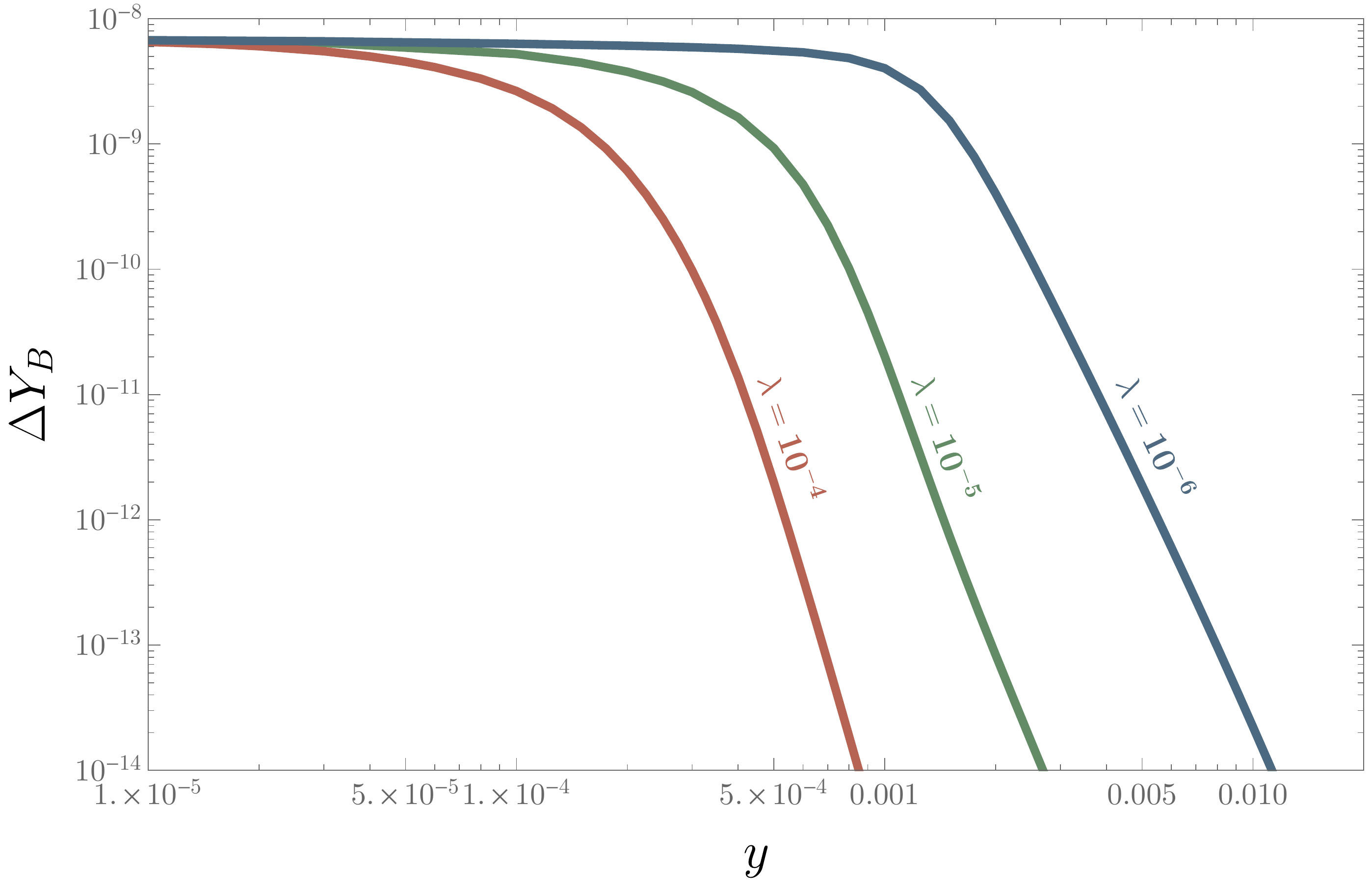}
                     \caption{
Baryon asymmetry including the full treatment of hidden-sector equilibration for $M_\phi=1$ GeV, $\Delta M=3\times10^{-8}$ GeV, SM Higgs coupling $F^{(I)}$, and the indicated values of $\lambda$.
    }
    \label{fig:asymmetry_fulleq_ydep}
\end{figure}

To determine the dependence of the asymmetry on the hidden-sector couplings, we fix $M_\phi$, $\Delta M$, and $F^{(I)}$ to the above values, and we vary $\lambda$ and $y$. The resulting baryon asymmetry is shown in Fig.~\ref{fig:asymmetry_fulleq_ydep}. For values of $y$ that are just large enough to equilibrate RHNs before oscillations, we recover the $y^{-10}$ power law dependence of the asymmetry. When $y$ is larger, however, we see a break in the power-law dependence for values of $y$ that correspond to internal equilibration within the hidden sector prior to $\phi$ coming into equilibrium with the SM. This break can be seen, for example, in the vicinity of $y=0.0015$ for $\lambda=10^{-5}$ in Fig.~\ref{fig:asymmetry_fulleq_ydep}, and the power law softens to between $y^{-6.5}$ and $y^{-7}$. This is because of the effect seen above where local equilibrium within the hidden sector suppresses, but does not completely halt, asymmetry generation. The asymmetry no longer has a simple power-law dependence on $\lambda$ either, since the asymmetry for any given benchmark point depends on an interplay of rates and temperatures involving both the SM and the hidden sector. Nevertheless, we still observe a steep suppression of the asymmetry as functions of both $\lambda$ and $y$.


\begin{figure}[t]
        \includegraphics[width=\columnwidth]{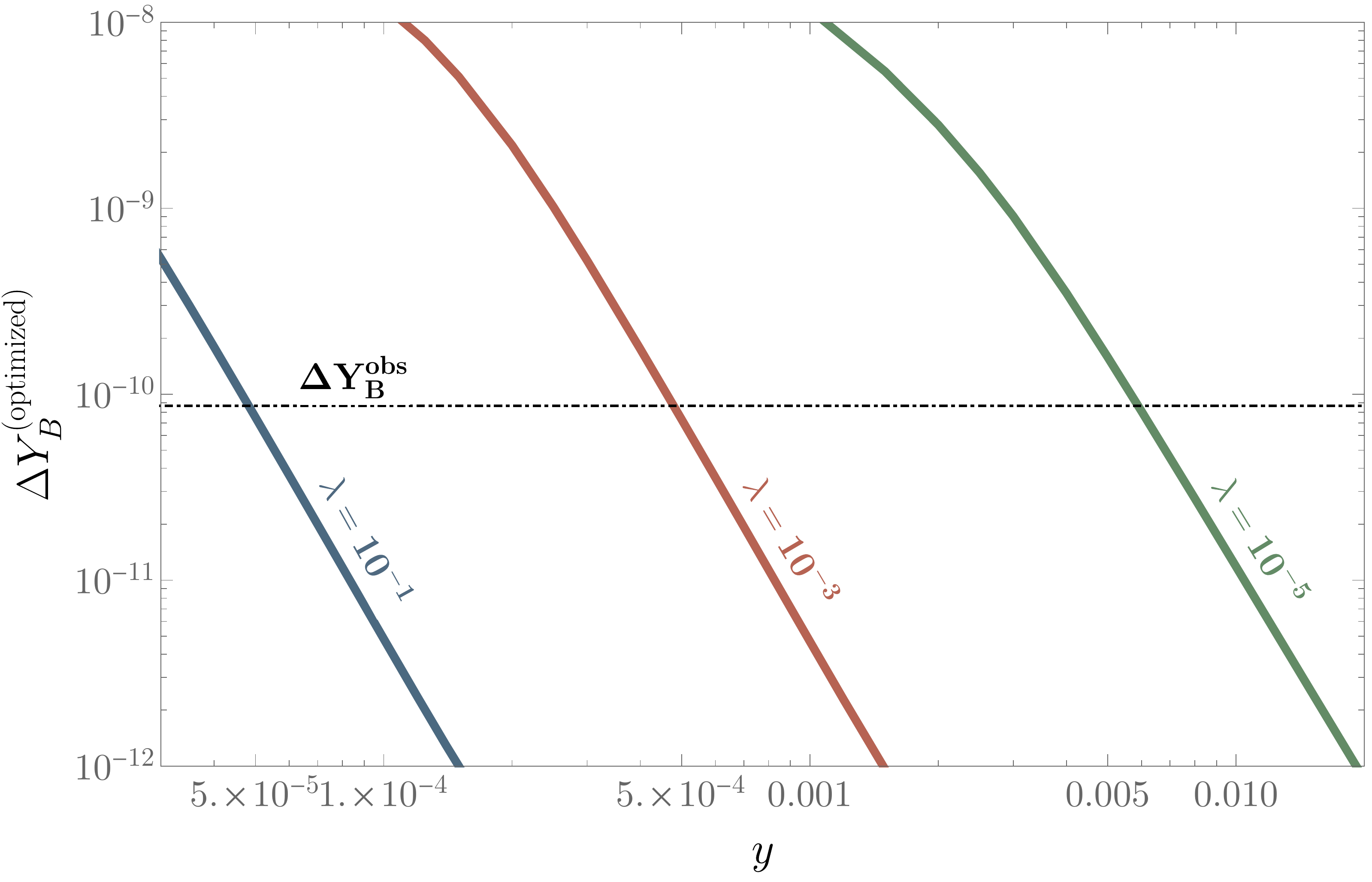}
                     \caption{
(Left) Baryon asymmetry as a function of $y$ obtained using the Yukawa texture $F^{(II)}$ from Eq.~\eqref{eq:second_yukawa} including the full treatment of hidden-sector equilibration. We have set $M_\phi=1\,\,\mathrm{GeV}$, and for each value of $y$ and $\lambda$ we optimized the mass splitting to give the maximum baryon asymmetry. The dot-dashed line indicates the observed baryon asymmetry. The ARS asymmetry corresponding to $y=0$ is $\Delta Y_B=2\times10^{-7}$.   }
    \label{asymmetry_fulleq_optimized}
\end{figure}


We finally turn to the optimal baryon asymmetry given a set of hidden-sector couplings $\lambda$ and $y$. We pursue a similar strategy as in Sec.~\ref{sec:viable_baryo_phiequilibrium}, where we choose the set of couplings $F^{(II)}$ given in Eq.~\eqref{eq:second_yukawa} that are enhanced relative to the na\"ive see-saw prediction so as to give the largest asymmetry without equilibrating the asymmetries in all three flavors of leptons. For each value of $y$ and $\lambda$, we choose the value of the mass splitting $\Delta M$ for which the baryon asymmetry is largest. We show the $y$-dependence of the optimized baryon asymmetry in Fig.~\ref{asymmetry_fulleq_optimized} for several values of $\lambda$ spanning various hierarchies for the hidden-sector couplings. At large $\lambda$, we recover the $y^{-4}$ power-law dependence seen in Sec.~\ref{sec:viable_baryo_phiequilibrium}, confirming that our earlier assumption that $\phi$ is always in equilibrium allowed us to predict the correct parametric dependence of the asymmetry. For smaller values of $\lambda$, the full equilibration of the hidden sector becomes important, although even in this case the power law is only marginally softer (by up to 10\% in the exponent).

Viable baryogenesis necessitates obtaining the observed value of $\Delta Y_B$. For $\lambda\gtrsim10^{-3}$, we obtain a consistent limit $y\sqrt{\lambda}\lesssim1.5\times10^{-5}$, which is nearly the same as our result from Sec.~\ref{sec:viable_baryo_phiequilibrium}. For smaller values of $\lambda$, it is slightly relaxed to $y\sqrt{\lambda}\lesssim2\times10^{-5}$.  Remarkably, these results are consistent with our simplistic perturbative analysis, as well as the analysis assuming that $\phi$ is always in equilibrium. This is in part due to the very steep suppression of the asymmetry with respect to couplings, such that even substantial changes in the asymmetry from variations in the assumptions underlying the hidden sector are compensated by minor adjustments to the couplings.

  \section{Phenomenology \& Implications for Leptogenesis}\label{sec:pheno}
  
  We have found that, for the case of a Higgs-portal scalar coupled to RHNs, viable baryogenesis requires $y\sqrt{\lambda}\lesssim2\times10^{-5}$ over a wide range of parameters. In this section, we investigate the phenomenological implications.
  
  At some level, $\phi$  inevitably mixes with the Higgs via loop-induced processes involving the RHNs and charged leptons. This is typically very small, being suppressed by the square of the Yukawa couplings, $F$, between the RHNs and SM Higgs. If $\phi$ gets a VEV ($v_\phi$), however, then mass mixing occurs at tree level:
  \be
  V &\supset & \frac{\lambda v_\phi v}{4}\,\phi h,
  \ee
where $h$ is the real, uneaten Higgs boson field. From measurements of the SM Higgs couplings and other collider probes, the strongest constraint on the mixing angle between $h$ and $\phi$  over the kinematic range of interest to us is $\theta\lesssim0.07$ \cite{Kozaczuk:2019pet,Carena:2019une}. In the limit of small mixing, the SM-dark Higgs mixing angle is
  \be
  \theta \approx \frac{\lambda v_\phi v}{4 (m_h^2-m_\phi^2)},
  \ee
 and we assume in the following that the two scalars are non-degenerate.
 
 $v_\phi$ is \emph{a priori} undetermined by the couplings $\lambda$ and $y$. We can identify two well-motivated possibilities:~in the first, $v_\phi$ is responsible for giving the RHNs mass\footnote{If this is the case, we should also consider the timing of the lepton-number-breaking phase transition and whether it was valid to use the zero-temperature RHN masses in our analysis. We address this point in Appendix~\ref{sec:thermal_masses}, showing that our general conclusions hold even in the case of vanishing tree-level RHN masses in the early universe.}, and consequently satisfies the relation $v_\phi = M_N/y$. This allows us to relate $v_\phi$ to existing model parameters. For the second possibility, we can imagine that $\phi$ is not the dominant source of mass for the RHNs, in which case $v_\phi$ is a free parameter. However, $m_\phi$ is in principle related to $v_\phi$ and the self-quartic coupling, $\lambda_{\rm s}$, by $m_\phi^2 = \lambda_{\rm s}v_\phi^2 / 3$, and since $\lambda_{\rm s}$ acquires radiative corrections from $\lambda$ the $\phi$ mass cannot be arbitrarily decoupled from $v_\phi$. Therefore, it is a well-motivated possibility that $v_\phi \sim$1--100 GeV, in which case it is straightforward to accommodate $\phi$ masses throughout the phenomenologically relevant range.

 There are at least three processes of phenomenological interest in probing  RHN couplings within the hidden sector:~SM Higgs boson decays to RHN pairs \cite{Graesser:2007yj,Graesser:2007pc,Shoemaker:2010fg,Cely:2012bz,Accomando:2016rpc,Caputo:2017pit,Mason:2019okp}, SM Higgs boson decays to $\phi$ pairs (followed by $\phi\rightarrow N_I N_I$) \cite{Nemevsek:2016enw}, and direct production of $\phi$ in heavy-quark meson decays, with subsequent decay to RHNs. We now consider each in turn.\\
  
  \noindent {\bf SM Higgs Decays to RHN pairs:}~The branching fraction summed over two RHN species is
  \be
  \mathrm{BF}(h\rightarrow N_I N_I) &\approx& \frac{\theta^2 y^2 m_h}{8\pi\Gamma_h},
  \ee
  where $\Gamma_h\approx 4$ MeV is the SM Higgs width.
  
\begin{figure}[t]
        \includegraphics[width=\columnwidth]{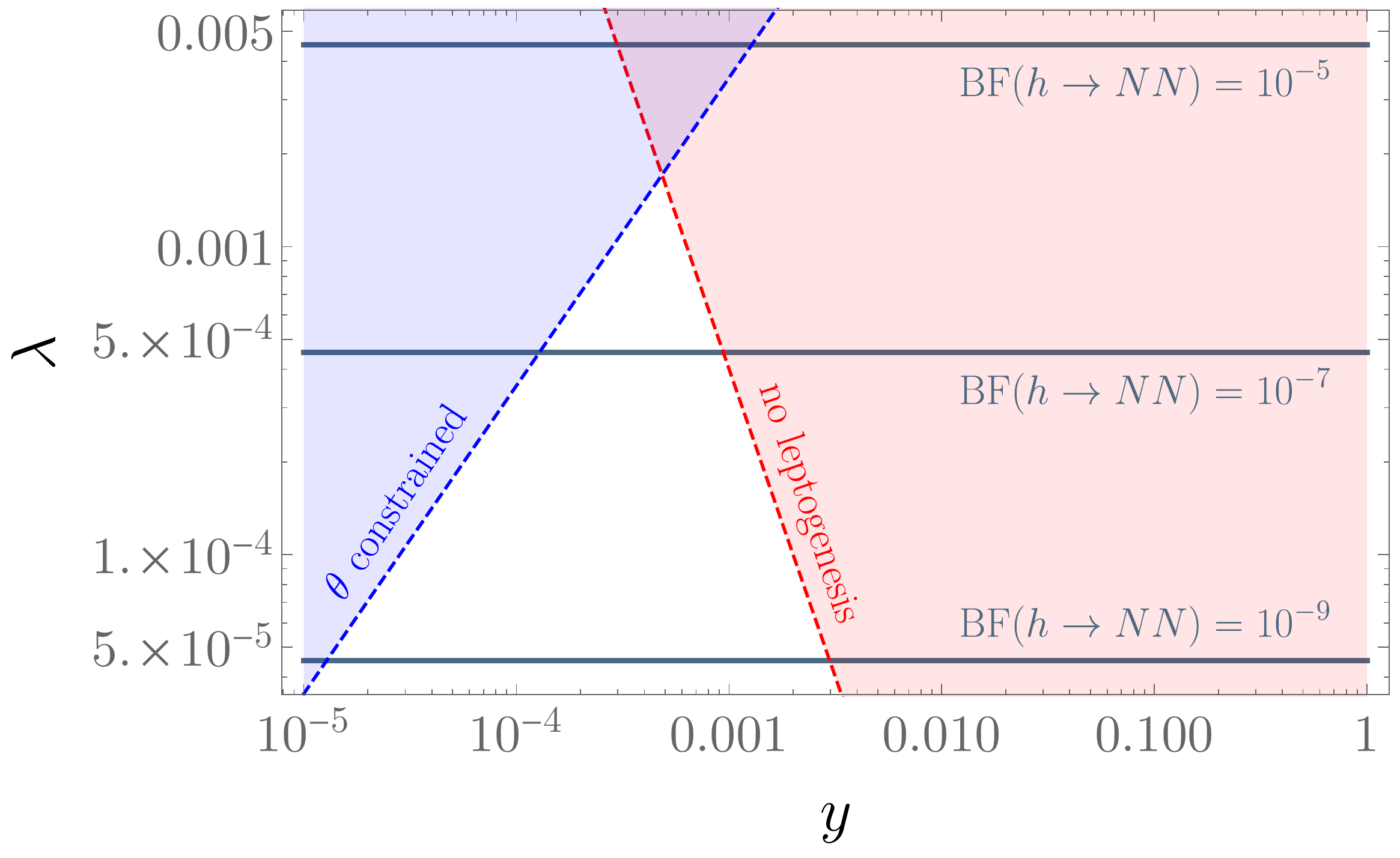}
                     \caption{
SM Higgs branching fraction to RHN pairs, summed over two RHN flavors, with $M_\phi=15$ GeV, $M_N=5$ GeV and assuming $M_N=yv_\phi$. The red shaded region corresponds to parameters incompatible with freeze-in leptogenesis, while the blue shaded region indicates where the mixing angle $\theta$ exceeds the level at which it is constrained by Higgs coupling measurements and direct searches.  }
    \label{higgsBF_l_ll_y}
\end{figure}

For the case in which the RHNs acquire their masses from $v_\phi$, the $y$-dependence cancels entirely to give
  \be
    \mathrm{BF}(h\rightarrow N_I N_I)&\approx& \frac{\lambda^2 M_{N_I}^2 v^2 m_h}{128\pi(m_h^2-m_\phi^2)^2\Gamma_h}.
  \ee
  In Fig.~\ref{higgsBF_l_ll_y}, we plot contours of this branching fraction as a function of $\lambda$ for $M_\phi = 15$ GeV, $M_N=5$ GeV, and assuming $v_\phi = M_N/y$. We indicate the region incompatible with leptogenesis, $y\sqrt{\lambda}\gtrsim2\times10^{-5}$, as well as the combination of couplings for which $\theta$ would exceed 0.07 and potentially conflict with constraints on the mixing angle from direct or indirect searches. In most of the blue shaded region, the mixing angle is large because $v_\phi\gtrsim v$.  
  

  Since we expect $\sim10^8$ Higgs bosons at the LHC \cite{LHCHiggsCrossSectionWorkingGroup:2011wcg}, we expect that in the most optimistic scenario it will be possible to achieve sensitivity to branching fractions $\sim10^{-7}$. There is consequently a small sliver of parameter space consistent with freeze-in leptogenesis and an LHC signal in SM Higgs decays, but for most of the branching fractions that can be probed in LHC searches, a discovery would strongly disfavor RHN involvement in freeze-in leptogenesis.
  

\begin{figure}[t]
        \includegraphics[width=\columnwidth]{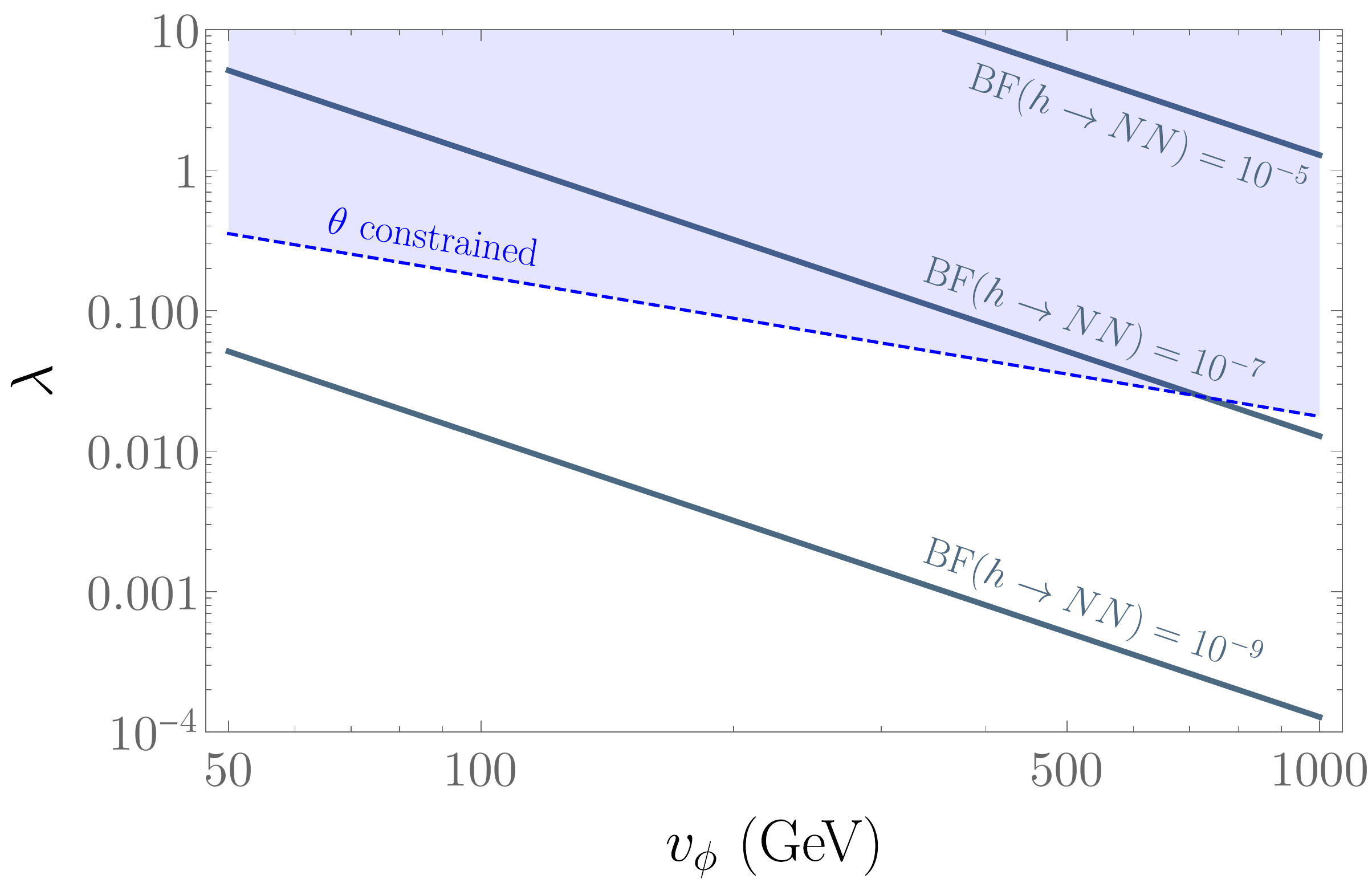}
                     \caption{
SM Higgs branching fraction to RHN pairs, summed over two RHN flavors, with $M_\phi=15$ GeV and $v_\phi$ treated as a free parameter (unrelated to RHN masses). The coupling $y$ is set to the largest value consistent with freeze-in leptogenesis, while the blue shaded region indicates where the mixing angle $\theta$ exceeds the level at which it is constrained by Higgs coupling measurements and direct searches.  }
    \label{higgsBF_vphi}
\end{figure}


Alternatively, if RHNs do not acquire mass through spontaneous symmetry breaking, then $v_\phi$ is a free parameter. In Fig.~\ref{higgsBF_vphi}, we show contours of the SM Higgs branching fraction to RHNs as a function of $v_\phi$ and $\lambda$. We fix $y$ by setting it to the largest value allowed by leptogenesis. Once again, we see a region of parameter space that is marginally testable with $v_\phi\gtrsim$ TeV, although there must be a hierarchy of hidden-sector parameters $v_\phi \gg M_\phi$. \\

  \noindent {\bf SM Higgs Decays to $\phi$ pairs:}~The decay $h\rightarrow\phi\phi$ occurs for all $m_\phi < m_h/2$ with approximate branching fraction (neglecting phase-space suppression)
  \be
  \mathrm{BF}(h\rightarrow\phi\phi) &\approx& \frac{\lambda^2 v^2}{128\pi m_h\Gamma_h}\\
  &\approx& 3\times10^{-8}\left(\frac{\lambda}{10^{-5}}\right)^2.
  \ee
  This rate is independent of both $y$ and $v_\phi$. However, if we want to test the coupling of $\phi$ to RHNs, we should also require $\phi\rightarrow N_I N_I$ decay. The rate we are therefore interested in is the product branching fraction summing over RHN flavors,
  \be\label{eq:hphiphi_productbf}
  \mathrm{BF}(h\rightarrow\phi\phi)\mathrm{BF}(\phi\rightarrow NN)^2 \quad\quad\quad\quad\quad\quad\quad\\
  \approx \frac{\lambda^2 v^2}{128\pi m_h\Gamma_h}\left(\frac{y^2}{y^2+\theta^2\beta m_f^2/v^2}\right)^2,
  \ee
  where $f$ is the heaviest fermion flavor to which $\phi$ can decay, and $\beta=3$ (1) is a color factor for decays to quarks (charged leptons). Note that for expediency we again neglect phase-space suppressions, which typically give $\mathcal{O}(1)$ corrections to these branching fractions.


\begin{figure}[t]
        \includegraphics[width=\columnwidth]{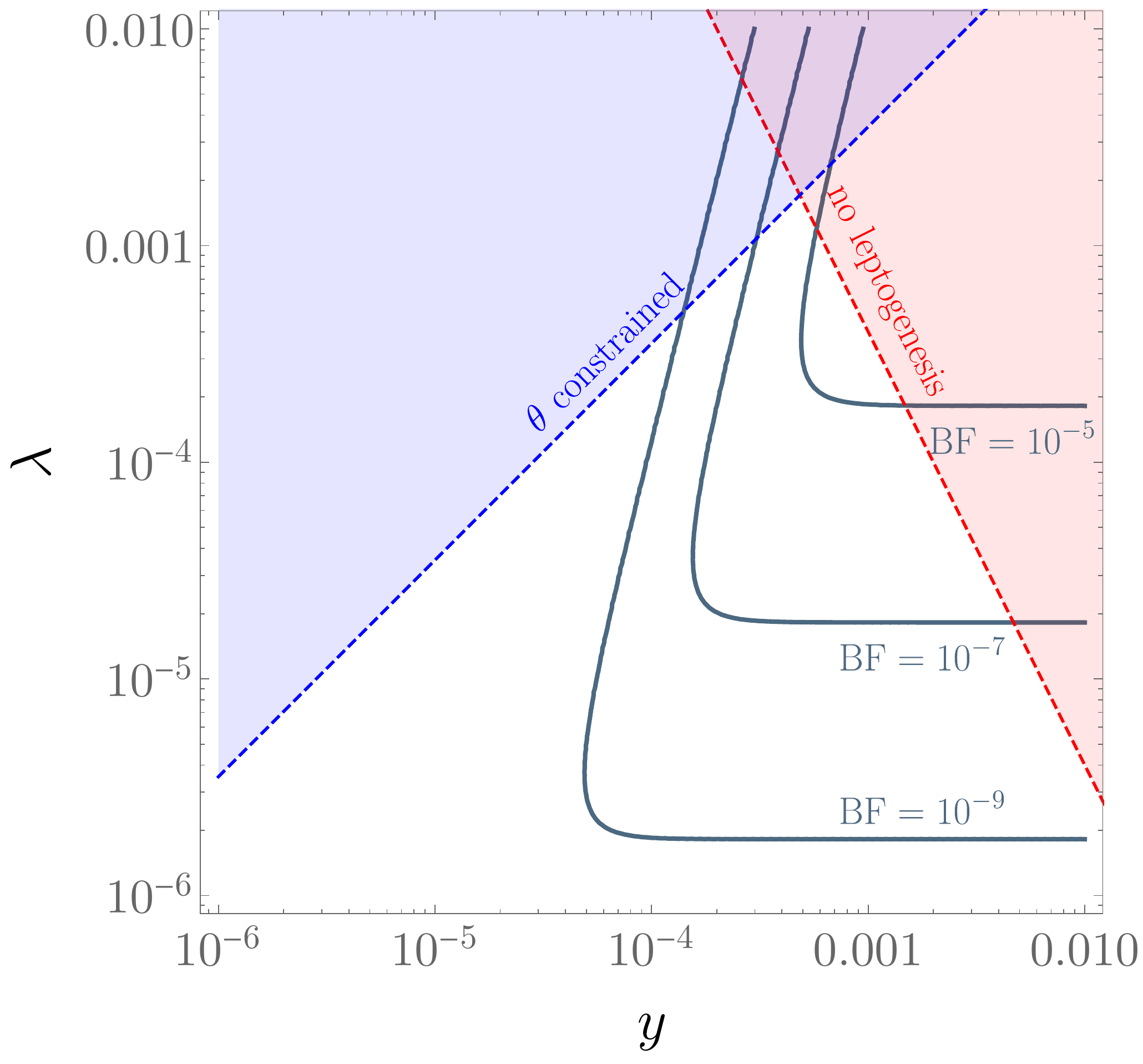}
                     \caption{
SM Higgs product branching fraction $\mathrm{BF}(h\rightarrow\phi\phi)\mathrm{BF}(\phi\rightarrow NN)^2$, with $M_\phi=15$ GeV, $M_N=5$ GeV, and $v_\phi=M_N/y$. The red and blue shaded regions are the same as in Fig.~\ref{higgsBF_l_ll_y}.  }
    \label{higgsBF_phi_y}
\end{figure}


We first consider the case where RHN masses originate from $v_\phi =  M_N/y$, showing our results in Fig.~\ref{higgsBF_phi_y} for a benchmark point with $M_\phi=15$ GeV and $M_N=5$ GeV. The prospects are somewhat more favorable for this signal than for direct $h\rightarrow NN$ decays, as branching fractions as large as $10^{-5}$ can be compatible with leptogenesis. This is still, however, a challenging target to reach given the difficulties of triggering and reconstructing exotic Higgs decays at the LHC \cite{Curtin:2013fra}.


\begin{figure}[t]
        \includegraphics[width=\columnwidth]{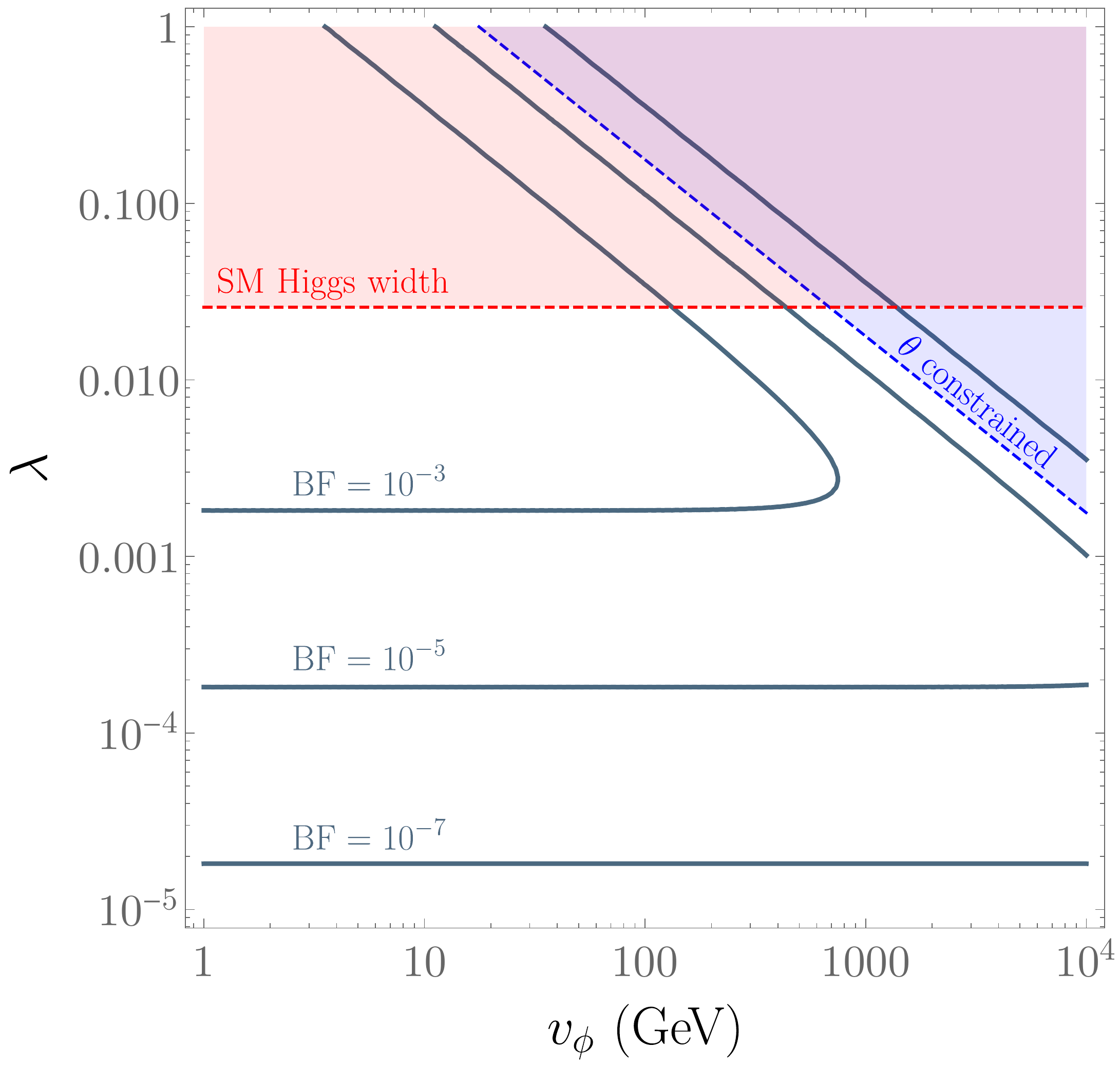}
                     \caption{
SM Higgs product branching fraction $\mathrm{BF}(h\rightarrow\phi\phi)\mathrm{BF}(\phi\rightarrow NN)^2$, with $M_\phi=15$ GeV, and $v_\phi$ treated as a free parameter (unrelated to RHN masses). The coupling $y$ is set to the largest value consistent with freeze-in leptogenesis. The red region indicates parameters that are ruled out by measurements of the SM Higgs width, while the blue region indicates where the mixing angle $\theta$ exceeds the level at which it is constrained by Higgs coupling measurements and direct searches.  }
    \label{higgsBF_phi_vphi}
\end{figure}


We next turn to the case where RHN masses have a separate origin and $v_\phi$ is a free parameter. We show our results in Fig.~\ref{higgsBF_phi_vphi}, and we see that leptogenesis is much less constraining of this scenario. The reason is that the mixing between $\phi$ and $h$ can be made arbitrarily small by taking $v_\phi \ll v$, while the branching fraction Eq.~\eqref{eq:hphiphi_productbf} is unaffected. Consequently, it is possible to take $y\ll\lambda$ and preserve the lepton asymmetry. In other words, the Higgs-portal coupling is large, leading to a significant $h\rightarrow \phi\phi$ rate, and as long as $\theta$ is appropriately small the $\phi$ particles still predominantly decay into RHNs even with a tiny value of $y$. 

In the small-mixing, large-$\lambda$ limit, the dominant model-independent constraints come from measurements of the SM Higgs width~\cite{CMS:2019ekd} such that the  Higgs branching fraction into $\phi$ is comparable to the total SM Higgs width. In practice, there can be stronger constraints from searches for $h\rightarrow \phi\phi$ but this depends on the RHN lifetime and mixing angle with particular lepton flavors and so we do not explicitly calculate these model-dependent constraints; a dedicated study is certainly merited. However, sensitivity to branching fractions $\gtrsim10^{-5}$ is potentially achievable, meaning that searches for $h\rightarrow\phi\phi\rightarrow 4N$ can likely probe parameter space motivated by leptogenesis.

It is worth noting that the above conclusion holds even if we take $v_\phi\rightarrow0$ such that $\phi$ is not responsible for lepton-number breaking. In this case, there is no mixing between $\phi$ and the SM Higgs and $\phi$ always decays 100\% to RHNs. It is therefore possible to get large $h\rightarrow\phi\phi\rightarrow 4N$ signals by taking $\lambda \gg y$ in the absence of mixing.\\

\noindent {\bf $B$-meson Decays to $\phi$:~}For $m_B > m_\phi+m_K$, the dominant production mechanism is $B\rightarrow X_s \phi$, with an inclusive rate of $\mathrm{BF}(B\rightarrow X_s \phi)\approx3.3\theta^2$ for $m_\phi\ll m_B$ \cite{Boiarska:2019jym}. The corresponding rate for the exclusive $B\rightarrow K\phi$ process is $\mathrm{BF}(B\rightarrow K \phi)\approx0.43\theta^2$. Given the sensitivity of existing LHCb searches \cite{LHCb:2016awg}, a sensitivity to branching fractions $\sim10^{-10}$ for high-efficiency, low-background searches seems feasible in the near future.


\begin{figure}[t]
        \includegraphics[width=\columnwidth]{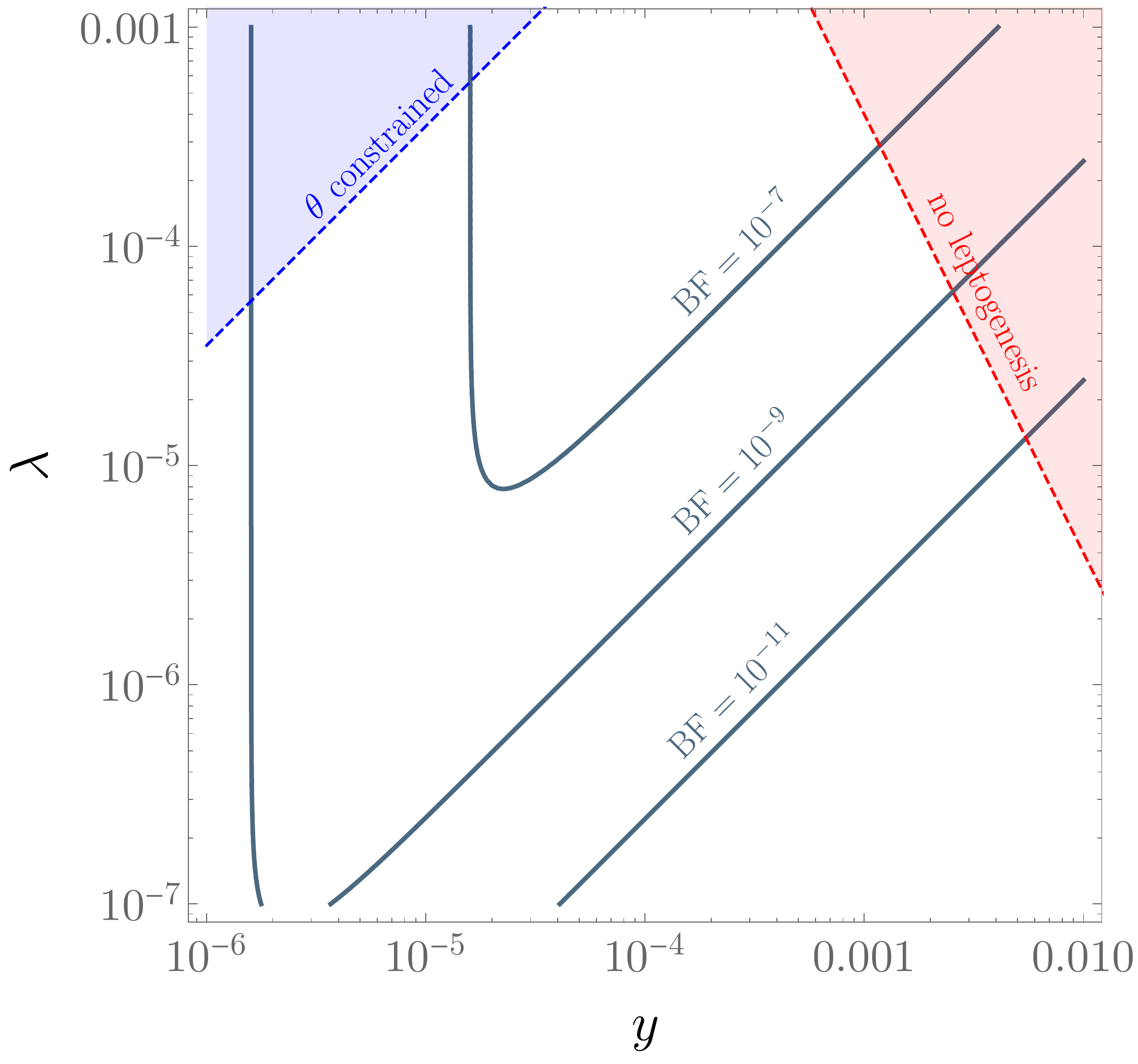}
                     \caption{
$B$-meson product branching fraction $\mathrm{BF}(B\rightarrow K\phi)\mathrm{BF}(\phi\rightarrow NN)$, with $M_\phi=2$ GeV, $M_N=0.5$ GeV, and $v_\phi=M_N/y$. The red and blue shaded regions are the same as in Fig.~\ref{higgsBF_l_ll_y}.  }
    \label{BBF_y}
\end{figure}


We show the $B\rightarrow K\phi,\,\phi\rightarrow NN$ product branching fraction as a function of the hidden-sector couplings in Fig.~\ref{BBF_y} under the assumption $v_\phi = M_N/y$. We consider a benchmark with $M_\phi=2$ GeV and $M_N=0.5$ GeV. We see that there is a relatively wide range of testable parameter space, although both $y$ and $\lambda$ must be very small. An appreciable mixing results because $v_\phi \gg v \gg m_\phi$, implying a significant hierarchy of scales in the hidden sector.


\begin{figure}[t]
        \includegraphics[width=\columnwidth]{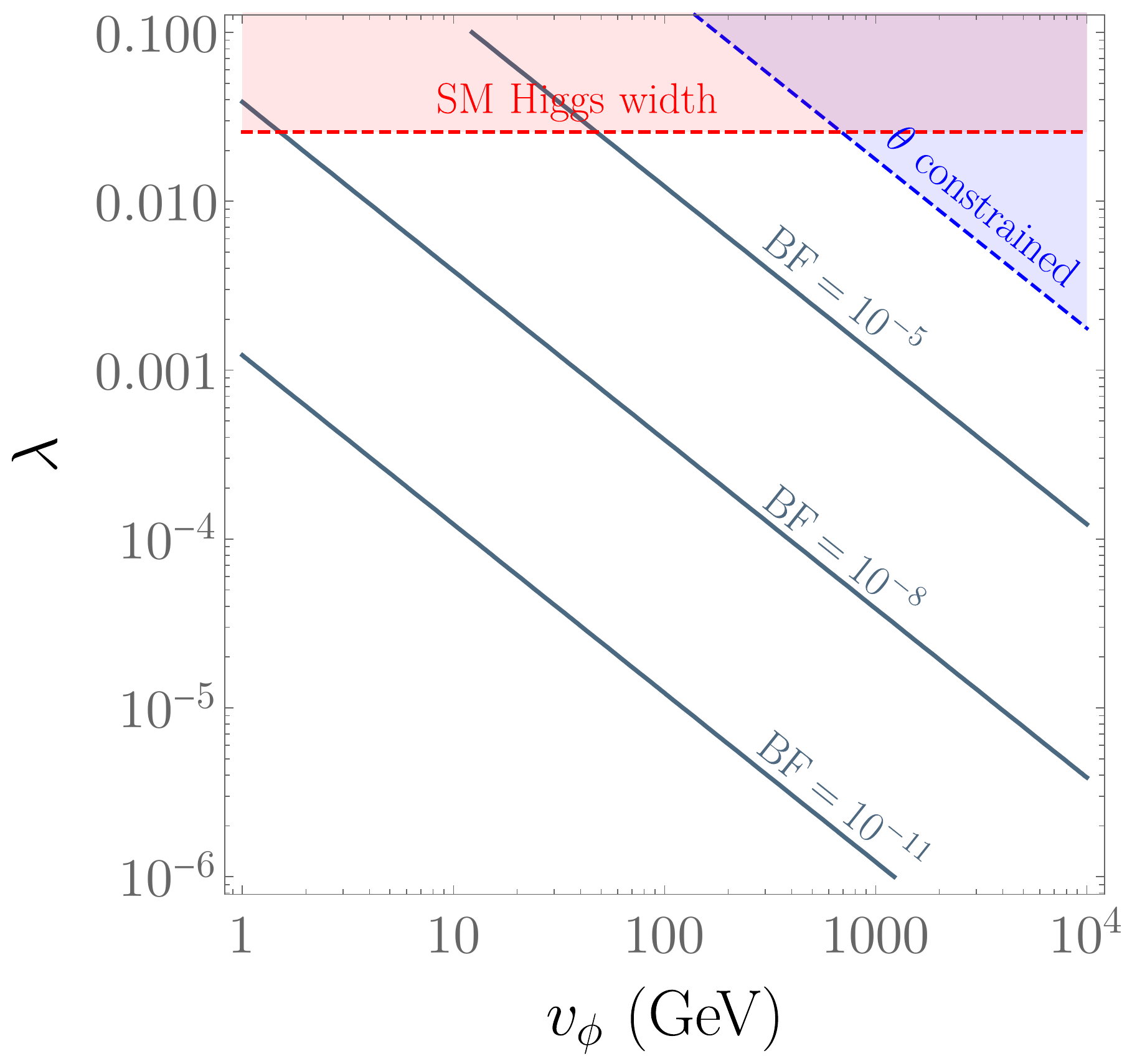}
                     \caption{
$B$-meson product branching fraction $\mathrm{BF}(B\rightarrow K\phi)\mathrm{BF}(\phi\rightarrow NN)$, with $M_\phi=2$ GeV, $M_N=0.5$ GeV, and $v_\phi$ treated as a free parameter (unrelated to RHN masses). The red and blue shaded regions are the same as in Fig.~\ref{higgsBF_phi_vphi}.  }
    \label{BBF_vphi}
\end{figure}


Finally, we consider the hypothesis where $v_\phi$ is a free parameter unrelated to RHN masses, and we plot the $B$ decay branching fractions in Fig.~\ref{BBF_vphi}. It is evident that a substantial parameter space is accessible to experiment while simultaneously being consistent with the observed baryon asymmetry. Part of the reason for this is that the heaviest available SM fermions into which $\phi$ can decay are strange quarks and muons, both of which have tiny masses. This means that a substantial branching fraction of $\phi\rightarrow NN$ occurs even with large $\theta$ and small $y$ to maintain the viability of leptogenesis. Furthermore, the large number of $B$ mesons expected at LHCb and $B$-factories allow even tiny branching fractions to be proved in the near future.\\

  \noindent {\bf Summary:}~The consistency of observable phenomenological signatures with leptogenesis depends strongly on the particular model and search mode. We find in general that it is very challenging to  accommodate a large $h\rightarrow NN$ decay rate while simultaneously satisfying SM Higgs coupling observations and  the observed baryon asymmetry through freeze-in leptogenesis. Similarly, there is only a narrow part of parameter space that can be tested at the LHC in $h\rightarrow\phi\phi,\,\phi\rightarrow NN$ decays provided $\phi$ is responsible for the generation of RHN masses. By contrast, if RHN masses are unrelated to the $\phi$ VEV, large signals can be accommodated in $h\rightarrow\phi\phi,\,\,\phi\rightarrow NN$, and seeing such a signal would strongly point towards a hidden sector with $\lambda\gg y$ for consistency with leptogenesis. If a large width were observed for $\phi$, this would disfavor leptogenesis mediated by the RHNs. Finally, we find that observable signals in $B\rightarrow K\phi,\,\phi\rightarrow NN$ are readily consistent with leptogenesis, although this channel is only relevant for $\phi$ masses below the $B$ mass.
 
 There are various ways to more carefully compare hidden-sector phenomenology with freeze-in leptogenesis. For example, we have neglected direct $\phi$ production modes such as $pp\rightarrow\phi$ in gluon fusion, which could be an order of magnitude larger than SM-Higgs-mediated production modes for light $M_\phi$. Furthermore, the overlap of phenomenologically accessible parameter spaces with those of successful leptogenesis could be more precisely determined with dedicated analyses of the signals and backgrounds for each production mode, decay mode, and lifetime. Nevertheless, our analysis provides a clear indication of which signals could falsify leptogenesis and which signals are largely consistent with the observed asymmetry; we leave a dedicated  study to future work.

  \section{Conclusions}
  We have performed a comprehensive study of the implications for freeze-in leptogenesis of hidden-sector interactions involving right-handed neutrinos. In particular, we have analytically derived the suppression of the lepton asymmetry due to early equilibration of RHNs from hidden-sector interactions. We have also conducted a numerical study for a particular model including two RHNs and a dark scalar, $\phi$. The resulting baryon asymmetry is significantly suppressed provided the RHNs come into equilibrium prior to the RHN oscillation time, and we have derived a bound from freeze-in leptogenesis on the hidden-sector couplings that is robust for different coupling hierarchies and equilibration timescales. 
  
  We further considered the phenomenological implications of our leptogenesis results, studying possible signatures of the $\phi$-RHN interaction at high- and low-energy colliders and mapping the couplings consistent with freeze-in leptogenesis into  $\phi$ and RHN production rates. We have found that an observation of the decay $h\rightarrow NN$ would likely conflict with the requirement of obtaining the observed asymmetry through freeze-in leptogenesis, while other exotic Higgs and $B$-meson decays are consistent with freeze-in leptogenesis. Our work informs the compatibility of freeze-in leptogenesis with different experimental searches of interest.
  
Given the concordance of our numerical findings with the analytic results derived for the general case in Sec.~\ref{sec:newinteractions}, we expect that our results directly extend to related models such as a $Z'$ coupled to the $\mathrm{U}(1)_{B-L}$ current and, hence, coupled to the RHNs  as well as dictated by anomaly cancellation. Such a model also requires a dark Higgs in order to generate Majorana masses for the RHNs, and if anything the constraints on the couplings should be even more strict compared to the simple model we have studied.

Finally, we remark that in recent years  there have been additional refinements to the quantum kinetic equations for leptogenesis to account for changes in rates in the broken electroweak phase and the gradual process of sphaleron decoupling, among other improvements (see \emph{e.g.,} Ref.~\cite{Klaric:2021tdt}). These works have also highlighted the possible role of freeze-out leptogenesis in low-scale RHN models \cite{Hambye:2016sby,Hambye:2017elz,Klaric:2021tdt,Drewes:2021nqr}. We have briefly touched on this point in Appendix  \ref{sec:freezeout}, and generally expect our results to qualitatively hold even with these improvements to the calculation of the lepton asymmetry; however, a more comprehensive study is warranted to determine the precise implications for the interplay between the hidden-sector phenomenology and viable leptogenesis.
  
\acknowledgments

We are grateful to Marco Drewes, Carlos Tamarit, and Dave Tucker-Smith for helpful conversations, and to Carlos Tamarit for  feedback on the manuscript. This work is supported by the U.S. National Science Foundation under Grant PHY-1820770.

\appendix

\section{Parametrization of Right-Handed Neutrino Couplings}\label{app:CI_param}

We assume that there are two RHNs, which results in the lightest SM neutrino being massless. The couplings between the RHNs, SM neutrinos, and SM Higgs can be parametrized following Casas and Ibarra \cite{Casas:2001sr},
\be
F &=& \frac{\sqrt 2i}{v}U_\nu\sqrt{m_\nu}R\sqrt{M_N},
\ee
where $m_\nu$ ($M_N$) is a $3\times3$ diagonal matrix of SM neutrino masses ($2\times2$ diagonal matrix of RHN masses), and $v=246\,\,\mathrm{GeV}$ is the SM Higgs VEV. $U_\nu$ is the PMNS matrix \cite{Pontecorvo:1957qd,Maki:1962mu} with Majorana phase $\eta$ and Dirac phase $\delta$,
\begin{widetext}
\small
\be
U_\nu &=&\left(\begin{array}{ccc} 
c_{12}c_{13} & e^{-i\eta}c_{13}s_{12} & s_{13}e^{-i\delta} \\
{}-c_{23}s_{12}-e^{i\delta}c_{12}s_{13}s_{23} & e^{-i\eta}\left(c_{12}c_{23}-e^{i\delta}s_{12}s_{13}s_{23}\right) & c_{13}s_{23} \\
s_{12}s_{23} -e^{i\delta}c_{12}c_{23}s_{13}& e^{-i\eta}\left(-e^{i\delta}c_{23}s_{12}s_{13}-c_{12}s_{23}\right) & c_{13}c_{23}
\end{array}\right),
\ee
\normalsize
\end{widetext}
where $c_{13}=\cos\theta_{13}$, etc. $R$ is an orthogonal RHN mixing matrix with complex angle $\omega$ given by
\be
R &=& \left(\begin{array}{cc} 
0 & 0 \\
\cos\omega & \sin\omega \\
-\sin\omega & \cos\omega \end{array}\right).
\ee
When the imaginary part of $\omega$ is large, then $\cos\omega \sim \cosh\omega$ and $\sin\omega \sim i\sinh\omega$, and we see that the Yukawa couplings grow exponentially even though their contributions to the SM neutrino masses are fixed due to cancellations among various terms. This can be the result of approximate lepton number symmetries that also make the RHN masses degenerate \cite{Shaposhnikov:2006nn}.

\section{Quantum Kinetic Equations for Leptogenesis}\label{app:be}
\subsection{Standard ARS Terms}\label{app:standard_ARS}
There have been extensive studies and refinements of the quantum kinetic equations for the evolution of the RHN density matrices and lepton flavor asymmetries in ARS leptogenesis (\emph{e.g.,} \cite{Asaka:2005pn,Canetti:2012kh,Hernandez:2016kel,Hambye:2017elz,Abada:2018oly,Eijima:2018qke}). For the standard ARS terms, we  use  the form and notation of the quantum kinetic equations and rates from Ref.~\cite{Abada:2018oly}. The equations determine the time evolution of the RHN density matrices, expressed as the dimensionless ratio $R_N = n_N/n_N^{\rm eq}$, and of the lepton flavor asymmetries in the anomaly-free quantities $B/3-L_\alpha$, which are expressed in terms of the corresponding chemical potential divided by the temperature, $\mu_{\Delta \alpha}$ for flavor $\alpha$. The washout terms depend not on the  $B/3-L_\alpha$ charge but on the actual asymmetry in lepton doublets, $\mu_\alpha$. The two quantities are related by the susceptibility matrix, $\chi$:
\be\label{eq:susceptibility}
\mu_\alpha &=& 2\sum_\beta \chi_{\alpha\beta} \mu_{\Delta \beta},\\
\chi &=& {}-\frac{1}{711}\left(\begin{array} {ccc} 
257 & 20 & 20 \\
20 & 257 & 20 \\
20 & 20 & 257
\end{array} \right).
\ee
Note the relative minus sign in $\chi$, which encodes the fact that there is a relative minus sign between the $L_\alpha$ and $B/3-L_\alpha$ charges. This makes the washout terms for $\mu_{\Delta_\alpha}$ have a positive coefficient when expressed in terms of $\mu_\alpha$. 

The ARS quantum kinetic equations  can be expressed in terms of the dimensionless time, $z=T_{\rm ew}/T$, giving \cite{Abada:2018oly}
\begin{widetext}
\small
\be
\frac{dR_N}{dz} &=& i\left[R_N,W_N\right] + 3iz^2\left[R_N,r\right] - \mathcal{C}^{(0)}\left\{R_N,W_N\right\} + 2\mathcal{C}^{(0)}W_N + \mathcal{C}^{(\mathrm{w.o.1})}o_\mu + \frac{1}{2}\mathcal{C}^{(\mathrm{w.o.2})}\left\{o_\mu,R_N\right\},\label{eq:full_YN_ARS}\\
\frac{32T_{\rm ew}}{M_0}\frac{d\mu_{\Delta\alpha}}{dz} &=& {}-\mathcal{C}^{(0)}\left(FR_NF^\dagger-F^*R_{\overline{N}}F^{\rm T}\right)_{\alpha\alpha}
+\mathcal{C}^{(\rm w.o.1)}\left(FF^\dagger\right)_{\alpha\alpha}\mu_\alpha+\frac{\mathcal{C}^{(\rm w.o.2)}}{2}\left(FR_NF^\dagger+F^*R_{\overline N}F^{\rm T}\right)_{\alpha\alpha}\mu_\alpha,
\ee
\normalsize
\end{widetext}
where
\be
W_N &=& \frac{\pi^2M_0}{144\zeta(3)T_{\rm ew}}F^\dagger F,\\
o_\mu &=& \frac{\pi^2 M_0}{144\zeta(3)T_{\rm ew}}F^\dagger\mu F,\\
r &=& \mathrm{diag}\left(0,\frac{\pi^2 M_0\Delta M_{21}^2}{108\zeta(3) T_{\rm ew}^3}\right),
\ee
are scattering and oscillation parameters for the ARS Yukawa couplings\footnote{Note that, through the rest of the paper, we have used Maxwell-Boltzmann statistics but the collision terms in this set of quantum kinetic equations have been derived assuming ultra-relativistic Fermi-Dirac RHNs. Since the differential equation is expressed in terms of $R_N$, the only effect of this is a $\sim10\%$ shift in scattering and oscillation rates relative to the Maxwell-Boltzmann predictions, which has a negligible effect on our conclusions. Therefore, for simplicity we use the form of the rates as presented in Ref.~\cite{Abada:2018oly} without attempting to correct them.}. The $R_{\overline{N}}$ evolution equation is the same as for $R_N$, but with $F\rightarrow F^*$ and $\mu\rightarrow -\mu$. 

The relationship between the dimensionless chemical potential and the yield for the $B/3-L_\alpha$ charges is
\be
\Delta Y_{\alpha} &=& \frac{45}{6\pi^2g_{*\mathrm{s}}}\,\mu_{\Delta\alpha},
\ee
which accounts for the sum over $\mathrm{SU}(2)$ indices in SM leptons. The values of $\mathcal{C}^{(i)}$ are
\be
\mathcal{C}^{(0)} &=& \frac{72\zeta(3)}{\pi^2 T}\langle\tilde\Gamma_h\rangle \approx 0.106,\\
\mathcal{C}^{(\mathrm{w.o.1})} &=& \frac{144\zeta(3)}{\pi^2 T}\langle\tilde\Gamma_{\rm w.o.1}\rangle \approx 0.114,\\
\mathcal{C}^{(\mathrm{w.o.2})} &=& \frac{144\zeta(3)}{\pi^2 T}\langle\tilde\Gamma_{\rm w.o.2}\rangle \approx 0.0526.
\ee
In the above, we have neglected scale-dependence of SM couplings and lepton-number-violating interactions (which we return to in Appendix \ref{sec:freezeout}), and a thermal average has been taken over RHN momenta (see Appendix \ref{app:momentum_averaging}). 

When solving the quantum kinetic equations  for large RHN mass splittings (corresponding to early oscillation times),  the numerical solution of the differential equations  can take a  very long time due to the  rapid oscillations occurring at late times. For parameter points for which this is an issue, we choose a cutoff time well after oscillations start and at which point the flavor asymmetries have flattened out, and feed the output of the differential equations at the cutoff time into a new set where we set the off-diagonal components of $R_N$ and $R_{\overline{N}}$ to zero and remove the oscillation terms from the quantum kinetic equations \cite{Shuve:2014zua} (we do not use this approach in the freeze-out calculations in Appendix \ref{sec:freezeout}, for which this assumption is not valid). To check the validity of this approach, we vary the cutoff time by a factor of 50\%, finding that for suitably chosen cutoff times the effect on the final asymmetry is below 1\%. This procedure is particularly robust when the RHNs equilibrate prior to oscillation, in which case the asymmetry very rapidly approaches a constant value.

\subsection{New RHN Interactions Assuming $\phi$ Always in Equilibrium}

We now include the effects of the model from Sec.~\ref{sec:darkhiggs} assuming that the dark scalar, $\phi$, is always in thermal equilibrium. In this case, we do not need to track the $\phi$ abundance and the number of quantum kinetic equations remains the same as for ARS. However, we add a collision term in the $R_N$ and $R_{\overline{N}}$ equations to account for $\phi\leftrightarrow N_IN_I$ processes. The evolution equations are modified as
\small
\be
\frac{dR_N}{dz} &=& -\frac{2Y_\phi^{\rm eq}\langle \Gamma_{\phi\rightarrow N_IN_I}\rangle}{zH(z)Y_N^{\rm eq}}\left(R_N^2-\mathbb{I}\right) + \mathrm{ARS},
\ee
\normalsize
where we use the thermally averaged width from Eq.~\eqref{eq:width_thermalavg}, and we include analogous extra terms for $R_{\overline N}$. There is, in principle, also a contribution to the oscillation terms from the thermal masses of $N_I$ induced by $\phi$, but since in Sec.~\ref{sec:phi_in_eq} we assumed flavor-universal couplings to $\phi$ this contributes a universal phase of no physical consequence. There is no modification to the  equation for the evolution of the lepton flavor asymmetries. \vspace{0.1cm}

\subsection{Full Boltzmann Equations for Hidden-Sector Equilibration}\label{app:be_full_hseq}
We begin with the number-density equations, which are derived using the standard collision term (\emph{e.g.}, Ref.~\cite{Kolb:1990vq}). The non-dimensionalized number-density Boltzmann equations are:
\begin{widetext}
\small
\be
\frac{dY_\phi}{dz} &=& {}-\frac{2s}{zH}\left[\langle\sigma(\phi\phi\rightarrow HH^*)v\rangle_{T_\phi}\, Y_\phi(t)^2 - \langle\sigma(\phi\phi\rightarrow HH^*)v\rangle_T \,Y_\phi^{\rm eq}(T)^2 \right] \\
&&{}-\frac{2}{zH}\sum_I\left[\langle\Gamma_{\phi\rightarrow N_IN_I}\rangle_{T_\phi}Y_\phi(t) - \langle\Gamma_{\phi\rightarrow N_IN_I}\rangle_{T_N}\,Y_\phi^{\rm eq}(T_N)\left(\frac{Y_{N_I}(t)}{Y_N^{\rm eq}(T_N)}\right)^2\right]\nonumber\\
&& {}-\frac{2s}{zH}\sum_I\left[\langle\sigma(\phi\phi\rightarrow \overline{N}_IN_I)v\rangle_{T_\phi}\,Y_\phi(t)^2-\langle\sigma(\phi\phi\rightarrow \overline{N}_IN_I)v\rangle_{T_N}\,Y_\phi^{\rm eq}(T_N)^2\left(\frac{Y_{N_I}(t)}{Y_N^{\rm eq}(T_N)}\right)^2\right]\nonumber,\\
\frac{dY_{N_I}}{dz} &=& \frac{2}{zH}\left[\langle\Gamma_{\phi\rightarrow N_IN_I}\rangle_{T_\phi}Y_\phi(t) - \langle\Gamma_{\phi\rightarrow N_IN_I}\rangle_{T_N}\,Y_\phi^{\rm eq}(T_N)\left(\frac{Y_{N_I}(t)}{Y_N^{\rm eq}(T_N)}\right)^2\right]\\
&&{}+\frac{s}{zH}\left[\langle\sigma(\phi\phi\rightarrow \overline{N}_IN_I)v\rangle_{T_\phi}\,Y_\phi(t)^2-\langle\sigma(\phi\phi\rightarrow \overline{N}_IN_I)v\rangle_{T_N}\,Y_\phi^{\rm eq}(T_N)^2\left(\frac{Y_{N_I}(t)}{Y_N^{\rm eq}(T_N)}\right)^2\right]\nonumber.
\ee
\normalsize
\end{widetext}
The thermally averaged rates in the above equation are, in the relativistic limit,
\small
\be
\langle\sigma(\phi\phi\rightarrow HH^*)v\rangle_T &=& \frac{\lambda^2}{128\pi T^2},
\ee
\be
\langle\sigma(\phi\phi\rightarrow \overline{N}_IN_I)v\rangle_T &=& \frac{1.50y^4}{64\pi T^2}\log\left(0.850\frac{T}{\overline{M}_\phi}\right),
\ee
\be
\langle\Gamma_{\phi\rightarrow N_I N_I}\rangle_T &=&  \frac{y^2\overline{M}_\phi^2}{64\pi T}.
\ee
\normalsize
Note that the $\phi\phi\rightarrow \overline{N}_IN_I$ total cross section has a $t$-channel singularity in the massless-$\phi$ limit, which gives rise to the observed logarithmic behavior in $\overline{M}_\phi/T$.

The energy-weighted Boltzmann equations have the form
\be
\dot{\rho} + 4H\rho &=& C_E.
\ee
The collision terms for the energy-weighted Boltzmann equation, $C_E$, have an additional factor of the energy for the relevant species appearing under the integral. For example, the collision term from the energy-weighted $\phi$ Boltzmann equation for the process $\phi\phi\rightarrow HH^*$  is (neglecting quantum statistical enhancement/blocking factors)
\begin{widetext}
\small
\be
C_E^{\phi\phi\rightarrow HH^*} &=& {}-\int\,d\Pi_{\phi_1}\,d\Pi_{\phi_2}\,d\Pi_H\,d\Pi_{H*}\,(2\pi)^4\delta^4\left(\sum p^\mu\right)\frac{1}{2}(E_{\phi_1}+E_{\phi_2})\left\langle|\mathcal{M}_{\phi_1\phi_2\rightarrow HH^*}|^2\right\rangle\left(f_{\phi_1}f_{\phi_2}-f_H^{\rm eq}f_{H^*}^{\rm eq}\right),
\ee
\normalsize
\end{widetext}
where the symmetry factor of $1/2$ accounts for the interchange of the $\phi$ momenta in the integral and we have written the integrand in a form that makes the $\phi_1\leftrightarrow\phi_2$ symmetry manifest. Similarly, for the elastic scattering process $\phi H\rightarrow \phi H$, the collision term is
\begin{widetext}
\small
\be
C_E^{\phi H\rightarrow \phi H} &=& {}-\int\,d\Pi_{\phi_1}\,d\Pi_{\phi_2}\,d\Pi_{H_1}\,d\Pi_{H_2}\,(2\pi)^4\delta^4\left(\sum p^\mu\right)\frac{1}{2}(E_{\phi_1}-E_{\phi_2})\left\langle|\mathcal{M}_{\phi_1 H_1\rightarrow \phi_2 H_2}|^2\right\rangle\left(f_{\phi_1}f^{\rm eq}_{H_1}-f_{\phi_2}f_{H_2}^{\rm eq}\right),
\ee
\normalsize
\end{widetext}
where the factor of $1/2$ accounts for a symmetry where we simultaneously interchange $\phi_1\leftrightarrow\phi_2$ and $H_1\leftrightarrow H_2$ which results from the fact that a single collision term includes both forward and reverse processes (again, we write it in the form where this symmetry is manifest in the integrand). The other collision terms for $\phi\rightarrow N_I N_I$, $\phi\phi\rightarrow \overline{N}_IN_I$, and $\phi N_I\rightarrow \phi N_I$ can be determined in an analogous fashion. In the above expressions, 
\be
d\Pi_X &=& \frac{g_X\,d^3p_X}{(2\pi)^32E_X}
\ee
is the Lorentz-invariant phase space for species $X$, $g_X$ is the number of degrees of freedom for $X$, and the squared matrix element is averaged over all  spins and $\mathrm{SU}(2)$ charges for initial and final states.

We wish to express our energy-weighted Boltzmann equations in terms of the evolution of the dimensionless quantities $Y_\phi$, $Y_N$, $w\equiv T_\phi/T$, and $u\equiv T_N/T$, all of which are invariant under Hubble expansion for constant $g_*$. Using our ansatz Eq.~\eqref{eq:noeq_ansatz}, we find that
\be
\rho_\phi(t) = \frac{n_\phi(t)}{n_\phi^{\rm eq}(T_\phi)}\rho_\phi^{\rm eq}(T_\phi) = 3T_\phi n_\phi(t)
\ee
in the relativistic limit and for Maxwell-Boltzmann statistics. This allows us to re-write the change in energy density as
\be
\dot{\rho}_\phi + 4H\rho_\phi &=& 3sT_{\rm ew}H\left(Y_\phi \frac{dw}{dz}+w\frac{dY_\phi}{dz}\right).
\ee
This form manifestly shows that, when the collision terms vanish, the equilibrium configuration $dw/dz=dY_\phi/dz=0$ is a valid solution to the energy-weighted Boltzmann equations. 

Having evaluated all of the collision integrals, we can now write the full, dimensionless form of the energy-weighted Boltzmann equation for $\phi$:
\begin{widetext}
\small
\be
Y_\phi\frac{dw}{dz}+w\frac{dY_\phi}{dz} &=& {}-\frac{s}{3T_{\rm ew}H}\left[\langle\sigma(\phi\phi\rightarrow HH^*)vE_\phi\rangle_{T_\phi}\,Y_\phi(t)^2 - \langle\sigma(\phi\phi\rightarrow HH^*)vE_\phi\rangle_{T}\,Y_\phi^{\rm eq}(T)^2\right]\\
&&{}-\frac{sY_H^{\rm eq}(T)Y_\phi(t)}{3T_{\rm ew}H}\langle\sigma(\phi H\rightarrow\phi H)vE_\phi\rangle_{T_\phi}\left(w-1\right)\nonumber\\
&& {}-\frac{2\overline{M}_\phi}{3T_{\rm ew}H}\sum_I\,\Gamma_{\phi\rightarrow N_IN_I}\left[Y_\phi(t) - Y_\phi^{\rm eq}(T_N)\left(\frac{Y_{N_I}(t)}{Y_N^{\rm eq}(T_N)}\right)^2\right]\nonumber\\
&&{}-\frac{s}{3T_{\rm ew}H}\sum_I\left[\langle\sigma(\phi\phi\rightarrow \overline{N}_IN_I)vE_\phi\rangle_{T_\phi}\,Y_\phi(t)^2 - \langle\sigma(\phi\phi\rightarrow \overline{N}_IN_I)vE_\phi\rangle_{T_N}\,Y_\phi^{\rm eq}(T_N)^2\left(\frac{Y_{N_I}(t)}{Y_N^{\rm eq}(T_N)}\right)^2\right]\nonumber\\
&&{}-\frac{2sY_\phi(t)}{9T_{\rm ew}H}\sum_I\,Y_{N_I}(t)\langle\sigma(\phi N_I\rightarrow \phi N_I)vE_\phi\rangle_{T_\phi}\left(\frac{w}{u}-1\right).\nonumber
\ee
\normalsize
\end{widetext}

In the relativistic limit, the rates in the above equation are given by
\small
\be
\langle\sigma(\phi\phi\rightarrow HH^*)vE_\phi\rangle_T &=& \frac{\lambda^2}{32\pi T},
\ee
\be
\langle\sigma(\phi H\rightarrow\phi H)vE_\phi\rangle_T &=& \frac{\lambda^2}{64\pi T},
\ee
\small
\be
\langle\sigma(\phi\phi\rightarrow \overline{N}_IN_I)vE_\phi\rangle_T &=& \frac{2.98y^4}{32\pi T}\log\left(\frac{1.09T}{\overline{M}_\phi}\right),
\ee
\be
\langle\sigma(\phi N_I\rightarrow \phi N_I)vE_\phi\rangle_T &=& \frac{y^4}{256\pi T}.
\ee
\normalsize

The corresponding energy-weighted Boltzmann equation for $N$ has a similar form:
\begin{widetext}
\small
\be
Y_{N_I}\frac{du}{dz}+u\frac{dY_{N_I}}{dz} &=& \frac{\overline{M}_\phi\Gamma_{\phi\rightarrow N_IN_I}}{3T_{\rm ew}H}\left[Y_\phi(t) - Y_\phi^{\rm eq}(T_N)\left(\frac{Y_{N_I}(t)}{Y_N^{\rm eq}(T_N)}\right)^2\right] \\
&&{}+\frac{s}{6T_{\rm ew}H}\left[\langle\sigma(\phi\phi\rightarrow \overline{N}_IN_I)vE_\phi\rangle_{T_\phi}\,Y_\phi(t)^2 - \langle\sigma(\phi\phi\rightarrow \overline{N}_IN_I)vE_\phi\rangle_{T_N}\,Y_\phi^{\rm eq}(T_N)^2\left(\frac{Y_{N_I}(t)}{Y_N^{\rm eq}(T_N)}\right)^2\right]\nonumber\\
&&{}+\frac{sY_\phi(t)Y_{N_I}(t)}{9T_{\rm ew}H}\langle\sigma(\phi N_I\rightarrow \phi N_I)vE_\phi\rangle_{T_\phi}\left(\frac{w}{u}-1\right).\nonumber
\ee
\normalsize
\end{widetext}

\subsection{Quantum Kinetic Equations Including Hidden-Sector Equilibration}\label{app:qke_full_hseq}
When interfacing the above Boltzmann equations with the quantum kinetic equations for leptogenesis, we have to deal with the fact that (prior to full equilibration) there exists a population of RHNs produced through hidden-sector interactions at temperature $T_N$, and another population produced from SM Higgs decay and scattering at temperature $T$.  We denote the former abundance by $Y_{\tilde{N}}$ and the latter abundance by $Y_N=Y_N^{\rm eq}(T)R_N$, and we denote the geometric mean temperature by $\overline{T}\equiv \sqrt{TT_N}$. Because $CP$ is conserved in the hidden sector, we can assume that the RHN and anti-RHN abundances are the same in the $\tilde{N}$ sector.  The ARS quantum kinetic equations for $R_N$ and $\mu_{\Delta \alpha}$ become
\begin{widetext}
\small
\be
\frac{dR_N}{dz} &=& i\left[R_N,W_N\right] + 3iz^2\left[R_N,r\right] - \mathcal{C}^{(0)}\left\{R_N +\frac{Y_{\tilde{N}}}{uY_N^{\rm eq}(T)}\mathbb{I},W_N\right\} + 2\mathcal{C}^{(0)}W_N + \mathcal{C}^{(\mathrm{w.o.1})}o_\mu \label{eq:full_eq_YN_ARS}\\
&&{}+ \frac{1}{2}\mathcal{C}^{(\mathrm{w.o.2})}\left\{o_\mu,R_N+\frac{Y_{\tilde{N}}}{uY_N^{\rm eq}(T)}\mathbb{I}\right\}-\frac{2}{zH}\langle\Gamma_{\phi\rightarrow N_I N_I}\rangle_{\overline{T}}\,\frac{Y_\phi^{\rm eq}(\overline{T})}{Y_N^{\rm eq}(\overline{T})^2}\,Y_{\tilde{N}} R_N\nonumber\\
&&{}-\frac{s}{zH}\langle\sigma(\phi\phi\rightarrow N_I \overline{N}_I)v\rangle_{\overline{T}}\,\frac{Y_\phi^{\rm eq}(\overline{T})^2}{Y_N^{\rm eq}(\overline{T})^2}\,Y_{\tilde{N}}R_N,\nonumber\\
\frac{32T_{\rm ew}}{M_0}\frac{d\mu_{\Delta\alpha}}{dz} &=& {}-\mathcal{C}^{(0)}\left(FR_NF^\dagger-F^*R_{\overline{N}}F^{\rm T}\right)_{\alpha\alpha}
+\mathcal{C}^{(\rm w.o.1)}\left(FF^\dagger\right)_{\alpha\alpha}\mu_\alpha\label{eq:full_eq_asymmetry_ARS}\\
&&{}+\frac{\mathcal{C}^{(\rm w.o.2)}}{2}\left(FR_NF^\dagger+F^*R_{\overline N}F^{\rm T}+\frac{2Y_{\tilde N}}{u Y_N^{\rm eq}(T)}FF^\dagger\right)_{\alpha\alpha}\mu_\alpha.\nonumber
\ee
\normalsize
\end{widetext}
As before, the $R_{\overline{N}}$ evolution equation is the same as for $R_N$, but with $F\rightarrow F^*$ and $\mu\rightarrow -\mu$.
Because we are assuming that the hidden sector is dominantly heated through $\phi$  and not directly through SM Higgs interactions with $N$, the influence of the couplings $F$ between the SM Higgs and the RHNs is negligible in determining the evolution of the hidden sector abundances and temperature. We can then plug in the solutions to the hidden-sector Boltzmann equations into Eqs.~\eqref{eq:full_eq_YN_ARS}--\eqref{eq:full_eq_asymmetry_ARS} to determine the impact on leptogenesis.

\section{Momentum Averaging and Neutrino Oscillations}\label{app:momentum_averaging}

A significant approximation underlying our quantum kinetic equations is the use of momentum averaging. In other words, we have assumed that the density matrix is $Y_N(k,t)_{IJ} = (R_N)_{IJ} Y_N^{\rm eq}(T) f_N(k,T)$, where $f_N$ is the Maxwell-Boltzmann distribution if using classical statistics or the Fermi-Dirac distribution if using quantum statistics, and in either case encapsulates the full dependence on $k$. We can then integrate over $k$ to obtain a set of momentum-averaged differential equations. In practice, this amounts to using $\int \,dt\,\langle E_2-E_1\rangle_T$ as the oscillation phase. 

We now assess the validity of this approximation for our study of the suppression of asymmetry from RHN equilibration. One concern is that our perturbative treatment in Sec.~\ref{sec:newinteractions} uses the same momentum-averaging procedure as the quantum kinetic equations and consequently there is a single oscillation time, $z_{\rm osc}$, for the entire population of RHNs. In reality, however, there is a separate oscillation time for each momentum mode given by
\be
z_{\rm osc}(q) &=& \left(\frac{6qT_{\rm ew}^3}{\Delta M_{21}^2M_0}\right)^{1/3},
\ee
where $q\equiv k/T$ is the co-moving RHN momentum. The thermal-averaging procedure replaces $\langle 1/q\rangle\rightarrow 1/2$ predicted by Maxwell-Boltzmann statistics in the oscillation phase to obtain the momentum-averaged $\langle z_{\rm osc}\rangle$ given in Eq.~\eqref{eq:zosc_averaged}.

The momentum-dependent $z_{\rm osc}(q)$ complicates our earlier prediction that the asymmetry is suppressed provided $z_{\rm osc}\gtrsim z_{\rm eq}$. Now, we see that \emph{for every mass splitting and equilibration time}, there is a population of RHNs that complete one oscillation and experiences no such suppression, while the remainder of the RHN population has its contribution to the asymmetry suppressed by equilibration\footnote{We note that each momentum mode will also equilibrate at a slightly different time, although we neglect this effect in the illustrative calculation to follow. We expect that the scattering rate for a small momentum mode to be larger than one with $k\sim T$, so if anything our assumption of equal equilibration times will slightly exaggerate the effects of thermal averaging of the oscillation phase.}. A more correct calculation of the asymmetry would compute the contribution to the asymmetry of each momentum mode, and then perform the sum over momenta weighted by the RHN momentum distribution.

It is difficult to   solve the full momentum-dependent quantum kinetic equations for the baryon asymmetry. We can, however, straightforwardly compute the momentum-dependent asymmetry perturbatively to estimate the inaccuracies of our momentum-averaging procedure. We follow Ref.~\cite{Shuve:2020evk}, which computed the correct momentum-averaged asymmetry for the case where a massive scalar decays into oscillating singlets, and we replace the dominant tree-level scalar mass in that case with the temperature-dependent SM Higgs mass ($m_H^2 \equiv \kappa T^2$, where $\kappa\approx0.39$ is determined from SM couplings \cite{Hambye:2017elz}) relevant for ARS leptogenesis. The asymmetry factor $\mathcal{A}(z)$, correctly averaged over momentum and accounting for the non-thermal momentum spectrum of RHNs produced from SM Higgs decays, is
\be
\mathcal{A}^{\rm (full)}(z_{\rm eq}) &=& \frac{\sqrt{\kappa}}{2K_1(2\sqrt{\kappa})}\int_0^\infty\,dq\,\frac{e^{-q-\kappa/q}}{q^2}\\
&&{}\int_0^{z_{\rm eq}}\,dz_2\,\int_0^{z_2}\,dz_1\,\sin\left[\frac{z_2^3-z_1^3}{z_{\rm osc}(q)^3}\right].
\ee
If we replace $z_{\rm osc}(q)$ with the momentum-independent version from Sec.~\ref{sec:newinteractions}, the $q$-integral gives 1 and we recover the earlier expression for $\mathcal{A}(z_{\rm eq})$, Eq.~\eqref{eq:osc_ARS}.

\begin{figure}[t]
        \includegraphics[width=\columnwidth]{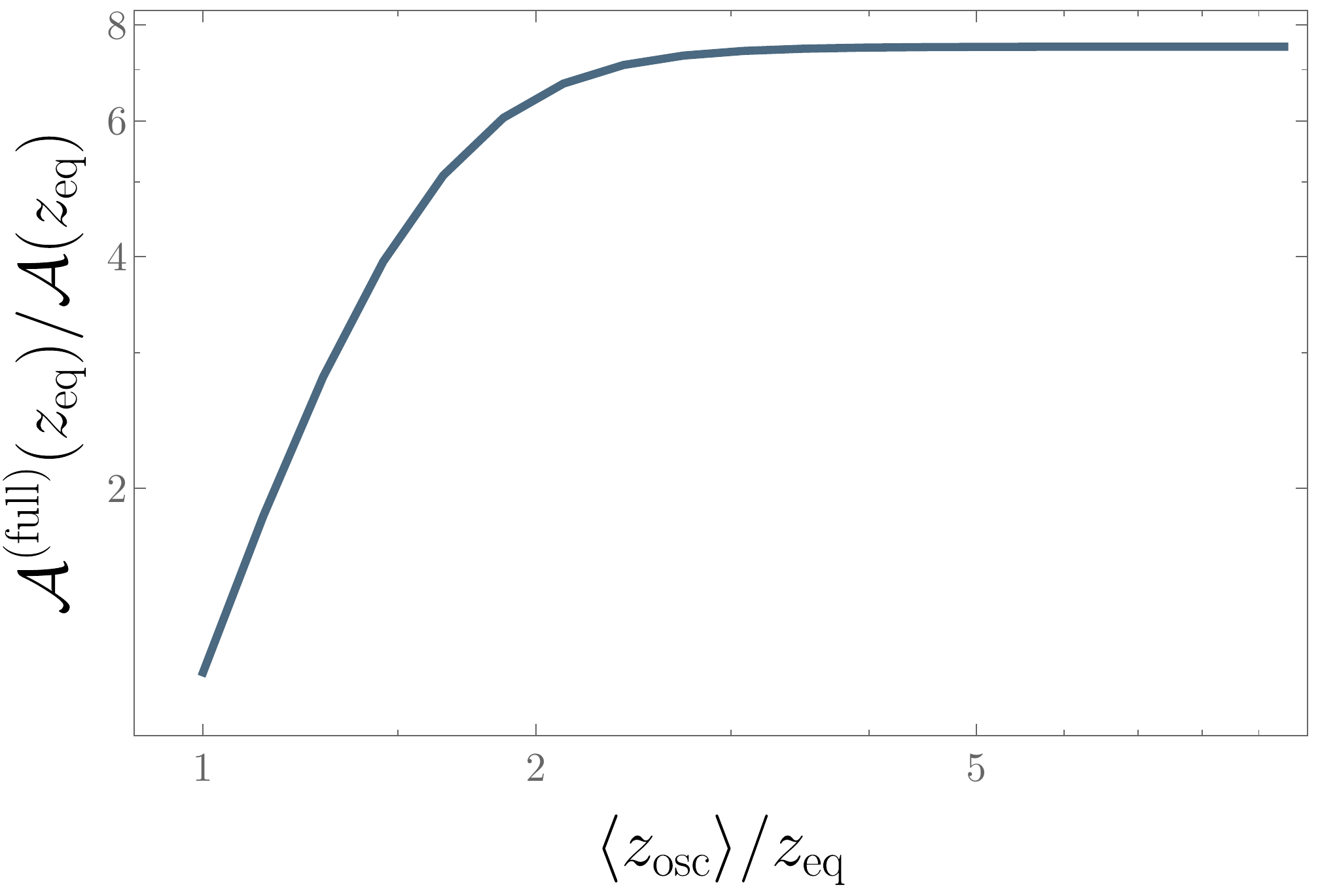}
                     \caption{
Ratio of the asymmetry factor with correct momentum-averaged asymmetry, $\mathcal{A}^{\rm (full)}(z)$, compared to the  
corresponding factor $\mathcal{A}(z)$ from Eq.~\eqref{eq:perturbative_asymmetry} with na\"ive momentum averaging of the oscillation phase. We plot
this ratio as a function of $\langle z_{\rm osc}\rangle/z_{\rm eq}$ for $z_{\rm eq}=0.01$, although the curve looks identical for other values of $z_{\rm eq}$. We find that
in the limit where oscillations occur after equilibration, the correct asymmetry is a factor of 7.5 larger than  the na\"ive averaging prediction, whereas the two methods agree within 15\% for $\langle z_{\rm osc}\rangle = z_{\rm eq}$. }
    \label{thermal_averaging}
\end{figure}

We compute the ratio of $\mathcal{A}^{\rm (full)}(z_{\rm eq})$ to that of Eq.~\eqref{eq:osc_ARS}, which was derived using the na\"ive averaging of the oscillation phase. We show our results in Fig.~\ref{thermal_averaging}. We find that for $\langle z_{\rm osc}\rangle= z_{\rm eq}$, which we found gave the optimal baryon asymmetry, the  asymmetries from the two methods agree within 15\%. For the regime of significant asymmetry suppression, $\langle z_{\rm osc}\rangle \gg z_{\rm eq}$, we find that the na\"ive momentum-averaging of the phase under-estimates the true asymmetry by a factor of 7.5. However, in this regime the asymmetry scales like $y^{-10}$, and so this change in the asymmetry only shifts the value of the coupling leading to a specified asymmetry suppression by about 20\%. Thus, even though our primary results in the paper use the na\"ive momentum averaging of the phase, our results for the coupling magnitudes needed for successful leptogenesis still hold both qualitatively and quantitatively up to 20\% differences. The full implementation and solution of the momentum-dependent quantum kinetic equations is left for future work.


 \section{Thermal-Mass Effects in Oscillations} \label{sec:thermal_masses}
  
  In complete models, we might expect the RHN masses to originate from spontaneous symmetry breaking of lepton number after $\phi$ gets a VEV. We restrict ourselves to models of spontaneous breaking of a discrete symmetry so that we don't need to consider the additional effect of low-mass Goldstone bosons, which would presumably further accelerate the RHN equilibration process.
  
  Unless $\phi$ is highly decoupled from the SM, we might expect that thermal contributions to the $\phi$ potential will lead to a restoration of  lepton number symmetry at high temperatures. If this is the case,  then RHNs will not have tree-level masses prior to the lepton-number-breaking phase transition, and consequently oscillations induced by the tree-level masses only occur after the phase transition. If the RHN equilibration time $z_{\rm eq}$ occurs after the phase transition, this is largely irrelevant because the oscillation phase goes like $z^3$ and is dominated by the latest times immediately prior to equilibration. If equilibration happens \emph{before} the phase transition, however, then our parametric estimates assuming that RHNs have their zero-temperature masses are incorrect.
  
If the RHNs have vanishing tree-level masses, their Hamiltonians are dominated by finite-temperature effects. In particular, the RHNs acquire an effective potential through interactions with the SM Higgs as well as with $\phi$. The former cannot lead to the generation of an asymmetry since the effective potential is aligned with the interaction basis, and consequently there is no interference of propagating energy eigenstates. The interactions with $\phi$, however, are presumably aligned with the RHN zero-temperature mass basis and the resulting finite-temperature potential can lead to oscillations. Following the methods of Ref.~\cite{Weldon:1982bn}, and assuming for concreteness that $\phi$ is in equilibrium but $N_I$ are not, we have computed the effective potential for $N_I$, finding
\be
V^{\rm eff}_{I} &=& \frac{y_I^2T^2}{24k}
\ee
for $k\gg V_I^{\rm eff}$ and where we have disregarded a $y_I$-independent momentum term. Defining $\Delta y^2 \equiv y_2^2 - y_1^2$, we can now write the oscillation phase as
\be
\sin\left[\int_{t_1}^{t_2}\,dt\left(E_2-E_1\right)\right] &=& \sin\left(\int_{t_1}^{t_2}\,dt\,\frac{\Delta y^2 T^2}{24k}\right)\\
&=& \sin\left[\frac{\Delta y^2 M_0}{24T_{\rm ew}(k/T)}\left(z_2-z_1\right)\right].\nonumber
\ee
For Maxwell-Boltzmann statistics, $\langle T/k\rangle = 1/2$, and hence thermally averaging the phase gives
\be
\sin\left[\int_{t_1}^{t_2}\,dt\left\langle E_2-E_1\right\rangle\right] &=& \frac{\Delta y^2 M_0}{48T_{\rm ew}}(z_2-z_1).
\ee
Assuming $z_2\gg z_1$, the dimensionless oscillation time for which this phase equals unity is now
\be
z_{\rm osc} &=& \frac{48T_{\rm ew}}{\Delta y^2 M_0}.
\ee

Following the perturbative calculations of Sec.~\ref{sec:asym_suppr} and Refs.~\cite{Asaka:2005pn,Hambye:2017elz,Shuve:2020evk}, the asymmetry is proportional to a factor
\be
\mathcal{A}(z) &=& \int_0^z\,dz_2\int_0^{z_2}\,dz_1\,\sin\left[\frac{z_2-z_1}{z_{\rm osc}}\right],
\ee
which accounts for integrating over the collision terms dressed by the oscillation phase. As in Sec.~\ref{sec:asym_suppr}, we assume that the RHNs come into equilibrium at a time
\be
z_{\rm eq} = \frac{T_{\rm ew}}{a_N y^2 M_0},
\ee
where $y$ is the coupling bringing the RHNs into equilibrium and $a_N$ is a dimensionless prefactor. If $z_{\rm eq} < z_{\rm osc}$, then equilibration occurs before oscillations begin and the asymmetry is suppressed. We can estimate the asymmetry suppression by cutting the integrals off at $z_{\rm eq}$ and assuming a small oscillation phase,
\be
\mathcal{A}(z_{\rm eq}) &= & \int_0^{z_{\rm eq}}\,dz_2\int_0^{z_2}\,dz_1\,\sin\left[\frac{z_2-z_1}{z_{\rm osc}}\right]\\
&\approx& \frac{\Delta y^2 T_{\rm ew}^2}{288 a_N^3 y^6 M_0^2}.
\ee
If we compare to Eq.~\eqref{eq:ARS_cutoff}, it seems at face value that the situation has improved:~the asymmetry suppression ``only'' scales inversely with the sixth power of the coupling compared to the tenth power with a tree-level mass splitting! The asymmetry also appears to be less suppressed by inverse powers of $M_0$. 

However, we now see that the optimal asymmetry is essentially unchanged from before. The largest asymmetry occurs if $z_{\rm osc}\approx z_{\rm eq}$. For a given value of $y$, this allows us to solve for the squared difference in couplings $\Delta y^2$ in terms of other parameters. The resulting optimized asymmetry factor is
\be
\mathcal{A}(z_{\rm eq})^{\rm (optimized)} &=& \frac{T_{\rm ew}^2}{6 a_N^2M_0^2 y^4}.
\ee
Comparing with our earlier perturbative result, Eq.~\eqref{eq:optimized_asymmetry}, we find \emph{the same optimized asymmetry} as before (up to a pre-factor that differs by 10\%). In particular, the $y^{-4}$ suppression of the asymmetry is the same even if the difference in RHN energies originates from thermal effects rather than tree-level masses.

The physical reason for this result is that the largest possible value of the sine of the oscillation phase is 1, whereas the integration of the collision terms that determines the magnitude of the asymmetry is determined by the Hubble expansion rate at the equilibration time. In other words, the parameters of any theory can always be adjusted to give the optimal oscillation phase, but the magnitudes of the integrals over the production and annihilation times of $z$ are restricted by $H(z_{\rm eq})$, giving rise to the particular relations for the optimal asymmetry found above. This strongly suggests that the results we derived assuming non-zero tree-level masses of the RHNs should carry over  to arbitrary finite-temperature mass corrections.

  \section{Freeze-out Leptogenesis} \label{sec:freezeout}
Throughout the manuscript so far, we have focused on freeze-in leptogenesis, which occurs on the approach of the RHN distributions to equilibrium. There is, however, also a contribution to the asymmetry during the process of freeze out or, in other words, the departure of the RHN distribution from equilibrium at low temperatures. Freeze out is less sensitive to other interactions than freeze in:~indeed, in conventional thermal leptogenesis, an asymmetry is generated by the decays of RHNs for $T\sim M_N$ even if the RHNs start out with a thermal abundance.

For GeV-scale RHNs, the decays of non-relativistic RHNs do not contribute to the baryon asymmetry because they occur after the electroweak phase transition. However, the equilibrium RHN abundance still changes due to finite-mass effects even when highly relativistic, with
\be
\frac{dY_N^{\rm eq}}{dz} &\approx& -\frac{45}{4\pi^4 g_{*S}}\left(\frac{M_N}{T_{\rm ew}}\right)^2 z 
\ee
for Maxwell-Boltzmann statistics and taking $T\gg M_N$. Even if the RHNs are kept ``in equilibrium'' by some interaction, there is a small deviation from equilibrium that results from the non-zero value of $dY_N^{\rm eq}/dz$. The importance of freeze-out leptogenesis for GeV-scale RHNs has been emphasized and comprehensively studied in several recent works, which  established a continuity between what had until recently been considered distinct regimes of resonant freeze-out leptogenesis and ARS freeze-in leptogenesis \cite{Hernandez:2016kel,Hambye:2016sby,Antusch:2017pkq,Hambye:2017elz,Granelli:2020ysj,Klaric:2021tdt,Drewes:2021nqr}.

We consider the same scenario as in the rest of the paper with the RHNs coupling to a dark scalar, $\phi$. For the purpose of the current argument, we assume that $\phi$ is in equilibrium with the SM although our results can be extended to the more complicated case using the methods of Sec.~\ref{sec:phi_not_in_eq}. Assuming that the RHNs are predominantly produced through the interactions with $\phi$, the leading term in the Boltzmann equation for $N$ is
\be
\frac{dY_N}{dz} &=& -\frac{2\langle\Gamma_{\phi\rightarrow NN}\rangle Y_\phi^{\rm eq}}{zH (Y_N^{\rm eq})^2}\left[Y_N^2-(Y_N^{\rm eq})^2\right].
\ee
When the RHNs are close to equilibrium, this becomes
\be
\frac{dY_N}{dz} &\approx& -\frac{4\langle\Gamma_{\phi\rightarrow NN}\rangle Y_\phi^{\rm eq}}{zH Y_N^{\rm eq}}\left(Y_N-Y_N^{\rm eq}\right).
\ee
In this limit, $dY_N/dz \approx dY_N^{\rm eq}/dz$, and we get that the deviation of the RHN abundance from equilibrium is
\be\label{eq:departure_from_eq}
Y_N-Y_N^{\rm eq} &\approx& \frac{45 HY_N^{\rm eq}z^2}{16\pi^4g_{*S}\langle\Gamma_{\phi\rightarrow NN}\rangle Y_\phi^{\rm eq}}\left(\frac{M_N}{T_{\rm ew}}\right)^2.
\ee
Because $\langle\Gamma_{\phi\rightarrow NN}\rangle \propto y^2$, we see that the deviation from equilibrium scales like $y^{-2}$:~the stronger the hidden-sector forces, the closer the RHN abundance is to equilibrium.

What are the implications for leptogenesis? Since the $CP$-violating source is proportional to $Y_N-Y_N^{\rm eq}$ \cite{Weinberg:1979bt}, this means that a stronger coupling within the hidden sector will quadratically suppress the lepton flavor asymmetries produced from freeze-out leptogenesis. In the absence of the coupling to $\phi$, the RHNs are kept in equilibrium by the much smaller coupling $F$ to the SM Higgs, and so the expected asymmetry suppression is $\sim F^2 / y^2$. Given that typical values of $F$ are in the vicinity of $10^{-7}$, then the lepton asymmetry from freeze-out leptogenesis is suppressed by many orders of magnitude even for quite small hidden-sector couplings.

There is another source of suppression:~the lepton asymmetry source term depends on the off-diagonal components of the RHN density matrix, which are exponentially damped by scattering with $\phi$. Much like in Sec.~\ref{sec:phi_neq_lepto_results}, the net production of off-diagonal components of the RHN density matrix from SM Higgs decays and scattering offsets the destruction from scattering into $\phi$, leading to a quasi-steady-state where the off-diagonal components of $R_N$ are further suppressed by powers of the coupling $y$. The combination of these two effects leads to a severe suppression of the asymmetry when RHN interactions with $\phi$ are in equilibrium at a low scale, even though the asymmetry source is different from what we considered in freeze-in leptogenesis.

To quantify these effects, we perform a  numerical study for some benchmark points for which freeze-out leptogenesis gives rise to a viable baryon asymmetry in the minimal model. By turning on the coupling $y$ to the scalar $\phi$, we determine the suppression of the asymmetry as a function of $y$. The asymmetry arising from freeze-out leptogenesis can be isolated by assuming an initial condition of $(R_N)_{IJ}=\delta_{IJ}$, which eliminates any freeze-in contribution. This is also a reasonable initial condition in the case that the RHNs equilibrate with a hidden sector at some temperature $T\gg T_{\rm ew}$. In Ref.~\cite{Klaric:2021tdt}, it was found that the observed baryon asymmetry can be achieved in the $\nu$MSM for $M_N\gtrsim10$ GeV with two RHNs, while in Ref.~\cite{Drewes:2021nqr} it was found that the freeze-out contribution could account for the observed baryon asymmetry for RHN masses as low as 3 GeV. The departure from equilibrium in Eq.~\eqref{eq:departure_from_eq} is more pronounced at larger RHN masses, leading to a viable baryon asymmetry at larger RHN masses. As with resonant leptogenesis, the asymmetry is maximized for small mass splittings $\Delta M$ such that the oscillation time is comparable to the Hubble time at the electroweak phase transition.

To take into account the processes contributing to freeze-out leptogenesis, we need to include lepton-number-violating (LNV) collisions in our quantum kinetic equations; we use the LNV rates from Ref.~\cite{Abada:2018oly}. Additionally, the quantum kinetic equations Eq.~\eqref{eq:qke_for_lepto} were derived assuming that $Y_N^{\rm eq}$ is a constant, but we must take into account the fact that $dY_N^{\rm eq}/dt\neq0$ to obtain the departure from equilibrium that drives freeze-out leptogenesis. This can be readily accommodated by replacing 
\be
\frac{dR_N}{dt}&\rightarrow& \frac{dR_N}{dt}+ \frac{R_N}{Y_N^{\rm eq}}\,\frac{dY_N^{\rm eq}}{dt}
\ee
 on the left-hand side of the quantum kinetic equations. We have checked that, with an initial condition $(R_N)_{IJ}=\delta_{IJ}$, we obtain a non-zero baryon asymmetry from freeze out if we make the modifications described here, but get zero baryon asymmetry if we use the original form of the quantum kinetic equations from Eq.~\eqref{eq:qke_for_lepto}.

\begin{figure}[t]
        \includegraphics[width=\columnwidth]{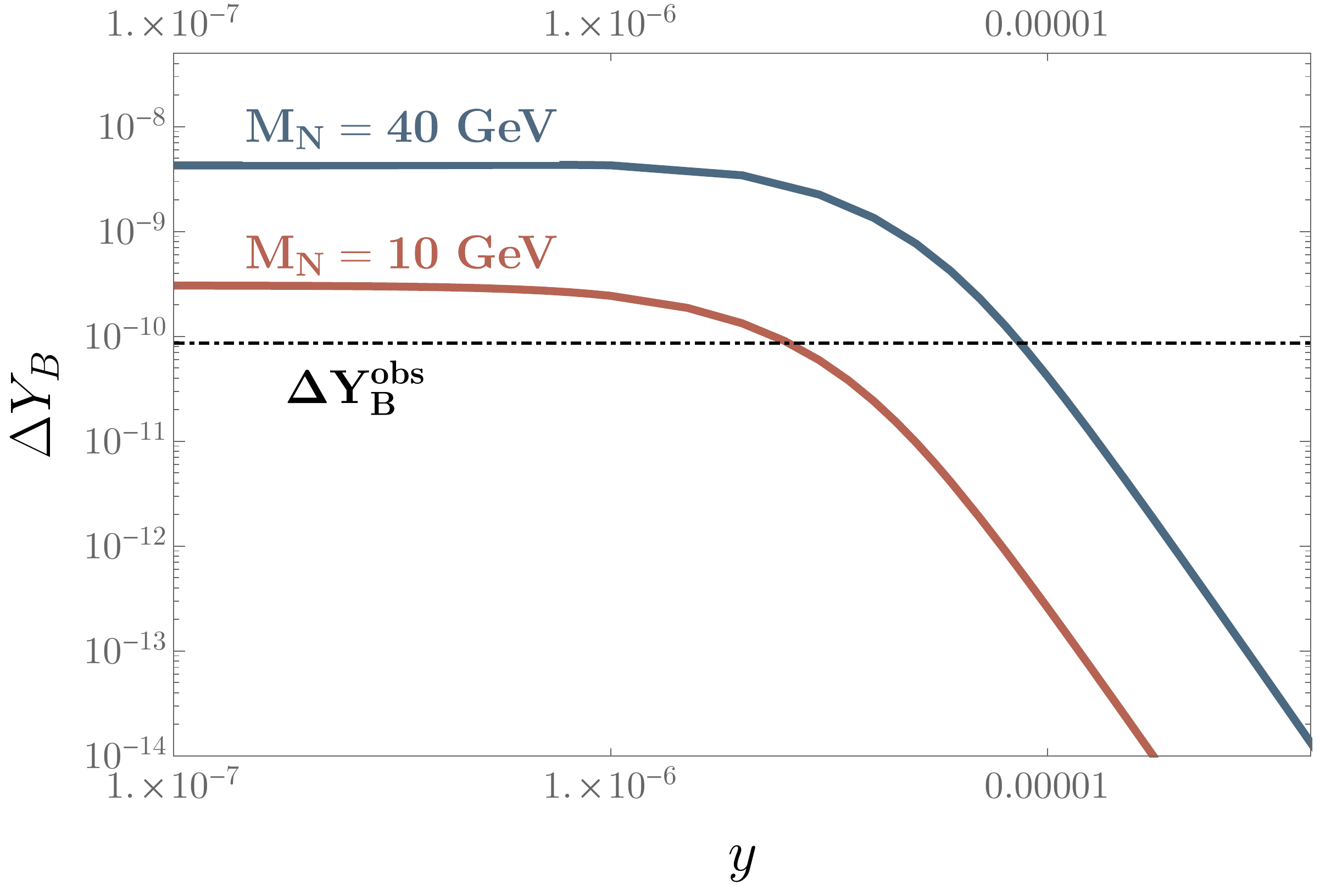}
                     \caption{
Baryon asymmetry from freeze-out leptogenesis as a function of $y$ with initial condition $(R_N)_{IJ}(0)=\delta_{IJ}$. We fix $\lambda=0.1$, $M_\phi=10$ GeV, and $\Delta M_N=2\times10^{-11}$ GeV, and the Yukawa coupling $F^{(I)}$ from Eq.~\eqref{eq:first_yukawa}. }
    \label{asymmetry_freezeout}
\end{figure}

We fix the Yukawa coupling texture to approximately that of $F^{(I)}$ in Eq.~\eqref{eq:first_yukawa}, although the overall scale of the Yukawa couplings is determined as a function of $M_N$ according to the Casas-Ibarra parametrization. Furthermore, we take $\mathrm{Im}\,\omega\approx 0.7$, which optimizes the asymmetry for large $M_N$. Fixing $\lambda=0.1$, $M_\phi=10$ GeV, and $\Delta M=2\times10^{-11}$ GeV (which is close to the optimal value), we compute the freeze-out baryon asymmetry as a function of $y$ for two choices of RHN mass:~$M_N=10$ GeV and $M_N=40$ GeV. We show our results in Fig.~\ref{asymmetry_freezeout}. It is evident that for tiny values of $y$ we obtain a viable baryon asymmetry through the freeze-out mechanism for both masses. However, for $y\gtrsim10^{-6}$ there is a sharp fall off in the asymmetry, with an approximate $y^{-5.8}$ power-law dependence due to a combination of suppressed deviation from equilibrium and damping of the off-diagonal elements of the RHN density matrix. This is less severe than the asymmetry suppression of freeze-in leptogenesis but not by much, and is still of sufficient magnitude as to render baryogenesis non-viable for $y\gtrsim10^{-5}$ depending on the precise value of $M_N$. Unlike for freeze-in, the asymmetry cannot be substantially enhanced by varying $\Delta M$.


The effects of the hidden sector can be substantially mitigated if $ M_\phi\gg T_{\rm ew}$. For freeze in, any interactions that bring the RHNs into equilibrium over the entire cosmic history prior to asymmetry generation greatly suppresses the asymmetry, and thus $M_\phi$ must be very heavy to give a viable lepton asymmetry (as seen in Fig.~\ref{asymmetry_mass_dependence}). In freeze-out leptogenesis, however, the bulk of the asymmetry is generated close to $T_{\rm ew}$, and as long as the interactions are out of equilibrium at this time the asymmetry is not suppressed. This is seen in Eq.~\eqref{eq:departure_from_eq} from the fact that the deviation from equilibrium of the RHN abundance is inversely proportional to $Y_\phi^{\rm eq}\sim e^{-M_\phi/T}$ for $T\ll M_\phi$, and hence we can get a large departure from equilibrium by taking $T_{\rm ew}\ll M_\phi$. In quantitative terms, we find that if $M_\phi$ is larger than about 10 TeV,  freeze-out baryogenesis can occur for essentially any perturbative value of $y$, while having viable baryogenesis with $y\sim10^{-5}$ requires $M_\phi\gtrsim3$ TeV. As a result, making $\phi$ heavy provides a more substantial loophole for avoiding asymmetry suppression in freeze-out baryogenesis, although if this is the case $\phi$ is likely not to be within kinematic reach of existing or near-future experiments.

Finally, we remark that our study of freeze-out leptogenesis suggests that the constraints from leptogenesis on the hidden-sector couplings are comparable to those from our study of freeze-in leptogenesis. However, we have not performed a comprehensive study, in part because the asymmetry in freeze-out leptogenesis is dominantly produced during the electroweak crossover and consequently details of rates and sphaleron decoupling in the broken phase become important \cite{Klaric:2021tdt}. We do not expect those refinements to dramatically change the range of allowed couplings, but the question merits a dedicated study that is beyond the scope of the current work and its focus on freeze-in leptogenesis.

\bibliography{biblio}

\end{document}